

\def\ssvsk{\vskip 0.3cm}
\def\svsk{\vskip 0.5cm}

\def\sssk{\hskip 1pt}
\def\ssk{\hskip 3pt}
\def\msk{\hskip 20pt}
\def\lsk{\hskip 0.5cm}

\def\para{\par\noindent}

\def\rea{{\rm I\!\! R}}

\def\noi{\noindent}
\def\t#1{\tilde #1}
\def\wt#1{\widetilde #1}
\def\ca#1{{\cal#1}}
\def\cid{_{\circ}}
\def\bud{_{\bu}}
\def\ci{\circ}
\def\bu{\bullet}

\def\vfi#1{\varphi_{#1}}

\def\K{K\"ahler~}
\def\simod{$\sigma$-model~}
\def \first{$1^{st}\sssk$}
\def\second{$2^{nd}\sssk$}
\def\se{stress-energy tensor~}
\def\mcf{Maurer-Cartan forms~}
\def\mce{Maurer-Cartan equations~}
\def\nt{Nijenhuis tensor~}
\def\ni{Nijenhuis~}
\def\kk{Kaluza-Klein~}
\def\cy{Calabi-Yau~}
\def\dol{Dolbeaut cohomology~}


\def\zbar{\bar z}

\def\pa{\partial}
\def\derp{\partial}
\def\derm{\bar\partial}

\def\varia#1{\sssk \delta \!\sssk #1}

\def\sopra#1#2{{\raise 0.8 ex
\hbox{$
{{\scriptscriptstyle \,_{#2}}	\atop \displaystyle{#1}}
$}}
}
\def\sotto#1#2{{\lower 0.8 ex
\hbox{$
{{\displaystyle \,{#2}}	\atop \scriptscriptstyle{#1}}
$}}
}

\def\twomat#1#2#3#4{\left(\matrix{#1&#2\cr #3&#4\cr}\right)}
\def\stwomat#1#2#3#4{\left[\matrix{#1&#2\cr #3&#4\cr}\right]}
\def\twovec#1#2{\left(\matrix{#1\cr #2\cr}\right)}

\def\sp#1#2{{#1\brack#2}}
\def\comm#1#2{[\sssk #1\sssk,\sssk #2\sssk ]\sssk }
\def\acomm#1#2{\{\sssk #1\sssk ,\sssk #2\sssk \}\sssk }

\def\un{{\bf 1}}
\def\unhat{\hat{\bf 1}}
\def\tr{{\rm Tr}\sssk}
\def\bra#1#2{{#1\brack #2}}

\def\si#1{\sigma^{#1}}


\def\emenoduef{e^{-2 \Phi}}

\def\e2pi{e^{2i\pi}}
\def\em2pi{e^{- 2\pi i}}
\def\ep4{e^{i \pi /4}}
\def\emp4{e^{-i \o{\pi}{4}}}

\def\o#1#2{{#1\over#2}}

\def\unmezzo{{1 \over 2}}
\def\1su4p{{1 \over 4 \pi}}
\def\meno1su4p{\o {-1}{4 \pi}}

\def\caR{{\cal R}}

\def\cM{{\cal M}}
\def\epsd#1#2#3{\sssk \epsilon_{#1 #2 #3}\sssk}
\def\epsu#1#2#3{\sssk\epsilon^{#1 #2 #3}\sssk}
\def\epsild#1#2#3#4{\sssk\epsilon_{#1 #2 #3 #4}\sssk}

\def\xp{z}
\def\xm{\zbar}


\def\raduek{\sqrt{\o 2k}}
\def\raduekp{\sqrt{\o 2{k+2}}}

\def\emenrad{e^{- \raduek t}}
\def\emenradt{e^{- \raduek \t t}}
\def\emenradp{e^{- \raduekp t}}
\def\emenradpt{e^{-\raduekp \t t}}
\def\okp#1{\o{#1}{\sqrt{k+2}}}

\def\theory#1#2{$(#1,#1)_{#2,#2}$}
\def\theo#1#2{(#1,#1)_{#2,#2}}

\def\f#1#2#3{f_{#1 #2 #3}}
\def\Ga{\sssk\Gamma\sssk}
\def\Gai#1#2{\sssk\Gamma_{#1 #2}\sssk}

\def\ginv{\sssk g^{-1}\sssk}
\def\Ome#1{\sssk \Omega^{#1}\sssk}
\def\Omet#1{\sssk{{\tilde \Omega}^{#1}}\sssk}

\def\V#1{ V^{#1}}
\def\Vp#1{V_{+}^{#1}}
\def\Vm#1{V_{-}^{#1}}
\def\Vt#1{{\tilde V^{#1}}}

\def\Vz#1{V_z^{#1}}
\def\Vzb#1{V_{\zbar}^{#1}}

\def\Vtzb#1{\t V^{#1}_{\zbar}}

\def\omepm{\omega^{\pm}}
\def\omep{\omega^{+}}
\def\omem{\omega^{-}}
\def\omer{\omega^{\rm R}}
\def\omepmi#1#2{\omega^{\pm}_{#1 #2}}
\def\omepi#1#2{\omega^{+}_{#1 #2}}
\def\omemi#1#2{\omega^{-}_{#1 #2}}

\def\omerid#1#2{\omega^{\rm R}_{#1 #2}}

\def\omepti#1#2#3{\omega^{+}_{#1 #2 #3}}
\def\omemti#1#2#3{\omega^{-}_{#1 #2 #3}}
\def\Rpm#1#2{R^{\pm}_{#1 #2}}
\def\Rp#1#2{R^{+}_{#1 #2}}
\def\Rm#1#2{R^{-}_{#1 #2}}
\def\curv#1#2#3#4{R^{#1 #2}_{\hskip 6pt #3 #4}}
\def\delpm{\sopra {\nabla} {\pm}}
\def\delp{\sopra {\nabla} {+}}
\def\delm{\sopra {\nabla} {-}}
\def\delr{\sopra{\nabla}{\rm R}}


\def\cGd{\ca G}
\def\cGbd{\overline{\ca G}}
\def\cG#1{\cGd^{#1}}
\def\cGb#1{\cGbd^{#1}}
\def\cGtd{\wt{\ca G}}
\def\cGbtd{\wt{{\overline{\ca G}}}}

\def\cpmJ#1{{\sopra {{\cal J}^{#1}}{\!\pm} } }
\def\cmJ#1{{\sopra {{\cal J}^{#1}}{\! -} } }
\def\cpJ#1{{\sopra {{\cal J}^{#1}}{\! +} } }
\def\cpmJi#1#2#3{{\sopra {{\cal J}^{#1}_{#2 #3}}{\!\pm} } }
\def\cmJi#1#2#3{{\sopra {{\cal J}^{#1}_{#2 #3}}{\! -} } }
\def\cpJi#1#2#3{{\sopra {{\cal J}^{#1}_{#2 #3}}{\! +} } }
\def\hJ#1{{\hat {\cal J}}^{#1}}
\def\hJi#1#2#3{{\hat {\cal J}}^{#1}_{#2 #3}}
\def\tJ#1{\t{\ca J}^{#1}}
\def\tJi#1#2#3{\t{\ca J}^{#1}_{#2 #3}}
\def\cJ#1{{\cal J}^{#1}}
\def\cJi#1#2#3{{\cal J}^{#1}_{#2 #3}}
\def\sxy#1#2{s^{#1}_{\hskip 3pt #2}}
\def\spxy#1#2{s^{#1}_{+\sssk #2}}
\def\axy#1#2{a^{#1}_{\hskip 3pt #2}}
\def\apxy#1#2{a^{#1}_{+\sssk #2}}
\def\svec#1{{\vec s}^{#1}}
\def\spvec#1{{\vec s}^{#1}_{+}}
\def\avec#1{{\vec a}^{#1}}
\def\apvec#1{{\vec a}^{#1}_{+}}

\def\su{$SU(2)~$}
\def\suR{$SU(2) \times \rea~ $}
\def\susei{SU(6)}
\def\sutre{SU(3)}
\def\sudue{SU(2)}
\def\eottop{E_8'}


\def\Gm{G^{-}}
\def\Gpmi#1#2{G^{\pm}_{#1#2}}
\def\Gpi#1#2{G^+_{#1#2}}
\def\Gmi#1#2{G^-_{#1#2}}

\def\Rdue{R^{(2)}}
\def\omedue{\omega^{(2)}}
\def\cT{{\cal T}}
\def\Tbu{T^{\bu}}
\def\Tbul#1{T^{\bu}_{#1}}
\def\Tci{T^{\circ}}
\def\Tcir#1{T^{\circ}_{#1}}
\def\Ppp#1{\Pi^{#1}_{+}}
\def\Pmm#1{\Pi^{#1}_{-}}
\def\z#1{\zeta^{#1}}
\def\c#1{\chi^{#1}}

\def\Mtar{{\cM}_{\rm target}}
\def\intb{\int_{\partial \cM}}
\def\intm{\int_{\cM}}
\def\misura{dz\sssk d\zbar}

\def\lai#1{{\lambda}^{#1}}
\def\mui#1{{\mu}^{#1}}
\def\lati#1{{\t\lambda}^{#1}}
\def\muti#1{{\t\mu}^{#1}}
\def\ps#1{\psi^{#1}}
\def\psit#1{{\t\psi}^{#1}}

\def\ho#1#2{$h^{#1,#2}$}
\def\Gt{\wt G}

\def\kvec#1{{\bf k}_{#1}}
\def\kvect#1{\t{{\bf k}}_{#1}}
\def\contr#1#2{i_{#1} #2}


\def\bchar#1#2{B_{#1}^{(#2)}}

\def\disf{\partial_{I^*} \phi}
\def\djsf{\partial_{J^*} \phi}
\def\4v{V^a \null V^b \null V^c \null V^d \epsilon_{abcd} }
\def\3vd{V_b \null V_c \null V_d \epsilon^{abcd} }
\def\duevu{V^c \null V^d \epsilon_{abcd} }
\def\duevd{V_c \null V_d \epsilon^{abcd} }
\def\vabu{V^a \null V^b}

\def\Zia{Z^I_a}
\def\Zisa{Z^{I^*}_a}
\def\Zjsa{Z^{J^*}_a}
\def\gij{ g_{I{J^*}} }

\def\dif{\partial_I \phi}
\def\palla{ {\otimes} }

\def\epsiu{\epsilon^{abcd} }
\def\kuno{\kappa_1}
\def\kdue{\kappa_2}

\def\dzb{\partial_{\zbar}}
\def\modg{\sqrt{\vert g \vert}}
\def\gssb{g_{S \bar S}}
\def\usf{\over f}
\def\new{{\rm new}}
\def\old{{\rm old}}

\def\um{{\scriptstyle{\o 12}}}
\def\eipx{e^{\left [ i \, k \, \cdot \, X(z,\bz) \right ]}}
\def\pmutilde{{\tilde P}^{\mu}(\bz)}

\def\bz{{\bar z}}
\def\un{{\bf 1}}

\def \vertex#1#2{V_{{#1}}^{{#2}}\,\left (\, k, \, z, \, {\bar z} \,
 \right )}
\def \spinalph{ e^{\um \, \phi^{sg}(z)} \, S_{\alpha}\,(z)}
\def \spinaldot{ e^{\um \, \phi^{sg}(z)} \,
S^{\dot \alpha}\,(z)}




\tolerance 10000
\font\small=cmr7
\magnification=1200

\null
\hskip 12cm \vbox{
\hbox{SISSA 159/92/EP}
\hbox{IFUM/431/FT}
\hbox{September 1992}}
\vskip 1.2cm
\centerline{\bf GRAVITATIONAL INSTANTONS IN HETEROTIC STRING THEORY:}
\vskip 0.2cm
\centerline{\bf THE H-MAP AND THE MODULI DEFORMATIONS}
\vskip 0.2cm
\centerline{\bf OF (4,4) SUPERCONFORMAL THEORIES
\footnote*{\it Work
supported in part by Ministero dell'Universit\`a e
della Ricerca Scientifica e Tecnologica
}}
\vskip 0.6cm
\centerline{\bf  Marco Bill\' o,  Pietro Fr\`e }
\vskip 0.2cm
\centerline{\sl SISSA - International School for Advanced Studies}
\centerline{\sl Via Beirut 2, I-34100 Trieste, Italy}
\vskip 0.1cm
\centerline{\sl and I.N.F.N. sezione di Trieste }
\vskip 0.2cm
\centerline{\bf Luciano Girardello }
\vskip 0.1cm
\centerline{\sl Dipartimento di Fisica, Universit\'a di Milano}
\centerline{\sl Via Celoria 16, I-20133 Milano, Italy}
\vskip 0.1cm
\centerline{\sl and I.N.F.N. sezione di Milano }
\vskip 0.2cm
\centerline{\bf  Alberto Zaffaroni }
\vskip 0.2cm
\centerline{\sl SISSA - International School for Advanced Studies}
\centerline{\sl Via Beirut 2, I-34100 Trieste, Italy}
\vskip 0.1cm
\centerline{\sl and I.N.F.N. sezione di Trieste }

\vskip 1.2cm

\centerline{\bf ABSTRACT}
\vskip 0.3truecm
We study the problem of  string propagation
in  a general instanton background for the case of the complete heterotic
superstring.
We define the concept of generalized
HyperK\" ahler manifolds and we relate it to (4,4) superconformal theories.
We propose a generalized h-map construction that predicts a universal $SU(6)$
symmetry for the modes of the string excitations moving in an instanton
background.
We also discuss
the role of abstract $N$=4 moduli and, applying it  to the particular  limit
case
of the solvable \suR instanton found by Callan et al. we show that it admits
deformations
and corresponds to a point in a 16-dimensional moduli space. The geometrical
characterization of the other spaces in the same moduli-space remains an
outstanding
problem.
\vfill\eject


\centerline{\bf 1. Introduction}
\vskip 0.3cm
Gravitational instantons (for a review see [1]) have an intrinsic interest and
furthermore may provide a mechanism
for the non perturbative breaking of local supersymmetry [3].
Their study,  in the context of string theory,  is a part of a more
general study, namely the identification of the Conformal Field Theories
corresponding to the geometries under consideration and the study
of their properties.

Significant advances have been recently made in identifying conformal
field theories
related with black-hole space-times in unphysical dimensions ($D$=2)
[4] and some insights  have also been obtained on the physical case [5].
A limit case of instanton conformal field theory has been
discussed in a seminal paper by Callan, Harvey and Strominger [6].

So far this issue has not been addressed in a full-fledged heterotic
superstring framework.
Indeed, like the case of the Witten black-hole
[4] the investigation
has been limited to a discussion of the \simod on the selected
space-time,
ignoring the internal degrees of freedom of the superstring and
the question of modular invariance.

In this paper we consider the issue of gravitational instantons in the
context of the complete heterotic superstring.

The first basic point of our work is the proposal of a
{\sl generalized $h$-map},
according to which the propagation on an instanton background of a
heterotic superstring
compactified on a Calabi-Yau space is given by the tensor product
of three conformal
theories: a $c=6$ (4,4)-theory, a $c=9$ (2,2)-theory and the $c=11$
right-moving  current algebra of $SO(6)\; \times \; E_{8}{'}$. A general
prediction of our framework is that all particle modes in any
istantonic background of this type are classified
by an $SU(6)$ symmetry group. This prediction is absolutely
analogous to the prediction that, in Calabi-Yau compactifications,
massless particles  fall into $E_6$ representations.

We study in depth the relation between (4,4) world-sheet supersymmetry
and the self-duality (respectively antiselfduality) of the curvatures
$R(\omega_{R} \pm T)$, where T is the torsion and $\omega_{R}$ is
the Riemannian connection.
Spaces with this property correspond, as we show, to a
{\sl generalization of HyperK\" ahler manifolds } and have the proper
geometry to
describe axionic-dilatonic instantons. These configurations were
originally introduced
by D'Auria and Regge [7] long time ago and were more recently
rediscovered in string theory by Callan,
Harvey and Strominger [6] and by Rey [8]. Their distinguished feature is
asymptotic flatness which is an essential feature in order to utilize
the instanton
in any supersymmetry breaking mechanism \`a la Konishi et al [3]. Indeed,
for  the instanton to contribute to a scattering amplitude, the
asymptotic states must be the same in flat space and in the
instanton background. Such asymptotic flatness requires an
unsoldering of the Lorentz-bundle from the tangent bundle
which is indeed what the axionic and dilatonic fields combined
are able to realize.

The second basic point of our investigation concerns the
moduli-deformations of
the (4,4)-theory. We discuss the general characterization of the
moduli within an arbitrary
$c=6$ (4,4)-theory  and their use to construct the emission vertices of
the particle zero-modes appearing in the spectrum of the heterotic
superstring. In this discussion we utilize the $K_3$ example as a
guideline for our generalization. Actually $K_3$ is compact, non
asymptotically flat and HyperK\" ahler, yet both
HyperK\" ahler and generalized HyperK\" ahler manifolds lead to (4,4)
world-sheet
supersymmetry, so that the associated conformal field theory has
the same structure and the same properties for both kind of manifolds.
Incidentally this
is the very reason why we name generalized  HyperK\" ahler the
manifolds under consideration in the present paper.

Considering with special attention the limit case where
the   manifold  of the axion-dilaton instanton reduces to \suR [6], case  where
asymptotic flatness is lost, but conformal solvability is gained,
we discuss the
(4,4)-moduli of this specific model showing that they are four as
for flat space.
We explicitely exhibit the infinitesimal deformations of the metric and of
the torsion, finding apparently that the \suR instanton is a point in a
16-dimensional
space. In the same way, if one counts the deformations of the flat space
$\rea^4$ one finds that they depend on 16 parameters. However, although one
gets the correct counting of zero-modes, from the geometrical point of view
all the deformations can be reabsorbed by diffeomorphisms (any constant metric
can be diagonalized and rescaled to unity in this way). The same does not hold
true if the space has the topology of a torus, since global diffeomorphisms
must
respect the foundamental identifications. In the case of the instanton,
as we will see, in general the deformations are significative; however
further study is required to decide if someone (and which one) of the
deformations can be reabsorbed by diffeomorphisms for certain values of
the moduli.
The global characterization of this
moduli space and the geometric interpretation of the deformed space
is therefore still an outstanding problem.

 The paper is organized in seven sections plus an appendix.
Each section contains
 an introductory discussion of the topic developed therein, so we dwell
 no longer on these general remarks.

 In particular the general philosophy and the
 perspective of our proposal are discussed in  section 2, where
the generalized $h$-map is introduced and the role of the $N$=4
moduli is illustrated.

 Section 3 discusses asymptotically flat axion-dilaton instantons
from the viewpoint of the effective superstring lagrangian, using the
New minimal formulation
of  $N$=1 supergravity, which is the appropriate one for  heterotic
string derived  theories.

Section 4 describes supersymmetric \simod with dilaton-axion coupling in
the rheonomy framework.

Section 5 discusses the conditions for extended (4,4) world-sheet
supersymmetry
and introduces the notion of generalized  HyperK\" ahler  manifolds.

Section 6 discusses the conformal field-theory of the \suR model
and its moduli deformations.

Section 7 studies the deformed geometry of the \suR model.

Appendix A contains the complete list of emission vertices for all
massless particle zero-modes in an arbitrary (4,4) background.
\svsk


\def\susei{SU(6)}
\def\sutre{SU(3)}
\def\sudue{SU(2)}
\def\eottop{E_8'}
\def\kk{Kaluza-Klein~}
\def\cy{Calabi-Yau~}
\def\theo#1#2{(#1,#1)_{#2,#2}}
\def\bchar#1#2{B_{#1}^{(#2)}}
\def\bra#1#2{{#1\brack #2}}
\centerline{\bf  2. Gravitational instantons and (4,4)-superconformal
theories:}
\centerline{\bf the idea of a generalized h-map.}
\svsk
\noi{\bf The basic idea}
\svsk
We want to investigate the possibility of constructing consistent
heterotic string vacua where the usual $c=(6,4)$ conformal field-theory
(CFT) that represents four dimensional {f}lat space is replaced
by some new $c=(6,6)$
theory describing  string propagation on a non-trivial four-dimensional
geometry.
For many reasons, that  will become clear  in the sequel, particularly
appealing are
the possibilities offered by $c=6$ theories possessing $N=4$ world-sheet
supersymmetry. We focus on these theories.\para
The first part of our discussion is  somehow  heuristic: we
 use, as a guideline,  the analogy of the scheme we propose
with the procedure utilized to compactify string-theory
on 6-dimensional  manifolds of $SU(3)$ holonomy [9]. As it is well-known
[10], from the abstract point of view, this operation is  represented as the
replacement of
the $c=(9,6)$ theory, corresponding to six {f}lat dimensions, by a
\theory 92 conformal
theory \footnote*{\small From now on we use the notation
$\scriptstyle{(c_L,c_R)_{n_L,n_R}}$ to
mean a CFT of central charges $\scriptstyle{c_L (c_R)}$
in the left (right) sector,
possessing $\scriptstyle{n_L, n_R}$ left (right) supersimmetries.}\par
Let us briefly review  the process of this compactification, in order
to proceed  in analogy with it also for the space-time part.
\para
The ``initial'' situation is that  required by critical heterotic
string theory [11] in $d$=10, namely
the vacuum is a CFT of central charges $(15,26)$ that can be realized as
$$
(15,26)=(15,10)\oplus (0,16)
\eqno(2.1)
$$
The $(15,10)$ theory is generated by 10  left-moving $\oplus$ 10 right-moving
world-sheet
bosons, together with 10 left-moving  fermions: it represents the heterotic
\simod on {f}lat 10-dimensional  space. The $(0,16)$ theory is that generated
by 32
right-moving fermions describing the gauge group  $G_{gauge}$ degrees of
freedom, namely those of the Ka\v c-Moody algebra $\hat G_{gauge}$. The choice
of $G_{gauge}$
is determined by the enforcement of modular invariance and we
consider the version of the theory where $G_{gauge}=\eottop
\times E_8$. We
consider the 10-dimensional space to be split in a 6-dimensional internal
submanifold and a 4-dimensional space-time manifold. At the level of conformal
field-theories this
means:
$$
(15,26)=(6,4)\oplus (9,6)\oplus (0,16)
$$
The main point of the $h$-map construction [10] is the possibility of
considering heterotic string vacua where the above situation is modified as
follows:
$$
(15,26)=(6,4)\oplus (9,9)\oplus (0,13)
\eqno(2.2)
$$
This is commonly expressed by saying that six of the heterotic
fermions have been ``eaten up'' by the internal theory (which becomes
left-right symmetric); the remaining thirteen generate the current algebra of
$\eottop \times SO(10)$.\para
{}From the \kk viewpoint, one is considering a 10-dimensional manifold
with the following structure:
$$
\cM_{10}=\cM_6 \times\cM_4^{\rm {f}{l}{at}}
\eqno(2.3)
$$
The ``eating'' of six heterotic fermions is due to the 10-dimensional axion
Bianchi identity
 $dH=0$  which (at \first order) requires
$$
0=\tr F\wedge F - \tr R_{(6)}\wedge R_{(6)}
\eqno(2.4)
$$
and is solved by embedding the spin connection into the gauge
connection. In this way  the gauge group  is broken to the normalizer of  the
internal manifold holonomy  group
 ${\cal H}ol(\cM_6)$  .\para
In the particular case of manifolds with
$\sutre$ holonomy (\cy manifolds),  the residual gauge group is $E_6 \otimes
\eottop$, as it follows from the maximal subgroup embedding:
$$
E_6\times\sutre \ssk\longrightarrow\ssk E_8
\eqno(2.5)
$$
 Thus \kk analysis shows that the massless fields on $\cM_4^{\rm
{f}lat}$ are organized in $E_6$ representations.  From the
abstract point of view, the case of $\sutre$  holonomy  corresponds
to the particular case of the decomposition (2.2) in
which the internal theory has $(2,2)$-supersymmetries:
$$
(15,26)=(6,4)\oplus\theo 92\oplus (0,13)
$$
One can show [10] that the $U(1)$ current
appearing in the $N=2$ algebra and the $SO(10)$ currents of the
heterotic fermions combine, together  with suitable spin fields, to yield the
current algebra of $E_6$, in due agreement with the maximal subgroup embedding:
$$
SO(10)\times U(1)\ssk\longrightarrow\ssk E_6
\eqno(2.6)
$$
Hence the emission vertices of  the 4-dim fields are
organized in  $E_6$-representations as it is required by \kk analysis.\para
The question of consistency of these compactified theories and, in
particular, the question of their (1-loop) modular invariance is better
addressed by looking at their construction from a
different viewpoint. Consider a modular-invariant type II superstring vacuum:
for what concernes central charges we have:
$$
(15,15)=(6,6)\oplus(9,9) \eqno(2.7)
$$
the $(6,6)$-theory corresponding to {f}lat 4-dim space and the $(9,9)$-theory
describing some non-trivial ``internal''  manifold. One shows that the
``$h$-mapped''  heterotic vacuum, obtained by replacing,  in  the partition
function of (2.7),
the subpartition function of the two right-moving transverse fermions with that
of $2+24$
fermions (generating a $\eottop\times SO(2+8)=\eottop\times
SO(10)$  current algebra) is also modular invariant.\para
When the internal theory has $N=2$ supersymmetry, the fundamental
implication of modular invariance is the projection onto
 odd-integer charge states with respect to the  diagonal  $U(1)$ group
obtained  by summing  the  $U(1)$ of the  $N=2$
algebra with the SO(2) generated by the transverse  space-time fermions. This
is
just the rephrasing in the present context of the GSO projection [10,13].\par
Let's now consider an extension of the above described mechanism.
We start from the  conformal field-theory describing the heterotic string
compactified on a \cy manifold,
$$
(15,26)=(6,4)\oplus\theo 92\oplus (0,13)
$$
and we let the four-dimensional theory eat four  of the heterotic fermions, so
that
$$
(15,26)=(6,6)\oplus\theo 92\oplus (0,11)
\eqno(2.8)
$$
The remaining heterotic fermions generate a  current algebra
$\eottop\times SO(6)$.\para
{}From the geometrical \simod point of view,  what we have done is to consider
a
target space of the form
$$
\cM_{10}=\cM_6 \times\cM_4
$$
where $\cM_6$ is still a manifold of $\sutre$ holonomy but  $\cM_4$  is no
longer
{f}lat space. Condition (2.4) extends to
$$
0=\tr F\wedge F - \tr R_{(6)}\wedge R_{(6)}- \tr R_{(4)}\wedge R_{(4)}
$$
which can be solved by embbeding also the holonomy group ${\cal H}ol(\cM_4)$
into the gauge group.\para
In particular consider the case where ${\cal H}ol(\cM_4)\subset \sudue$: this
happens for gravitational instantons, whose curvature is either self-dual or
antiself-dual. In this situation the gauge group is broken  to $\susei$, as it
follows from the
maximal subgroup embedding:
$$
\susei\times\sutre\times\sudue\ssk\longrightarrow\ssk  E_8
\eqno(2.9)
$$
{}From the abstract viewpoint, this is reproduced if the $c=(6,6)$ theory
possesses a (4,4) supersymmetry:
$$
(15,26)=\theo 64\oplus\theo 92\oplus (0,11)\eqno(2.10)
$$
Indeed the $U(1)$ current of the $N=2$ algebra associated with $\cM_6$, the
$\sudue$ currents of the $N=4$ algebra associated with $\cM_4$  and
the $SO(6)$ currents of the heterotic fermions combine  together with suitable
spin fields to yield the $\susei$-current algebra, according to the maximal
embedding
$$
U(1)\times\sudue\times SO(6)\ssk\longrightarrow\ssk \susei
\eqno(2.11)
$$
Thus,  on this background, the emission vertices for particle-modes (both
massive and massless) are organized in $\susei$-representations, as it
is
requested by \kk analysis.
\svsk
\noi{\bf The issue of modular invariance}
\svsk
 As we already recalled, in the case of compactifications on \cy manifolds,
that is by means of a
\theory 92 theory, one  starts from a type II  modular invariant partition
function, corresponding to a CFT:
$$
(15,15)=(6,6)\oplus (9,9)
$$
The $(6,6)$ part, corresponding to 4-dimensional {f}lat space,  contains the
world-sheet
bosons $X^{\mu}, \t X^{\mu}$ and  the fermions $\ps{\mu}, \psit{\mu}$ and
the  complete partition function has the structure:
$$
Z_{tot}=\sum_{i,\bar\imath} Z_{i,\bar\imath}^{(9,9)}
Z(X^{\mu},\t X^{\mu}) \sssk\bchar{i}{4}\sssk
\left(\bchar{\bar\imath}{4}\right)^*
\sssk\bchar{i}{-2}\sssk\left(\bchar{\bar\imath}{-2}\right)^*
\eqno(2.12)
$$
where
\item{\sl i)}{$Z(X^{\mu},\t X^{\mu})$ is the usual partition function
for the four free bosons,}
\item{\sl ii)}{$\bchar{i}{4}$ are the $SO(4)$-characters in which we can
organize the partition function for the  four free fermions $\ps{\mu},
\psit{\mu}$ ( the index $i$ taking the values $0,v,s,\bar s$ , for the
scalar, vector and spinor conjugacy class, respectively),}
\item{\sl iii)}{$\bchar{i}{-2}$ are the partition functions for the
superghosts,}
\item{\sl iv)}{$Z_{i,\bar\imath}^{(9,9)}$ is the partition function
for the internal theory which couples to reps $(i,\bar\imath)$ of the
space-time $SO(4)$ and of the superghosts.}
\svsk
The reason why we have denoted as $\bchar{i}{-2}$ the superghost partition
function  becomes clear from the following considerations.  If the superghosts
have boundary conditions $\bra ab$, ($\bra
ab=\bra 00,\bra 01,\bra 10,\bra 11$) their partition function can be
computed to be[12]:
$$
Z^{sg}\bra ab (\tau |z)=\o{\eta(\tau)}{\theta\bra ab(\tau |z)}= \o
1{Z\bra ab(\tau |z)}
\eqno(2.13)
$$
which is exactly the reciprocal of the partition function for two free
fermions with spin-structure $\bra ab$ \footnote*{\small This reciprocity holds
only
at
genus g=1. For higher genera it is amended by a phase factor that amounts to a
correct
assignment of spin statistics[12]. In all known constructions if one fixes
1-loop
modular
invariance plus spin statistics, higher loop modular invariance is also
ensured. We assume that this will go through also in our construction.}
\para
Since the superghosts are forced by world-sheet supersymmetry to have the same
spin-structure as the space-time
fermions, dealing with the theory described by eq.(2.12)
one can use the cancellation of
the superghost partition function with the partition function for two
fermions.
Instead of (2.12) one can simply write
$$
Z_{tot}=\sum_{i,\bar\imath} Z_{i,\bar\imath}^{(9,9)}
Z(X^{\mu},\t X^{\mu})\sssk \bchar{i}{2}\sssk
\left(\bchar{\bar\imath}{2}\right)^*
\eqno(2.14)
$$
that is, one considers only the transverse fermions.\para
The  $h$-map construction of the associated heterotic theory is based on an
isomorphism between the $SO(2n)$-characters  and  those of $SO(2n+24)$ or
$\eottop\times SO(2n+8)$. It works  as follows. The action of the
modular transformations $S$ and $T$ on the characters of $SO(2n)$ in
the basis labeled by $0,v,s,\bar s$ is given by
$$
\eqalign{
\bchar{i}{2n}\ssk&\sopra {\longrightarrow}T\ssk T^{(2n)}_{ij}
\bchar j{2n}\cr
\bchar{i}{2n}\ssk&\sopra {\longrightarrow}S\ssk S^{(2n)}_{ij}
\bchar j{2n}
         }
\eqno(2.15)
$$
where
$$
\eqalign{
T^{(2n)}&= {\rm diag}(1,1,e^{i n\o {\pi}{4}},e^{i n\o {\pi}{4}})
e^{-i n\o{\pi}{12}}\cr
S^{(2n)}&=\left(\matrix{1 & 1 & 1 & 1\cr 1 & 1 & -1 & -1 \cr
1 & -1 & e^{i\o{n\pi}{12}} & -e^{i\o{n\pi}{12}}\cr
1 & -1 & -e^{i\o{n\pi}{12}} & e^{i\o{n\pi}{12}}\cr}\right)
         }
\eqno(2.16)
$$
The  isomorphism is realized by
$$
\eqalign{
T^{(2n)}&=M\sssk T^{(2n+24)}\sssk M\cr
S^{(2n)}&=M\sssk S^{(2n+24)}\sssk M
         }
\eqno(2.17)
$$
where the idempotent matrix $M$ is given by:
$$
M=\left(\matrix{ \matrix{0&1\cr1&0\cr} & {\bf 0} \cr
{\bf 0} & \matrix{-1&0\cr 0&-1\cr} \cr}\right)
$$
 It interchanges the scalar and the vector characters, besides
flipping the sign of the spinor and antispinor characters. The characters of
$\eottop\times SO(2m)$ transform
as those of $SO(2m+16)$, so that the isomorphism permits also to reach
these groups.\para
Due to (2.17), if one replaces the two right-moving transverse $\psit {\mu}$
fermions
with 26 heterotic
fermions that generate the gauge group $\eottop\times SO(10)$, by taking proper
account of the matrix $M$, the resulting theory has
a modular invariant partition function.\par
 In eq.(2.12) we  made  no explicite use of the
cancellation between superghosts and longitudinal fermions for the following
reason.
We wanted to enphasize the possibility of constructing, out of the $Z^{sg}
\bra ab$ and by means of combinations analogous to those used for the free
fermions, new characters labeled by an index  $i=0,v,s,\bar
s$, whose modular transformations are very similar to those in
eq.(2.16).\para
Indeed, in analogy with the characters  of $2n$ fermions let the superghost
characters be:
$$
\eqalign{
\bchar 0{-2} =\o 1{Z\bra 00} +\o 1{Z \bra 01}\ssk &;\ssk
\bchar v{-2} =\o 1{Z\bra 00} -\o 1{Z \bra 01}\cr
\bchar s{-2} =\o 1{Z\bra 10} +\o 1{Z \bra 11}\ssk &;\ssk
\bchar {\bar s}{-2} =\o 1{Z\bra 10} -\o 1{Z \bra 11}
         }
\eqno(2.18)
$$
Eq.(2.18) is obtained from the definition of the $\bchar i{2n}$ characters
[10,13] by the replacement  $(Z\bra ab)^n\longrightarrow 1/Z\bra ab$.
 Using the
modular transformations of the $Z\bra ab (\tau)$, already utilized  to obtain
eq.(2.16) , we find :
$$
\eqalign{
\bchar{i}{-2}\ssk&\sopra {\longrightarrow}T\ssk T^{(-2)}_{ij}
\bchar j{-2}\cr
\bchar{i}{-2}\ssk&\sopra {\longrightarrow}S\ssk S^{(-2)}_{ij}
\bchar j{-2}
         }
\eqno(2.19)
$$
where
$$
\eqalign{
T^{(-2)}&= {\rm diag}(1,1,\emp4,\emp4) e^{i\o{\pi}{12}}\cr
S^{(-2)}&=\left(\matrix{1 & 1 & 1 & 1\cr 1 & 1 & -1 & -1 \cr
1 & -1 & e^{-i\o{\pi}{12}} & -e^{-i\o{\pi}{12}}\cr
1 & -1 & -e^{-i\o{\pi}{12}} & e^{-i\o{\pi}{12}}\cr}\right)
         }
\eqno(2.20)
$$
Formally, these matrices are obtained from those in eq.(2.16) by setting
$2n=-2$, which explains the chosen notation. Moreover it is manifest
that we can use the $h$-map isomorphism to substitute the characters of
the superghosts with those of 22 heterotic fermions with gauge group
$\eottop\times SO(-2+8)=\eottop\times SO(6)$.\par
Consider now a modular invariant type II vacuum in which the
$c=(6,6)$ part represents a four-dimensional space with non trivial geometry.
The
partition function of such a theory is
$$
Z_{tot}=\sum_{i,\bar\imath} Z_{i,\bar\imath}^{(9,9)}\sssk
Z_{i,\bar\imath}^{(6,6)} \sssk \bchar{i}{-2}\sssk
\left(\bchar{\bar\imath}{-2}\right)^*
\eqno(2.21a)
$$
$Z_{i,\bar\imath}^{(6,6)}$ being the partition function for the $(6,6)$
theory which couples to the characters $(i,\bar\imath)$ of the
superghosts.\para
Although the $SO(4)$ characters  have disappeared from the game,
we can still perform the $h$-map construction of an associated modular
invariant
heterotic theory. The result is just of the form (2.8). If, in addition we
choose a
space-time with $\sudue$ holonomy, the result is of the form (2.10).
After $h$-map the partition function (2.21a) becomes
$$
Z_{tot}=\sum_{i,\bar\imath} Z_{i,\bar\imath}^{(9,9)}\sssk
Z_{i,\bar\imath}^{(6,6)} \sssk \bchar{i}{-2}\sssk
\left(\bchar{\bar\imath}{\eottop\times SO(6)}\right)^*
\eqno(2.21b)
$$

\noi{\bf The vertex operators}
\svsk
 The next step in the analysis of the heterotic theory (2.21b) is the
construction
of the corresponding vertex operators. This construction  summarizes many
aspects of the theory under discussion. Furthermore physical amplitudes are
expressed in terms of the vertex correlators, so that knowledge of the vertices
is an essential ingredient to
extract any physical information.
\par
In the case where $\cM_4$ is flat space, the emission vertex for any
particle-field
has the general form:
$$\vertex {\epsilon}{\cdot\cdot,\bu}~=~\Phi_{\epsilon}(z,\bz)\;
e^{ik\cdot X(z,\bz)}\;
\Psi^{\cdot\cdot}(z,\bz)
\; \Lambda^{\bullet}(z,\bz)$$
where $~\Phi_{\epsilon}(z,\bz)\; e^{ik\cdot X(z,\bz)} $,
$\Psi^{\cdot\cdot}(z,\bz)$ and
$\Lambda^{\bullet}(z,\bz)$ are conformal fields respectively belonging
to the theories
$(6,4)$, $(9,9)_{(2,2)}$ and $(0,13)$.  The last two factors
determine the internal quantum numbers $(\cdot\cdot,\bullet)$  of the
particle one considers. The first factor, instead, determines its
space-time character, namely its spin, its polarization ${\epsilon}$,
and its momentum $k$.
The compound $\Phi_{\epsilon}(z,\bz) \; e^{ik\cdot X(z,\bz)} $ is
the conformal field-theory corresponding of a pure-state wave-function
$\psi_{k,\epsilon}(X)$ satisfying the wave equation:
$$ \Delta \psi_{k,\epsilon}(X)~=~m^2  \; \psi_{k,\epsilon}(X)
\eqno(2.22)$$
where $\Delta$ is  the relevant wave-operator  (Dirac, Rarita-Schwinger,
Einstein, Yang-Mills,...) and $m^2=k^2$ is the squared mass.
The reason why polarization and momentum are utilized to label this
part of the vertex
operator is that they are good quantum numbers in flat space.
Indeed on a flat background a complete set of solutions of (2.22) can
always be expressed in
terms of plane waves. Massless-particles have $k^2=0$ and are the zero-modes of
the wave operator $\Delta$ . When we deal with some non trivial space-time
geometry, the eigenfunctions of the operators $\Delta$ are no longer
plane-waves
and their spectrum is labeled by a new set
of quantum numbers replacing the momentum $k$ and the polarization $\epsilon$.
Correspondingly the compound  $~\Phi_{\epsilon}(z,\bz)\; \eipx$ is replaced by
suitable
operators ${\Theta_{i}}(z,\bz)$
of the $(6,6)_{4,4}$ theory. A finite number  of these operators correspond
to the zero modes of $\Delta$ and can be used to calculate the scattering of
massless particles in
the non-trivial background under consideration.\par
Another important  remarque concerns the moduli: the non-trivial space-time one
considers usually admits continuous deformations that preserve both its
topology and its holonomy. The parameters of these deformations are named
moduli and, from the CFT viewpoint, they correspond to suitable marginal
operators one can add to the 2-dimensional lagrangian preserving its
(4,4)-supersymmetry. Exactly as in Calabi-Yau
compactifications, the spectrum of zero modes for the various operators
$\Delta$
depends on the number of these deformations: furthermore the corresponding
vertex
must be thought as a function of the moduli.\par
In the construction of the relevant vertices we proceed in analogy with what
one does
for the internal dimensions. We
relate the counting and the group-theoretical indexing of the possible
conformal operators that possess the correct dimensions and charges to the
 counting of zero-modes for the fields appearing in the low-energy effective
supergravity, when this latter is expanded around the  particular background,
abstractly
described  by the  CFT under investigation. The procedure is like a \kk
compactification to zero dimensions.
On the other hand, in order to gain a more intuitive comprehension of the role
of the operators appearing in the \theory 64 theory, it is instructive to
compare
the vertices with those of  {f}lat space.
To this purpose it is useful to recall that {f}lat four-dimensional space
possesses an $N$=4 world-sheet supersymmetry. Hence we can recast
the operators appearing in the vertices in a form suitable of generalization
to any \theory 64.
As pointed out, in what follows we  try to estabilish a general procedure; yet
we choose to illustrate it in terms of an example,  namely using the
$K3$ manifold. The  convenience of this choice is manifest. Indeed we want to
proceed in analogy with \cy compactifications and
$K3$ is the unique  non-trivial  compact \cy space in four
dimensions. This  makes the analogy  closer. Furthermore
compactification on $K3$-surfaces has been extensively studied in the past
[14] and it is known to be  represented by a \theory 64 theory, which in
some points of moduli
space is even solvable, being given by a tensor product of $N=2$ minimal
models.
The knowledge of $K3$
cohomology, described by the Hodge diamond
$$
\matrix{ 1\cr {0\msk 0}\cr {1\msk 20\msk 1}\cr {0\msk 0}\cr 1\cr}
\eqno(2.23)
$$
makes the counting of the zero-modes easy yielding non-trivial
results that  can be compared with the CFT counting of vertices.
\par
On the contrary, for physical reasons, $K3$ is not the most appealing
possibility. It is a gravitational instanton, but it is compact. Our goal is to
extend the same techniques to four-dimensional instantons of the effective
lagrangian that are
 asymptotically {f}lat (this last feature seems to be realizable only with
torsion[7]).
An instantonic solution with the desired properties has
appeared frequently in the recent literature [6,8], and in later sections
we focus on it. Unfortunately an exact and solvable
$N$=4 SCFT corresponding to this solution is known only in a particular
limit in which the asymptotic {f}latness is lost.  However the theory remains
interesting in its own right  as a case study. We may also stress that one of
the results
of the present paper is the explicit deformation of this solution by means of
its moduli
that we discuss later on. In this way we are able to extend the
particular
background of [6] to  a class of solutions depending on certain parameters
whose
geometrical interpretation is still an open problem.
\svsk
In  order to analyse the vertex-operators for the zero-modes we need the field
content of the effective four-dimensional theory, which is a
matter-coupled $D$=4,$N$=1 supergravity arising from compactification
on the internal \cy manifold. This field content   is described  in the next
section.
\para
We begin with the $E_6$ charged fields given by the gauge multiplet
(gauge bosons and gauginos, transforming in the  78 representation  ), by \ho
21 WZ multiplets transforming in the 27 and \ho 11
transforming in the $\overline{27} $-representation (these Hodge numbers
being those of the compactified CY space).
 We consider the zero-modes of these fields in the classical
background provided by a $K3$ manifold. As already enphasized, we embed
the space-time spin connection  into the gauge connection,  breaking the gauge
group as follows:
$$
E_6\ssk\longrightarrow\ssk\susei\times\sudue
\eqno(2.24)
$$
To investigate the zero-modes
we must take into account  the branching of the representations of $E_6$
under (2.24).
\para
The adjoint representation is decomposed as
$$
78=(35,1)+(1,3)+(20,2)
\eqno(2.25)
$$
Consider the gaugino field. Its index in the adjoint of $E_6$ is
split accordingly to eq.(2.25); it also has a spinorial index
on $K3$. Thus the possible cases are :\para
\ssvsk\noi
$\bud\ssk\lambda^{A,}_{\alpha}\ssk$ $A$ being an index in the adjoint
(35) of $\susei$, $\alpha$ being the spinorial index.
The zero-modes are in correspondence
with the \dol $H^{0,q}$. Since the chirality is given by $(-1)^q$,
looking at the Hodge diamond (2.23) we see that there are two
zero-modes both of the same chirality.\para
\ssvsk\noi
$\bud\ssk\lambda^{,X}_{\alpha}\ssk$ $X$ in the adjoint of the
\su holonomy group of $K3$. The zero-modes should be related to the cohomology
groups $H^{0,q}(End T)$ of  Endomorphism-(of the tangent bundle)-valued
antiholomorphic forms.  By the explicit
realization of $K3$ as an algebraic surface one can evaluate the dimension of
this cohomology group, case by case.
\ssvsk\noi
$\bud\ssk\lambda^{a,x}_{\alpha}\ssk$ $a$ belongs to the 20 of $E_6$;
$x$ in the 2 of \su is the same as a contravariant holomorphic index
which can be lowered by means of the holomorphic $(2,0)$ form.
Because of the spinorial index, the zero-modes correspond to
$(1,q)$ harmonic forms.
We can therefore have just \ho 11$=20$ zero-modes with the opposite
chirality with respect to those in the adjoint of $\susei$
\ssvsk\noi
Consider then the gauge bosons. According to the decomposition (2.25) we
have:
\ssvsk\noi
$\bud\ssk A^{A,}_{\mu}\ssk$ $\mu$ can be a holomorphic or antiholomorphic
index. Since, due to the vanishing of \ho 10 and \ho 01 the holomorphic
$\susei$ bundle is trivial, there is no zero-mode of this kind.
\ssvsk\noi
$\bud\ssk A^{,X}_{\mu}\ssk$ Zero-modes are related to the \dol
$H^1(End T)$.
\ssvsk\noi
$\bud\ssk A^{a,x}_{\mu}\ssk$ Again, $x$ behaves as a holomorphic index
that can be lowered by the holomorphic $(2,0)$ form or by the metric
according to the necessity to obtain again an antisymmetric form.
Then the zero-modes can
be set in corrispondence with $(1,1)$ forms, for both the type
of ${\mu}$.
We have thus  2 \ho 11 = 40 zero-modes of this kind.
\ssvsk\noi
The 27 of $E_6$ is decomposed as
$$
27=(15,1)+(6,2)
\eqno(2.26a)
$$
Consider the fermion field belonging to any of the WZ multiplets  that
transform in the 27 representation (these are the charged fields paired to
the complex structure
deformations of the Calabi-Yau manifold).
The decomposition (2.26a) gives rise to the following cases:
\ssvsk\noi
$\bud\ssk\chi^{A,}_{\alpha}\ssk$ $A$ belonging to the 15 of $\susei$,
$\alpha$ the spinorial index. Zero-modes correspond to the \dol $H^{0,q}$
so there are two zero-modes of the same chirality
\ssvsk\noi
$\bud\ssk\chi^{a,x}_{\alpha}\ssk$ $a$ is in the 6 of $\susei$;
$x$ in the 2 of \su is like a holomorphic
index; once lowered by the (2,0) form the zero-modes are put into
correspondence with $H^{1,q}$ so that there are \ho 11 $=20$ modes,
of opposite chirality  with respect to the previous ones.
\ssvsk\noi
The possibilities for the scalars of these 27 families are:
\ssvsk\noi
$\bud\ssk\varphi^{A,}\ssk$ for which there is just \ho 00 $=1$ zero mode.
\ssvsk\noi
$\bud\ssk\varphi^{a,x}\ssk$ Lowering the index, the correspondence is
with $H^{0,1}$ and so no zero-modes exist
\ssvsk\noi
The $\overline{27}$ of $E_6$ decomposes as
$$
\overline{27}=(\overline{15},1)+(\bar 6,\bar 2)
\eqno(2.26b)
$$
For the $\overline{27}$-spinors we have, analogously to the 27-ones, two
zero-modes in the $\overline{15}$ and twenty in the $\bar 6$, of
opposite chiralities; for the $\overline{27}$-scalars, one mode in the
$\overline{15}$.
\ssvsk\noi
Consider now the fields of the gravitational multiplet\para
To look for the zero-modes of the graviton field, i.e. of the metric,
means to look for the solutions of the Lichnerowitz equation on $K3$,
which are known to be 58. This number is of course determined by the
cohomology of $K3$, and to this purpose a discussion is  necessary,  about the
separated counting of the metric and torsion zero-modes.
It goes as follows. From the $K_3$ Hodge diamond (2.23) we know that
$h^{2,0}=1$ and $h^{1,1}=20$. Let $\Omega_{ij}$ be the (2,0)-holomorphic
form
and let $g_{ij^{\star}}$ be the fiducial Ricci flat K\" ahler  metric
($i,j=1,2$) that,  for each K\" ahler class, is guaranteed to exist by the \cy
condition $c_{1}(K_3)=0$. Furthermore let $U_{ij^{\star}}^{(\alpha)}$ be a
basis for  the (1,1)-forms ($\alpha=0,1.....,19 $). A variation of the
reference metric which keeps it Ricci-flat is given by:
$$ g_{\mu\nu} \; \longrightarrow \; g_{\mu\nu}\;+\; \delta g_{\mu\nu}~~~~; ~~~~
\delta g_{\mu\nu}~=~\cases{ \delta g_{ij}\cr
                                                \delta g_{ij^{\star}}\cr
                                                \delta
g_{i^{\star}j^{\star}}\cr}\eqno(2.27)$$
 where $\delta g_{ij}$, $\delta g_{ij^{\star}}$  and $\delta
g_{i^{\star}j^{\star}}$ are harmonic tensors of the type specified by their
indeces. Hence we can immediately write:
$$  \delta g_{ij^{\star}}~=~c_{\alpha} \;  U_{ij^{\star}}^{(\alpha)}
\eqno(2.28)$$
where $c_{\alpha}$ are  20 real coefficients. They parametrize the deformations
of the K\" ahler class. On the other hand,  using the holomorphic 2-form, any
harmonic tensor with two antiholomorphic indeces $t_{i^{\star}j^{\star}}$ can
be written as  the following linear combination:
$$ t_{i^{\star}j^{\star}}~=~- d_{ \alpha}^{\star} \; {{1}\over{|| \Omega ||^2
}} \; {\overline \Omega}^{k}_{i^{\star}}  \;  U_{kj^{\star}}^{\alpha}
\eqno(2.29)$$
where raising and lowering of the indeces is performed by means of the fiducial
metric and where $d_{ \alpha}^{\star}$ are constant  complex coefficients.
Since $h^{2,0}=1$
it follows that, of the 20 independent linear combinations appearing in (2.29),
only one leads
to an antisymmetric $t_{i^{\star}j^{\star}}$; all the other combinations
produce
a symmetric tensor $t_{i^{\star}j^{\star}}$. Hence we can choose a basis of the
(1,1)-harmonic forms such that:
$$ {\overline \Omega}_{i^{\star}j^{\star}}~=~- \; {{1}\over{|| \Omega ||^2 }}
\; {\overline \Omega}^{k}_{i^{\star}}  \;  U_{kj^{\star}}^{0} \eqno(2.30a)$$
$$ - \; {{1}\over{|| \Omega ||^2 }} \; {\overline \Omega}^{k}_{i^{\star}}  \;
U_{kj^{\star}}^{a}~=~S^{a}_{i^{\star}j^{\star}}~=~S^{a}_{j^{\star}i^{\star}}
{}~~~~~~~~(a=1,.....,19) \eqno(2.30b)$$
The 19 symmetric tensors $S^{a}_{i^{\star}j^{\star}}$ provide a basis for the
expansion of the  antiholomorphic part of the metric deformation
$$\delta g_{i^{\star}j^{\star}}~=~d_{a} \; S^{a}_{i^{\star}j^{\star}}
\eqno(2.31)$$
The holomorphic part just is  the complex conjugate and it is expanded along
the complex  conjugate basis $S^{a}_{ij}$ : $\delta g_{ij}~=~d_{a} \;
S^{a}_{ij}$.
The  19 complex coefficients $d_{a}$ parametrize the complex structure
deformations
of the $K_3$ manifold.  Summarizing the 58 zero-modes of the metric emerge from
the following
counting:
$$ \# ~{\rm metric~zero-modes}~=
{}~h^{1,1}~+~2 \; \left ( \; h^{1,1} \; - \; 1 \; \right )\eqno(2.32)$$
This formula is just a consequence of $h^{2,0}=1$ and it has a meaning also for
non-compact manifolds,
like the instanton we consider later in this paper, as a counting of local
deformations.
For global deformations one has still to check if they can be reabsorbed by
diffeomorphisms.\para
In string-theory, the metric is not the only background field. We have also the
antisymmetric axion $B_{\mu\nu}$, whose curl $H_{\lambda\mu\nu}$ is identified
with the torsion $T_{\lambda\mu\nu}$, as we are going to see while discussing
the \simod formulation (see section 4). The zero-modes of the field
$B_{\mu\nu}$ are counted in a similar way to the case of the metric. From the
linearized field equation around the reference background, one concludes that
$\delta B_{ij}$,
$\delta B_{i^{\star}j^{\star}}$ and  $\delta B_{ij^{\star}}$ must be harmonic
tensors.
Because of the different symmetry  of the indices, this time we have:
$$ \delta B_{ij}~=~A  \; \Omega_{ij} \eqno(2.33a)$$
$$ \delta B_{ij^{\star}}~=~b_{\alpha} \; U^{\alpha}_{ij^{\star}}\eqno(2.33b)$$
where $A$ is a complex parameter and $b_{\alpha}$ are real parameters.
Hence we have 22 axion zero-modes that emerge from the following counting:
$$ \# ~{\rm axion~zero-modes}~=~h^{1,1}~+~2 \eqno(2.34)$$
Altogether there are $58\oplus 22 = 80 \; = 4 h^{1,1} $ zero
modes
of the field $g_{\mu\nu}+i B_{\mu\nu}$. In the next subsection we see
that
this counting agrees with the counting  of $N$=4 preserving marginal
operators in a
$(6,6)_{4,4}$-theory.
\vskip 0.1cm
The gravitino zero-modes are the zero-modes of the Rarita-Schwinger operator.
 Utilizing the standard trick of writing spinors as differential forms we
can  relate the number of these modes to the dimensions of the cohomology
groups.
Let
\def\Ga{\Gamma}
\def\om{\omega}
$$
\acomm {\Ga_{i}}{\Ga_{j}}= 0 ~~~~~;
{}~~~~~ \acomm {\Ga_{i^{\star}}}{\Ga_{j^{\star}}}=0
$$
$$
\acomm {\Ga_{i}}{\Ga_{j^{\star}}} =2\; g_{ij^{\star}} \eqno(2.35)
$$
be the Clifford algebra written in a well-adapted basis. A spin $\o{3}{2}$
field
$\psi_{\mu}$ can be written as follows:
$$ \psi_{i}~=(~\om_{i} \un \; + \; \om_{ij^{\star}} \Ga^{j^{\star}} \, + \,
\om_{ij^{\star}k^{\star}} \; \Ga^{j^{\star}k^{\star}} \; ) | \; \zeta \, > $$
$$\psi_{i^{\star}}~=(~\om_{i{\star}} \un \; + \; \om_{i^{\star}j^{\star}}
\Ga^{j^{\star}} \, + \,
\om_{i^{\star}j^{\star}k^{\star}} \; \Ga^{j^{\star}k^{\star}} \; ) | \; \zeta
\, > \eqno(2.36)$$
where the spinor $ | \; \zeta \, >$ satisfies the condition:
$$ \Ga_{i^{\star}} \; | \; \zeta \, >~=~\Ga^{i} \; | \; \zeta \,
>~=~0\eqno(2.37)$$
The field $\psi_{\mu}$ is a zero mode if the coefficients $\om_{....}$ in
(2.36) are
harmonic tensors.  Hence from $\om_{i}$ and $\om_{i^{\star}}$ we get
$h^{1,0}$ and $h^{0,1}$ zero-modes respectively. From $\om_{ij^{\star}}$
and
$\om_{i^{\star}j^{\star}}$ we obtain $h^{1,1}\;+\;h^{1,1}$ zero-modes.
Finally
$2 \; h^{1,2}$ zero-modes arise from $\om_{ij^{\star}k^{\star}} $ and
$\om_{i^{\star}j^{\star}k^{\star}}$. In view of the symmetries of the Hodge
diamond the total number of   zero-modes  for the gravitino field is given by
the formula
$$\# \; {\rm gravitino~zero-modes}~=~2 \; h^{1,1} \; + \; 4 \; h^{1,0}
\eqno(2.38)$$
In the case of $K_{3}$ the above number is $40$.
\ssvsk\noi
Finally, for the $E_6$ neutral WZ multiplets, the fermion has two zero-modes
of the same chirality (in correspondence with $H^{0,q}$), while the scalar
has just the (trivial) zero-mode corresponding to $H^{0,0}$.
\svsk
\noi{\bf The {\theory 64} theory and the moduli operators}
\svsk
A fundamental role is played by those fields of the $N$=4 theory
that can be identified with the abstract (1,1)-forms of the  associated
manifold
[13,sec.VI.10].
The $N$=4 algebra  contains the \se, four supercurrents and  three currents
$A^i$ that close
an $SU(2)_1$ algebra. In a (4,4)-theory there is a realization of these
operators both in the left and  in the right sector. The fields of the theory
are organized in representations  of $SU(2)_L \otimes SU(2)_R$. We denote by
$\Phi\sp{h,\t h}{J,\t J}^{m,\t m}$ a primary conformal
field with left and right dimensions $h,\t h$ and isospins $J, \t J$,
and with third components $m,\t m$.\para
Consider for example the left sector. The $SU(2)_1$can be bosonized in
terms of a single free boson $\tau (z)$:
$$
A^3=\o i{\sqrt{2}}\derp\tau\msk ;\msk A^{\pm}=e^{\pm i\sqrt{2}\tau}
\eqno(2.39)
$$
The spectral {f}low of the $N=2$ theories is extended to a ``multiplets of
spectral {f}lows'':
$$
\Phi{\sp hJ}^m =e^{im\sqrt{2}\tau}{\hat\Phi}^{(h-m^2)}\eqno(2.40)
$$
where ${\hat\Phi}^{(h-m^2)}$ is a singlet of \su of conformal weight
$h-m^2$.\para
For example a doublet of \su ,$\Psi\sp{1/2}{1/2}$,
made of an $N=2$ chiral and an antichiral field of weight $1/2$,
(note that the charge respect
to the $U(1)$ of the $N=2$ contained in the $N$=4 is twice the third
component of the isospin) in the NS sector is related by the spectral
{f}low (2.40) to an \su singlet in the R sector:
$$
\Psi\sp{1/2}{1/2}^{\pm\unmezzo}=e^{\pm i\o{\tau}{\sqrt{2}}}\Psi
\sp{1/4}{0}
\eqno(2.41)
$$
We use the convention of giving the same name to fields related
by spectral {f}low, distinghuishing them when necessary by their weight and
isospin.\para
As explained in [13], the $N$=4 analogues of the $(c,c)$
and $(c,a)$ fields of weight $(\unmezzo,\unmezzo)$, which play the role of
 ``abstract'' (1,1)- and (2,1)-forms in the \theory 92 theory,  is given by
those primary fields of the \theory 64 CFT that are of the form
$$
\Psi_A\sp{\um,\um}{\um,\um}
\eqno(2.42)
$$
and correspond to the lowest
components in a  short representation of the $N$=4 algebra. In (2.42) the index
$A$ runs on $ h^{1,1} $ values.  Focusing on the left sector a short
representation is made of the following set of fields
$$
\Psi\sp{1/2}{1/2}^a(z)\ssk ,\ssk \Phi\sp 10(z)\ssk ,\ssk \Pi\sp 10(z)
$$
satisfying the OPEs
$$
\eqalign{
&\cG a(z)\Psi^b(w) = {\epsilon^{ab}\Phi (w)\over z-w} + reg.\cr
&\cGb a(z)\Psi^b(w) = {\delta^{ab}\Pi (w)\over z-w}+ reg.\cr
&\cG a(z)\Phi (w)=\cGb a(z)\Pi (w)=0\cr
&\cGb a(z)\Phi (w) =2\epsilon^{ab}\partial\left({\Psi^b(w)
\over z-w}\right)+reg.\cr
&\cG a(z)\Pi (w) = -2\delta^{ab}\partial\left({\Psi^b(w)
\over z-w}\right)+reg.
          }
\eqno(2.43)
$$
where $\cG a(z),\cGb a(z),\sssk$ $a$=1,2 denote the supercurrents
organized in two  \su doublets. The fields $\Phi$ and $\Pi$
have dimension 1 and, being  the last components of an $N$=4 representation
(see
the last two of the OPEs (2.43)), when added (in suitable combinations
of the left and right sectors) to the Lagrangian they don't break its
$N$=4 invariance. We call them  the ``$N$=4 moduli''.\para
As already hinted,
the fields $\Psi_A\sp{\um,\um}{\um,\um}$ represent the abstract
(1,1)-forms on the manifold described by the \theory 64 -theory.
\par
As a first example of \theory 64 theory let's brie{f}ly consider that
associated with {f}lat space.  The $N$=4 algebra
(as an illustration we consider the left moving sector) is realized by the \se
$$
T(z)=-\unmezzo \derp X^{\mu}\derp X^{\mu}+\ps{\mu}\derp\ps{\mu}
\eqno(2.44a)
$$
by the supercurrents
$$
\eqalign{
G^0(z)&=\sqrt{2}\ps{\mu}\derp X^{\mu}\cr
G^x(z)&=\sqrt{2}\left(\hJ x\psi\right)^{\mu}\derp X^{\mu}
          }
\eqno(2.44b)
$$
and by the \su currents
$$
A^i(z)=-\o i2\ps{\mu}{\hJ i}_{\mu\nu}\ps{\nu}=
i(\ps 0\ps i+\unmezzo\epsu ijk\ps j\ps k)
\eqno(2.44c)
$$
The three tensors ${\hJ x}_{\mu\nu}$ are constant complex structures
and satisfy the quaternionic algebra; the role of their generalizations
to any $N$=4
theory will be discussed at length in section 5. For the moment it suffices to
know that they can be  explicitely  constructed, so that from their
expression (implicitely exhibited in eq.(2.44c)) we get
$$
\eqalign{
&G^1=\sqrt{2}\left\{\ps 0\derp X^1-\ps 3\derp X^2+\ps 2\derp X^3
-\ps 1\derp X^0\right\}\cr
&G^2=\sqrt{2}\left\{\ps 3\derp X^1+\ps 0\derp X^2-\ps 1\derp X^3
-\ps 2\derp X^0\right\}\cr
&G^3=\sqrt{2}\left\{-\ps 2\derp X^1+\ps 1\derp X^2+\ps 0\derp X^3
-\ps 3\derp X^0\right\}
         }
$$
In the left sector we can find two short representations , given by
$$
\eqalign{
\Psi_1\sp{1/2}{1/2}&=\twovec{\ps 0+i\ps 3}{\ps 2+i\ps 1}\ssk \cr
\Psi_2\sp{1/2}{1/2}&=\twovec{\ps 2-i\ps 1}{-(\ps 0-i\ps 3)}\ssk
         }
\eqalign{
&;\ssk\Phi_1\sp 10=-\derp X^2-i\derp X^1\ssk \cr
&;\ssk\Phi_2\sp 10=\derp X^0-i\derp X^3\ssk
         }
\eqalign{
&;\ssk\Pi_1\sp 10=-\derp X^0-i\derp X^3\cr
&;\ssk\Pi_2\sp 10=-\derp X^2+i\derp X^1
          }
\eqno(2.45a)
$$
Two analogous ones exist in the right sector. Multiplying
them in all possible
ways we obtain four abstract (1,1)-forms $\Psi_A\sp{1/2,1/2}{1/2,1/2}$.
For instance we can set:
$$
\eqalign{
\Psi_1\sp{\um,\um}{\um,\um}\;(z,\bz)~&=
{}~\Psi_1\sp{\um}{\um}\;(z) \;\Psi_1\sp{\um}{\um}\;(\bz)
\cr
\Psi_2\sp{\um,\um}{\um,\um}\;(z,\bz)~&=
{}~\Psi_1\sp{\um}{\um}\;(z) \;\Psi_2\sp{\um}{\um}\;(\bz)
\cr
\Psi_3\sp{\um,\um}{\um,\um}\;(z,\bz)~&=
{}~\Psi_2\sp{\um}{\um}\;(z) \;\Psi_1\sp{\um}{\um}\;(\bz)
\cr
\Psi_4\sp{\um,\um}{\um,\um}\;(z,\bz)~&=
{}~\Psi_2\sp{\um}{\um}\;(z) \;\Psi_2\sp{\um}{\um}\;(\bz)
            }
\eqno(2.45b)
$$
This number of $N=4$-moduli agrees with the Hodge diamond of {f}lat space
\footnote*{\small We refer by this to the Hodge diamond of the flat space
compactified to
a torus.}
$$
\matrix{ 1\cr {2\msk 2}\cr {1\msk 4\msk 1}\cr {2\msk 2}\cr 1\cr}
\eqno(2.46)
$$
According to (2.46) we have also two holomorphic 1-forms and two
antiholomorphic ones. At the level of CFT they are represented by operators of
the form $\Psi_{\ca A}\sp{\um,0}{ \um,0}$ and
$\Psi^{\star}_{{\ca A}^{\star}}\sp{0,\um}{0, \um}$, respectively.
The index $\ca A (\ca A^{\star})$ runs on 2=$h^{1,0}$ ( $h^{0,1}$ ) values.
The explicit expression of the  two (0,1)-forms can be taken to be
$$\eqalign
{\Psi^{\star}_{1^{\star}} \sp{ 0,\um}{ 0, \um}~&=~\un\sp{0}{0}(z)
\Psi_1\sp{\um}{\um}(\bz)\cr
\Psi^{\star}_{2^{\star}} \sp{ 0,\um}{ 0, \um}~&=~\un\sp{0}{0}(z)
\Psi_2\sp{\um}{\um}(\bz)\cr}\eqno(2.47)$$
The two (1,0)-forms have an analogous expression with the role of the left and
right-moving sectors interchanged.

Another interesting point is the identification of the spin fields with
the spectral {f}lows of the identity operator and of the lowest component of
the
short representations (2.45).
 This is a very important point because the spin fields  appear in  the fermion
emission vertices. If we are able to recast these vertices in an abstract
 \theory 64  language the extension from flat space to an instanton background
is guaranteed.  The gravitino emission vertex, for instance, that includes  the
proper gravitino and dilatino vertices,  in flat space has the following
expression [13, page 2063]:
 $$ \vertex {\mu}{\alpha}~=~\spinalph \derm \t X(\bz) e^{ik\cdot X(z,\bz)}
\un \twomat {3/8}{0}{-3/2}{0}\eqno(3.12a)$$
$$ \vertex {{\dot\alpha} \mu}{~}~=~\spinaldot \derm \t X(\bz) e^{ik\cdot
X(z,\bz)}
\un \twomat {3/8}{0}{3/2}{0}\eqno(3.12b)$$
the two formulae referring to the two possible chiralities.
The last operator in the
above formul\ae is a spectral flow of the identity in the internal theory.
In order to convert these expressions
to an abstract $N$=4 notation we need the interpretation of the operators
$S_{\alpha}\,(z)\;$ $\derm \t X(\bz) \;$ $ e^{ik\cdot X(z,\bz)} $ and
$S^{{\dot \alpha}}\,(z)
\;$ $\derm \t X(\bz)\;$ $e^{ik\cdot X(z,\bz)}$.
\para
To this effect we note that $\pmutilde~=~{\overline {\pa}} \, X^{\mu} \,
(z,\bz)$ is expressed by linear combinations of the operators $\t\Pi_1 \sp 10$,
$\t\Pi_2 \sp 10$,
$\t\Phi_1 \sp 10$ and $\t\Phi_2 \sp 10$, the right-sector counterparts
of those appearing in (2.45).
It remains to consider the spin fields.
The four free fermions can be bosonized in terms of two free bosons as
$$
\eqalign{
\ps 0\pm i\ps 3&=\pm c^{\pm}\,\, e^{\pm i\vfi 2}\cr
\ps 2\pm i\ps 1&=\pm c^{\pm}\,\, e^{\pm i\vfi 1}
         }
\eqno(2.49)
$$
where the signs and the cocycle factor($c^{\pm}=e^{\mp\pi p^1}$) are
arranged to reproduce the anticommutation properties of the fermions.
The \su currents of eq.(2.44c) can be reexpressed via eq.(2.49). In
particular
$$
A^{\pm}=\pm c^{\pm}\,\, e^{\pm i\vfi 2}e^{\mp i\vfi 1}
$$
However, we can rephrase all the algebra in terms of the vertex operators
$e^{\pm i\vfi 1}, e^{\pm i\vfi 2}$, eliminating the need of
preserving  anticommutation relations (these operators anticommute
 with themselves and  commute with each other). Then the \su currents
are simply given by
$$
\eqalign{
A^3&=\o i2 (\derp\vfi 2 -\derp\vfi 1)\cr
A^{\pm}&=e^{\pm i\vfi 2}e^{\mp i\vfi 1}
         }
\eqno(2.50)
$$
Comparison with the standard bosonized form (2.39) is immediate. We get:
$$
\tau=\o 1{\sqrt{2}}(\vfi 2-\vfi 1)
$$
so that the spectral {f}low of eq.(2.40) is rewritten as
$$
\Phi{\sp hJ}^m =e^{im(\vfi 2-\vfi 1)}{\hat\Phi}^{(h-m^2)}
\eqno(2.51)
$$
The fields $\Psi_1,\Psi_2$ of eq.(2.45), as doublets with respect to the
currents
(2.50) are given by
$$
\Psi_1=\twovec{e^{i\vfi 2}}{e^{i\vfi 1}}\ssk ;\ssk
\Psi_2=\twovec{e^{-i\vfi 1}}{e^{-i\vfi 2}}
\eqno(2.52)
$$
We can single out the spectral {f}low and find their Ramond partners:
$$
\Psi_1\sp{1/2}{1/2}=\twovec{e^{i\vfi 2}}{e^{i\vfi 1}}=\twovec{e^{\o i2
(\vfi 2-\vfi 1)}}{e^{-\o i2 (\vfi 2-\vfi 1)}} e^{\o i2(\vfi 2+\vfi 1)}
=spectral\ssk {f}low\sssk\cdot \Psi_1\sp{1/4}{0}
$$
$$
\Psi_2\sp{1/2}{1/2}=\twovec{e^{-i\vfi 1}}{e^{-i\vfi 1}}=
\twovec{e^{\o i2 (\vfi 2-\vfi 1)}}{e^{-\o i2 (\vfi 2-\vfi 1)}}
e^{-\o i2(\vfi 2+\vfi 1)} =spectral\ssk {f}low\cdot \Psi_2\sp{1/4}{0}
$$
Therefore we see that these R fields are just the two spin fields
of positive chirality. Indeed, once the fermions have been bosonized as in
eq.(2.49),
the spin fields, corresponding
to the weights of the $SO(4)$ spinor ($s$) and antispinor ($\bar s$)
representations,  are expressed as follows
$$
\eqalign{
+\sssk {\rm chirality}\ssk(s\sssk{\rm rep}):\lsk
&S^1=e^{\o i2\vfi 2+\o i2\vfi 1}\cr
&S^2=e^{-\o i2\vfi 2-\o i2\vfi 1}\cr
-\sssk {\rm chirality}\ssk(\bar s\sssk{\rm rep}):\lsk
&S^{\dot 1}=e^{\o i2\vfi 2-\o i2\vfi 1}\cr
&S^{\dot 2}=e^{-\o i2\vfi 2 +\o i2\vfi 1}
          }
\eqno(2.53)
$$
Finally note that the spin fields of negative  chirality form
a doublet under the $SU(2)_{L}$ and are related through spectral
{f}low to the identity operator:
$$
\twovec{S^{\dot 1}}{S^{\dot 2}}\sp{1/4}{1/2}=\twovec{e^{\o i2(\vfi 2
-\vfi 1)}}{e^{-\o i2(\vfi 2-\vfi 1)}}= spectral\ssk {f}low\sssk
\cdot\un\sp 00
\eqno(2.54)
$$
Comparing these results with equations (2.48) we see that, in flat space, the 8
gravitino zero-modes of positive chirality are given by the left spectral flow
of the abstract (1,1)-forms (2.45b), while the 8 zero-modes of negative
chirality are given by the left spectral flow  of the (0,1)-forms (2.47). In
both cases, the right-moving part of the operator is SUSY-transformed to the
last
multiplet-components. This is in perfect agreement with formula (2.38 ) and
with the Hodge diamond of
flat space (2.46). In the case of the $K_3$-manifold only the positive
chirality zero-modes are
present, since the analogues of the (0,1)-forms
(2.47) do not exist ($h^{1,0}=0$).
\ssvsk
This general discussion suffices to illustrate the idea of the generalized
$h$-map and of  the use of \theory 64 theories as a description of
gravitational
instantons backgrounds. In Appendix A we give an exhaustive  list of the
emission vertices for
for all particle zero-modes in a generic \theory 64 case.
\ssvsk
In the next section we begin the discussion of the instanton of [6]
starting from the effective supergravity Lagrangian.
\vfill\eject


\centerline{\bf 3. New Minimal N=1, D=4 Supergravity}
\centerline{\bf and Asymptotically f{lat}  Dilaton-Axion Instantons}
\svsk
The low-energy effective lagrangian of heterotic superstring theory is a
supergravity lagrangian. If the superstring is compactified on a 6-dimensional
Calabi-Yau manifold, then this effective lagrangian corresponds to that of an
$N$=1, $D$=4 supergravity  [15] which, when restricted to the bosonic fields,
has the following well known general form:
$${\cal L}^{(N=1)}_{Bosonic}~=~\sqrt{-g}\; \left [ \; {\cal R} \; -
\; g_{IJ^{\star}} \; \nabla_{\mu}z^{I}\; \nabla^{\mu}z^{J^{\star}}\; - \;
{{1}\over{4}} \; Re f_{\alpha\beta}(z) F_{\mu\nu}^{\alpha} \; F^{\beta \mu\nu}
\; - V(z,{\bar z}) \; \right ]-$$
$$- \; {{1}\over{8}} \; Im f_{\alpha\beta}(z)\; F^{\alpha}_{\mu\nu}
\; F^{\beta}_{\rho\sigma} \; \varepsilon^{\mu\nu\rho\sigma} \eqno(3.1)$$
In (3.1), besides the gravitational field, described by the metric
$g_{\mu\nu}$, one
has the gauge fields $A_{\mu}^{\alpha}$ belonging to the Lie algebra of a
suitable gauge
group $G_{gauge}$ and a set of complex scalar fields $z^{I}$ corresponding to
the bosonic
content of the Wess-Zumino scalar multiplets.  The kinetic term of these
scalars has
a \simod form in terms of a K\" ahler metric
$g_{IJ^{\star}}=\partial_{I}\partial_{J^{\star}}G(z,{\bar z})$ . The real K\"
ahler function
$G(z, {\bar z})$, besides determining the kinetic term, determines also the
scalar potential term, via the celebrated formula [15,16]:
$$ V(z,{\bar z})~=~4 \left ( \; g^{IJ^{\star}} \; \partial_{I}G \;
\partial_{J^{\star}}G \; - \;
3 \; \right ) \; e^{G} \; - \; g^{2} \left [ Re f_{\alpha\beta} \right ]^{-1}
\; {\cal P}^{\alpha} \; {\cal P}^{\beta} \eqno(3.2)$$
To be precise $G(z, {\bar z})$ is not exactly the K\" ahler potential of the
metric
$g_{IJ^{\star}}$, rather it is the norm squared
$$G(z,{\bar z})~=~K(z,{\bar z})\; + \;  ln | W(z) |^2 ~=~ln \; || W(z) ||^2
\eqno(3.3)$$
of a holomorphic section  $W(z)$ in a line bundle
${\cal L}$, whose first Chern class is the K\" ahler class
$\omega = ig_{IJ^{\star}} dz^{I}\wedge d{\bar z}^{J^{\star}}$ of that metric :
$$ \omega~=~\partial {\overline {\partial}}|| W ||^2 \eqno(3.4)$$
The holomorphic section $W(z)$ is named the superpotential and the hermitean
metric
$K(z,{\bar z})$ of this line bundle is the proper K\" ahler potential. In
addition, if the gauge group has a linear action $\delta
z^{I}=(T_{\alpha})^{I}_{J} \; z^{J} $ on the scalar
fields, then the contribution to the scalar potential (3.2) proportional to the
gauge
coupling constant $g^2$ is given in terms of Killing vectors prepotentials of
the form
$$ {\cal P}^{\alpha}~=~- \;i\; \partial_{i} G (T_{\alpha})^{I}_{J} \;
z^{J}\eqno(3.5) $$
When the action of the gauge group is non linear, then the expression of ${\cal
P}^{\alpha}$
is more complicated, but we shall not be interested in this case. Finally, the
gauge coupling function $f_{\alpha\beta}(z) $ is some holomorphic function with
$adjoint\otimes adjoint $ indices of the gauge group.
In the case of Calabi-Yau compactifications [9] of the heterotic string the
gauge group is
$E_6 \otimes {E_8}^{'}$ and the scalar multiplets (all neutral under
${E_8}^{'}$) are of
six different types [17,18]:
$$ z^{I}~=~\cases {\;S~=~{\rm dilaton-axion~field}\cr
\; {\cal M}^{a}~=~{\rm (2,1)-moduli}~{(a=1,......,h^{2,1})}\cr
\; {\cal M}^{i}~=~{\rm (1,1)-moduli}~{(i=1,......,h^{1,1})}\cr
\; {\cal C}^{a}~=~{\rm 27-charged~fields}~{(a=1,......,h^{2,1})}\cr
\; {\cal C}^{i}~=~{\rm {\overline
{27}}-charged~fields}~{(i=1,......,h^{1,1})}\cr
 \; {\cal Y}^{u}~=~{\rm non-moduli~singlets}~{(u=1,......,\# End(T))}\cr}
 \eqno(3.6)$$
in correspondence with the cohomological properties of the internal space,
dictated by
its Hodge numbers $h^{1,1}, h^{2,1}$ and by the number of deformations of its
tangent
bundle $\# End(T)$. Of particular relevance are the moduli-fields,
that describe the deformations of the compactified manifold, and their special
K\" ahler geometry.  Indeed,  to lowest order in the charged fields and
non-moduli singlets,
the general forms of the complete K\" ahler potential and  complete
superpotential are respectively given by:
$$
K~=~- \; log \left (  S + {\bar S}  \right ) \; +
{\hat K} ( {\cal M},{\overline {\cal M} }) \;
+
\; {\cal G}_{a b^{\star}}({\cal M},{\overline {\cal M}})\;{\cal C}^{a}\;
{\overline {\cal C}}^{b^{\star}}\; $$
$$+ \; {\cal G}_{ij^{\star}} ({\cal M},{\bar {\cal M}})\;
{\cal C}^{i}\;{\overline {\cal C}}^{j^{\star}}~+~....\eqno(3.7)$$
and
$$W~=~{{1}\over{3}} \; W_{abc}({\cal M}) \; {\cal C}^{a} \;{\cal C}^{b} \;{\cal
C}^{c} \;
           + \;
{}~{{1}\over{3}} \; W_{ijk}({\cal M}) \; {\cal C}^{i} \;{\cal C}^{j} \;{\cal
C}^{k} \;
{}~+~.................\eqno(3.8)$$
where ${\hat K} ( {\cal M},{\overline {\cal M} }) $ is the K\" ahler potential
of the
moduli-space and $W_{abc}({\cal M}) ,W_{ijk}({\cal M})$ are the Yukawa
couplings.
These quantities are related by the peculiar identities of special geometry.

Notwithstanding the
importance of these fields,  in the present paper, we are rather interested in
the  first
term of eq.(3.7), namely in the universal $S$-field that includes both the
dilaton and the axion. The structure of (3.7) implies that this field spans an
$SU(1,1)/U(1)$ coset manifold and that the total scalar manifold is the direct
product of this coset with
some other K\" ahler manifold ${\cal K}^{'}$. That this is the case follows
from very
general considerations we shall now review. Furthermore it is just the presence
of  $S$ that allows for the existence of instantonic solutions that are
asymptotically f{lat}
and not only locally asymptotically f{lat}. To this effect we recall that
according to a very
interesting mechanism discovered by Konishi et al [3], gravitational
instantons might induce a non-perturbative breakdown of supersymmetry via their
contribution to
the functional integral. An explicit calculation was in fact  performed in
[3], utilizing the Eguchi-Hanson metric [2]. The problem is that, for a
correct implementation of this mechanism, the instanton should be
asymptotically f{lat}. This is
not the case of the Eguchi-Hanson metric, for which asymptotic f{lat}ness is
local and not
global.
The problem of finding asymptotically f{lat} gravitational instantons was
considered
several years ago by D'Auria and Regge[7].  They realized that in  order
to reconcile
the self-duality of the curvature with asymptotic f{lat}ness one needs an
``unsoldering"
of the principal Lorentz-bundle from the tangent bundle. This can be achieved
by writing  gravity in first order formalism and coupling it to a pseudoscalar
field, whose derivative becomes the dual of the 3-index torsion. Indeed D'Auria
and Regge
proposed a certain configuration that realizes the desired
instanton and that is a solution of an ad hoc
constructed lagrangian.  As we are going to see, their configuration  is just
equivalent
to the dilaton-axion instanton discovered by Rey [8] to be  an exact
solution of the string derived Supergravity lagrangian (3.1) with K\" ahler
potential (3.7).
What D'Auria and Regge missed  in their action and had to simulate with an ad
hoc
interaction term was just the dilaton. Indeed their pseudo-scalar was nothing
else but  the axion.
In a certain limit the dilaton-axion instanton corresponds to an exactly
solvable (4,4) superconformal
theory that has been discovered by Callan [6].  In later sections of this
paper we   use this examples and its associated (4,4)-theory to illustrate our
ideas on the
generalized $h$-map, studying also the corresponding moduli deformations. In
the
present section we discuss the derivation of this instanton in the context of
the effective low-energy  lagrangian, enphasizing the role of the New Minimal
formulation of Supergravity.
\svsk
The key point here is the observation that, indipendently from the
compactification scheme the effective supergravity lagrangian should contain
the coupling of
 a linear multiplet $(\phi,\chi,B_{\mu \nu})$ that arises directly  via
dimensional reduction from the dilaton and  $B_{\mu \nu}$ field of the ten
dimensional effective theory. In four dimension, this multiplet can be
transformed into an ordinary
 WZ multiplet by a ``duality transformation'' relating the $B_{\mu \nu}$ field
strength to an axion field:
$$\nabla_{\mu} A = {1 \over 24 } {\epsilon_{\mu \nu \rho \sigma} \over \modg}
e^{-2\phi} H^{\nu \rho \sigma} \eqno (3.9)$$
As we just recalled in matter-coupled 4-dim supergravity the complex bosonic
matter fields are interpreted as  the coordinates $z^I$ of a K\"ahler manifold
${\cal K}$.
For a generic theory, and if we derive the action from the Old Minimal
off-shell
formulation of supergravity [19], the manifold ${\cal K}$ is arbitrary: we
recall  that the Old minimal formulation is characterized by the presence of a
scalar auxiliary field appearing in the SUSY-transformation rule of the
gravitino. On the other
hand if we adopt the New Minimal formulation [20], characterized by the
absence of
this  scalar auxiliary field, then ${\cal K}$ cannot be arbitrary: it  is
constrained  by
conditions that imply the existence of a coordinate frame where the  the K\"
ahler function has the following form
$$G= \alpha \log (z + \zbar) + \hat G (z^i , \zbar^{i^*})
\eqno(3.10)$$
the indices  being split  as  follows $\{z^I\}=\{z,z^i\}$ and  $\hat G (z^i ,
\zbar^{i^*})$ being an arbitrary K\" ahler function for the remaining scalar
fields $z^i $, once the special
field $z$ has been subtracted. The parameter $\alpha$ is any real constant. In
other words the existence of a New Minimal Formulation requires a factorization
(at least a
local one) of the scalar manifold into:
$$ {\cal K}_{scalar}~=~{{SU(1,1)}\over{U(1)}} \; \otimes \; {\cal K}^{'}
\eqno(3.11)$$
These results were derived in [21].  In the same paper, it was also shown
that the
conditions for the existence of a New Minimal formulation are the same
conditions that
guarantee the possibility of  duality-rotating one of the WZ-multiplets to a
linear multiplet $(\phi,\chi,B_{\mu \nu})$, via equation (3.9).

In view of these very general results, it follows that a superstring derived
supergravity, since it includes a linear multiplet, has necessarily a K\" ahler
function of the form (3.10), and admits a New Minimal formulation.
The second statement is further supported by the results of [22], showing
that in heterotic string theory one cannot construct an emission vertex for the
scalar auxiliary field.

Having clarified this crucial point we proceed to discuss the derivation of the
dilaton-axion instanton in supergravities characterized by a scalar manifold of
type (3.11).  Using the New-Minimal Lagrangian we retrieve  as an exact
solution the  Callan et al configuration [6], that is also of the same
form as the one considered by D'Auria and Regge in [7].  Performing the
generalized Weyl-transformation that
maps the New into the Old Minimal theory, the Callan instanton flows into the
Rey instanton, characterized by an exactly f{lat} metric and a singular dilaton
and axion.

Let us then go back to eq.(3.7)  and concentrate on the K\" ahler function
$ G(S,\bar S) = - \log (S + \bar S)$.
When we consider the theory in Minkowski spacetime the fields of the dilaton
multiplet
$$\eqalign{&S = f + i g \cr &\bar S = f - i g}\eqno (3.12)$$
($f$ representing the original dilaton, $g$ being the axionic field)
span the factorized $SU(1,1)/U(1)$ part of the scalar  manifold, according to
eq.(3.11). It turns out, however, that, while  performing the Wick rotation to
reach the Euclidean region, (due to the $\epsilon$ symbol appearing in the
duality
transformation eq.(3.9)),
it is  also necessary to perform a  Wick rotation on the  scalar manifold.
Eq.(3.12)  becomes
$$\eqalign{&S = f + g \cr &\bar S = f - g}\eqno (3.13)$$
 From now on we will consider the Euclidean case, since we  search for
an instantonic solution. However, for convenience, we  continue to use the same
``complex'' notation as before the rotation.\par
Restricting our attention to the bosonic sector of the theory, in the New
Minimal formulation, according to the results of [21], the
curvature two-forms
\footnote*{\small Through all the paper, we omit the wedge symbol for
the exterior product of forms}
$$\eqalign{&R^{ab} = d \omega^{ab} - \omega^a_{\enskip c}
\null \omega^{cb} \cr &R^a = D V^a  \cr &R^\palla = d{\cal A} }$$
(${\cal A}$ being the K\"ahler connection on the scalar manifold )
are parametrized  as follows:
$$\eqalign{
&R^{ab} = R^{ab}_{\hskip 10pt cd } V^c \null V^d \cr
&R^\palla = F_{ab} \vabu \cr
&R^a = \kdue \epsiu t_b V_c \null V_d \hskip 1cm (D_a t^a = 0)\cr
&d z^I = \Zia V^a
           }\eqno(3.14) $$
the parameter $\kappa_2$ being a free constant.
The fields in this formulation are obtained from those in the Old
Minimal one through a Weyl transformation,
$$V^a_{\new} = e^{\phi/2} V^a_{\old} \hskip 0.5cm \rightarrow
\hskip 0.5cm \Zia\null^{\new} = e^{- \phi/2} \Zia\null^{\old}
 \eqno(3.15a)$$
(for the bosonic fields).\par
In order for the transformation to be succesful, it is
required that
$$\phi = \log \dzb G \eqno(3.15b)$$
where $G(z,{\bar z})$ is given by equation (3.10).
The auxiliary fields are then expressed as
$$\eqalignno{
&t^{\new}_a = Im (\dif \ssk \Zia ) \hfill &(3.16a)\cr
&{\cal A}^{\new} = Im \ssk dG - (2 \kuno + 1) Im (\dif \ssk\Zia)  &(3.16b)
            }$$
$\kuno$ is a constant appearing in the New Minimal parametrization of
the fermionic curvatures for whose expression we refer to [21].
One sees that having  the dilatonic WZ multiplet in the game, we are
precisely in the situation of eq.(3.10) with  $\alpha=-1,z = S$.
Hence in the case
of the superstring effective lagrangian we obtain the identification
$$\phi=\log {1 \over S + \bar S} = \log {1 \over 2 f}\eqno(3.17)$$
The first order formulation of the bosonic New Minimal
lagrangian is given by
$$\eqalign{{\cal L}=&e^{- \phi} \biggl\{R^{ab}\null \duevu + 4 \kdue
t_a R_b \null
\vabu +(\dif\ssk \Zia + \disf\ssk \Zisa)R_b \null \duevd
+ \hfill \cr &+({2 \over 3} \gij -
\dif\ssk \djsf ) \left[ \Zia d z^{J^*} + \Zjsa d z^I \right]
\null \3vd +
\cr &- \left[\dif\ssk \partial_J \phi\ssk \Zia d z^J +
\partial_{I^*} \phi\ssk \djsf \ssk
Z^{i^*}_a d z^{j^*}\right] \null \3vd +\cr &+ \biggl[ -{1 \over 4}
({2 \over 3} \gij -\dif\ssk \djsf ) Z^I_r Z^{{J^*}r} +
{1 \over 8} \bigl(\dif
\ssk\partial_J \phi \ssk Z^I_r Z^{Jr} \cr &+
\partial_{I^*} \phi \ssk \djsf \ssk Z^{I^*}_r
Z^{{J^*}r}\bigr)+ {1 \over 2} \kdue^2 \ssk t_r t^r  -M \biggr]\4v \biggr\}
}\eqno(3.18) $$
where the scalar potential takes the new form
$M$=$-{2 \over 3} \bar e^2 \bigl(3 + \alpha - g^{i{j^*}} \partial_i
\hat G \enskip \partial_{j^*} \hat G \bigr)\cdot$ $ e^{\hat G}
{(z + \zbar)}^{\alpha}$, to be compared with eq.(3.2)
\par
Recalling that in $2^{nd}$ order formalism
$$R^{ab} \null \duevu = {1 \over 2} {\cal R} \modg d^4 x$$
where ${\cal R}$ is the curvature scalar, and
comparing the lagrangian in eq.(3.18) with
the effective action used by Callan et al.[6],
$$S={1 \over 2} \int \modg d^4 x e^{2 \Phi} \left( {\cal R} +...
\right)$$
we have the correspondence
$$\phi = -2 \Phi \eqno(3.19)$$
We can consistently search for a particular solution
in which only the dilaton and the
axion field are relevant, setting the other fields $z^i$ to constant
values $c^i$  such that
$$\partial_i M(c^i) = 0 \hskip 0.5cm ; \hskip 0.5cm M(c^i) = 0 \eqno(3.20)$$
We furthermore impose the radial ansatz
$$\eqalign{
&V^a = e^{-\lambda(r)} e^a \hskip 0.9cm
(r^2 = x_a x^a) \cr
&S=S(r) \hskip 0.3cm {\rm i.e.} \hskip 0.3cm f=f(r), g=g(r)
} \eqno(3.21)$$
Recalling that
$$\gssb= {1 \over (S + \bar S)^2} = {1 \over 4 f^2}$$
and using as f{lat} vierbeins the following ones
$$\cases {e^0 = dr \cr e^i =-(r/\sqrt{2}) \Ome i}$$
with $\Ome i$=$SU(2)$-Maurer-Cartan forms such that
$d\Ome i = -(\epsu ijk/\sqrt{2})\Ome j \Ome k$,
the variational equations obtained from the lagrangian (3.18) read:
\vskip 0.3cm\noi
$\bud$ Matter equations ($g- ,f-${\it variations respectively})
$$ \eqalignno{
&{g" \usf} - 2 \lambda ' {g' \usf} + {3 \over r}
{g' \usf} -
{g' \usf} {f' \usf} =0  &(3.22a) \cr
&-12 \lambda " + 12 (\lambda ')^2 -
36 {\lambda ' \over r}+\biggl[{f" \usf} - 2 \lambda '{ f' \usf} +
{3 \over r} {f ' \usf} - {1 \over 2} \biggl({f' \usf}\biggr)^2
\biggr] +\biggl({g' \usf}\biggr)^2 = 0 &(3.22b)
              }
$$
$\bud$  Einstein equations ($V^d-${\it variation})
$$
\eqalignno{
&8\lambda " - 4 (\lambda ')^2 + 16 {\lambda' \over r}
- 4 \lambda '
{f ' \usf} - { 8 \over r} {f ' \usf} +2 \biggl({f' \usf}\biggr)^2 +
\biggl({g' \usf}\biggr)^2 = 0  &(3.22c) \cr
&-12 (\lambda ')^2 +24 {\lambda '
\over r} +12 \lambda' {f' \usf} -{12 \over r} {f' \usf} -2\biggl(
{f' \usf} \biggr)^2 -\biggl({g' \usf}\biggr)^2 = 0 &(3.22d)
            }
$$
\vskip 0.1cm
\noi primes meaning derivatives with respect to $r$.
One sees that under the position
$\lambda ' = \um (f'/f)$
the Einstein equations reduce to a single expression,
$(f'/f)^2 -(g'/f)^2=0$, requiring
$$f' = \pm g' \eqno(3.23)$$
Inserting the above conditions into the matter equations, the following
solution is obtained:
$$\lambda = \log \left({ r/R_0 \over \sqrt{1+ \left(r/R_1 \right)^2}
}\right) \hskip 0.5cm ; \hskip 0.5cm f = {1 \over c}
{(r/R_0)^2 \over 1+\left(r/R_1 \right)^2} \eqno(3.24)$$
where $c,R_0,R_1$ are arbitrary constant,which clearly reproduces
the metric
configuration of the one-instanton solution of Callan et al.[6].
For the choice $c=2$ we have indeed
$$\eqalign{&V^a = e^{- \Phi} e^a \cr
&e^{-2 \Phi} = e^{-2 \lambda} =
{1 \over 2 f} = \left(R_0 \over R_1 \right) + \left(R_0^2 \over r^2
\right)}\eqno(3.25)$$
which gives the correspondence
$$\left(R_0 \over R_1\right)^2 = e^{-2 \Phi_0} \hskip 0.5cm ;
\hskip 0.5cm R_0^2 = n \eqno(3.26)$$

The configurations leading to the $SU(2) \times {\bf R}$ model,
associated with a solvable (4,4)-theory is
obtained in the limit $R_1\longrightarrow \infty$.\para
For example in this last case it's easily checked that also the
expression for the torsion agrees:
making  formula (3.16a) explicit we find
$$t_a = {1 \over 2} {1 \over S + \bar S} \left( \partial_a S -
\partial_a \bar S \right) = {1 \over S + \bar S} \partial_a g =
{\partial_a g \over 2f}$$
and inserting this in  the relation (3.23), we get
$ t_a =\um\partial_a \log{f}$, that is
$$\eqalign { &t_i =0  \cr &t_0 = {1 \over 2} V_0^r
(\log {f})' =
e^{\Phi} (\log {f})' = {r \over \sqrt{n}} {1 \over r} =
{1 \over \sqrt{n}} }$$
{}From the parametrization (3.14) we obtain thus
$$\eqalign{ &T^0 = 0 \cr
&T^i = \kdue {\epsilon_{ijk} \over \sqrt{n} }
V^j \null V^k }\eqno(3.27)$$
which agrees with the expression of the torsion for this particular
solution, as obtained by Callan et al.,with the choice $\kdue = 1$.
\svsk
It is then clear that the configuration
$$
\eqalign{
&ds^2=\emenoduef (dx)^2 \hskip 1.2cm\cr
&\emenoduef = A + {2k\over r^2}
         }\leftrightarrow\lsk
\eqalign{
&\V a=e^{-\Phi} e^a \cr
&\Phi=\log{(A+\o {2k}{r^2})^{-\o 12}}
         }
\eqno(3.28)
$$
$$H_{abc}=\o 13 \epsild abcd \derp_d\Phi$$
is an exact euclidean solution of the effective superstring lagrangian in the
New-Minimal
formulation. When transformed back to the Old-Minimal formulation, by means of
eq.s (3.15), this configuration becomes the dilaton-axion instanton found by
Rey [8]. This is obvious from the fact that the metric in the Old-Minimal
formulation
becomes the f{lat} one.
\svsk
\centerline{\bf 4. The rheonomic description of $\sigma$-models }
\centerline{\bf with dilaton and axion coupling}
\svsk
As we explained in the introduction, our basic goal is the study of
gravitational instanton configurations that correspond to exact solutions of
heterotic string theory, compactified on Calabi-Yau manifolds.  An example of
such a configuration that is an exact solution of the effective low-energy
lagrangian was discussed in the previous section: its basic feature is the role
played by the dilaton and axion fields.

We are now more demanding and we look for configurations that are exact
solutions of string theory. These configurations correspond to suitable c=6
superconformal
field-theories that describe the 4-dimensional space and  that can be adjoined
to  the  other  two conformal theories describing, respectively, the internal
space and the gauge degrees of freedom, namely a c=9 (2,2)-theory and the c=11
right-moving current algebra of $SO(6)\otimes E_8$.  This happens because of
the generalized $h$-map,  discussed in section 2.
Furthermore, as the characterizing feature of the internal theory is that of
being of
type (2,2), the characterizing feature of the instanton theory is that of being
of type
(4,4). This general result follows from the $SU(2)$-holonomy of the instanton
as discussed in section 2.  In the sequel the relation  between the
self-duality
of the curvature for the torsionful connection and the number  four of
world-sheet
supersymmetries will be analysed in more detail.

In any case, although a solution to order $O(\alpha^{'})$ of the equations of
motion
is not usually an all order solution, the lesson taught by the example of the
previous
section is that the dilaton and axion background fields are an essential part
of a stringy
gravitational instanton. Therefore, in order to discuss  the superconformal
field theory associated with instanton geometries,
we need to discuss the formulation of $\sigma$-models  with dilaton and axion
coupling.
To this effect we utilize
the rheonomy framework  [13].  In the first part of this section we consider
the bosonic
$\sigma$-model, in the second part we extend the construction to locally
supersymmetric
$\sigma$-models of (1,1) type.  Our results correspond to the generalization,
with dilaton
coupling, of the construction presented in [23]. Freezing the
two-dimensional gravitinos
one obtains the \simod action with global (1,1) supersymmetry, that can be
utilized
to discuss the structure of the corresponding superconformal theory.
The local construction, however, is essential
to obtain the stress-energy tensor and the two supercurrents (left and
right-moving).
In the type II version of string theory, these supercurrents are coupled to the
worldsheet
gravitinos.
After  the $h$-map to the heterotic string, only the left-moving current
corresponds
to a local world-sheet symmetry.
The right-moving supersymmetry ceases to be local and its role is the same
for X-space as it is for the internal compactified space, namely
it relates emission vertices of different particle modes. In the internal space
this leads
to remarkable consequences, in particular to the pairing between moduli fields
and
charged fields and to the special K\" ahler geometry of the moduli space. In
subsequent
sections we will discuss the analogue consequences for X-space.

After having established the formalism for (1,1) $\sigma$-models we shall
consider the
conditions under which the global supersymmetry of the same model is
accidentally extended to larger $N$. In particular
we shall consider the conditions for (4,4) global SUSY. This will be done in
section 5
 and, as we are going to see, in $d$=4 this  provides the link with instanton
geometry.
\svsk
\noi{\bf The Bosonic \simod}
\svsk
In corrispondence with a solution of the equations of motion derived from the
effective  bosonic lagrangian:
$$
{\ca L}_{eff} = \emenoduef (\caR - 4 D_{\mu} \Phi D^{\mu}
\Phi +...)
$$
that contains the metric $G_{\mu \nu}$ (i.e. equivalently the
vielbeins
$\V a$), the three form $H$ and the dilaton $\Phi$,
we write the action for the correspondent bosonic \simod
utilizing a geometric first order formalism:
$$
\eqalign{
&\1su4p\intb\V a(\Ppp{a} e^+ - \Pmm{a} e^-)+
\Ppp{a}\Pmm{a}
e^+ e^- - 2 \Phi \Rdue + p_{+} T^{+} + p_{-} T^{-} +
\cr
&+\1su4p \int_{\cM} H_{abc} \V a \V b \V c
        }
\eqno(4.1)
$$
Once rewritten in the \second order formalism, this action takes more
familiar form
$$
S=\meno1su4p \int_{\partial \cM} d\xp d\xm \left[G_{\mu \nu}(X) \derp
X^{\mu} \derm X^{\nu} + B_{\mu \nu}(X) \derp X^{\mu} \derm X^{\nu} \right]
\eqno(4.2)
$$
where $ds^2=G_{\mu \nu} dX^{\mu} \otimes dX^{\nu}=\V a \otimes \V a$ is
the target space line element
and the antisymmetric tensor $B_{\mu \nu}$ is such that
$H=2\sssk dB$.\para
Note that when the action is written as in
eq.(4.2)
it keeps no tracks of
the dilaton which  contributes only to the classical stress-energy
tensor of the model. This contribution is obtained in a simple way in the
\first
order formulation. In a similar way, when we consider the supersymmetric
extensions of the above model, the contributions of the dilaton to the
supercurrents
are also easily retrieved  from the \first order formulation.
\para
Let's briefly explain the somewhat unusual notations and the meaning of the
quantities appearing in eq.(4.1)[13,23]. In particular
 $e^+$ and $e^-$ are the vielbein on the
world-sheet $\partial \cM$, whose geometry is described by the structure
equations
$$
\eqalign{
&de^+ - \omedue e^+ =T^+\cr
&de^- + \omedue e^- =T^-\cr
&d\omedue =\Rdue
        }
\eqno(4.3)
$$
$\omedue, T^{\pm}, \Rdue$ are the two-dimensional spin connection,
torsion and curvature respectively.
 Classical conformal invariance of the model
allows the choice of the ``special conformal gauge'':
$$
e^+=d\xp \msk ; \msk e^-=d\xm \msk ; \msk \omedue=\Rdue=0
\eqno(4.4)
$$
where $z\!=\!x^0+x^1$ and $\bz\!=\!x^0-x^1$.
This is the choice we have used to obtain the \second order  form of the action
(2). More specifically ``after variation'' we can use eq.(4.4).
$\Pi^a_{\pm}, p_{\pm} $ are ``\first order fields'': they can be
reexpressed in terms of the usual dynamical fields upon use of the
equations obtained by varying in $\Pi^a_{\pm}, p_{\pm}, \omedue$.
$$
\eqalign{
&{\rm Varying \ssk in}\ssk \Pi^a_{\pm}\ssk: \lsk \Pi^a_{\pm}=V^a_{\pm}
\cr
&{\rm Varying \ssk in}\ssk \omedue\ssk: \lsk p_{\pm}=\mp 2 \partial_{\pm}
\Phi = \mp 2 \partial_a \Phi V^a_{\pm}\cr
&{\rm Varying \ssk in}\ssk p_{\pm}: \lsk T_{+} = T_{-} = 0
         }
\eqno(4.5)
$$
In the present formalism, the general recipe to obtain the components
of the
\se is to vary the action with respect to the w.s. vielbiein, defining
$$
\varia S =\o{-1}{2\pi} \int \cT_{+} \varia{e^+} + \cT_{-} \varia{e^-}
\eqno(4.6)
$$
and  to consider the expansion $\cT_{+} = \cT_{++} e^+ + \cT_{+-} e^-$
(and the analogous one for $\cT_{-}$).
 The conformal invariance of the model implies
 that $\cT_{+-} = \cT_{-+} =0$ and one defines the usual
holomorphic and antiholomorphic part of the \se to be
$$
T(\xp)=\cT_{++} \msk ; \msk \wt T(\xm)=\cT_{--}
\eqno(4.7)
$$
For the model described by the action (4,1) varying, for example, in $e^+$, we
obtain:
$$
\varia S=\o{-1}{2\pi} \left(-\unmezzo \right)\int (\V a \Ppp a - \Ppp a
\Pmm a
e^- ) \varia{e^+} + p_{+} d\delta\! e^+
$$
Substituting eqs.(4.5),  we obtain:
$$
T(\xp)=\cT_{++}=- \unmezzo \Vz a \Vz a - \derp \derp \Phi
\eqno(4.8)
$$
\svsk
In view of our specific interest in the instanton configuration of eq.(3.28),
let us consider the
above \simod in the case where $d$=4 and the metric, dilaton and axion fields
are
chosen
as follows:
$$
\eqalign{
&ds^2=\emenoduef (dx)^2 \hskip 1.2cm\cr
&\emenoduef =  {2k\over r^2}
         }\leftrightarrow\lsk
\eqalign{
&\V a=e^{-\Phi} e^a \cr
&\Phi=\log{\o r{\sqrt{2k}}}
         }
\eqno(4.9)
$$
$$\Phi=H_{abc}=\o 13 \epsild abcd \derp_d\Phi$$
The above eq.s correspond to the limit $A=0$ of (3.28). In this limit the
manifold has
the curious and somewhat unwanted topology of $R \otimes SU(2)$, which is not
asymptotically
f{lat}.  Asymptotic f{lat}ness is instead ensured when $A$ is non vanishing.
Yet as we are
going to see at $A=0$ the corresponding \simod defines a solvable conformal and
superconformal field-theory. Hence this limit is quite worth to be considered.
In (4.9) $\{e^a\}$ is a set of vielbein for the f{lat} 4-dimensional space,
$r$ being
a radial coordinate and the remaining 3 coordinates being the coordinates of a
3-sphere. Indeed we choose to write the f{lat} metric as follows
$$
dx^2=dr^2+\o{r^2}{2} \Ome i \otimes \Ome i
$$
where $\Ome i$ are the \mcf of \su which satisfy the equations
$$
d\Ome i=- \o{\epsu ijk}{\sqrt{2}} \Ome j \Ome k
$$
so that
$\unmezzo \Ome i \otimes \Ome i$ is the metric on the three-sphere of
unit
radius. The metric of the configuration (4.9) becomes
$$
ds^2=2k \o{dr^2}{r^2}+k \Ome i \otimes \Ome i
\eqno(4.10)
$$
Redefining the radial coordinate as follows: $t=\sqrt{2k} \log{(r/\sqrt{2k})}$
we obtain:
$$
ds^2 = dt^2 + k \Ome i \otimes \Ome i
\eqno(4.11)
$$
(showing that the singularity in (4.10) is a coordinate artifact), while
the dilaton is linear in the coordinate $t$:
$$
\Phi = \o t{\sqrt{2k}}
\eqno(4.12)
$$
In correspondence with eq.(4.11) we choose the vielbeins as follows:
$$
\eqalign{
&\V 0 = dt\cr
&d\V i = -\sqrt{k} \Ome i
         }
\eqno(4.13)
$$
The only non-zero components of H in the Maurer-Cartan basis
$\{\Ome i\}$ turn out to be
$$
H_{ijk}=\o 13 \epsd ijk (-\sqrt{k})^3 \o{\partial \Phi}{\partial t}=
-\o k3 \o{\epsd ijk}{\sqrt{2}}
\eqno(4.13)
$$
(note that, with our choice of \mcf, $\sqrt{2}\epsd ijk $ are just
the structure constants of \su). \para
The \simod action corrisponding to the configuration we have described is
$S=S_t+S_{\rm WZW}$ where
$$
\eqalign{
&S_t=\1su4p \intb dt~(\Pi_{+} e^+ - \Pi_{-} e^-)
+\Pi_{+} \Pi_{-} e^+ e^- -\raduek t \Rdue + p_{+} T^+ + p_{-} T^- \cr
&S_{\rm WZW}=\o 1{4\pi}\intb -\sqrt{k}\Ome i (\Ppp i e^+ -
\Pmm i e^-)+ \Ppp i \Pmm i e^+ e^- -\o 13 \o k{4\pi} \intm
\o{\epsd ijk}{\sqrt{2}} \Ome i\Ome j\Ome k
        }
\eqno(4.14)
$$
Once rewritten in \second order formalism, these two actions take the simpler
form
$$
\eqalign{
&S_t=\meno1su4p \intb d\xp d\xm ~\derp t \derm t \cr
&S_{\rm WZW}=\o{-k}{4\pi} \intb d\xp d\xm ~\Ome i_{+} \Ome i_{-} -
\o 13 \o k{4\pi}\intm \o {\epsd ijk}{\sqrt{2}} \Ome i \Ome j
\Ome k
         }
\eqno(4.15)
$$
$S_{\rm WZW}$ is the correct expression for the action of the WZW model
realized at level $k$, and corresponds  to a CFT of central charge
$$
c_{\rm WZW}=\o {3k}{k+2}
\eqno(4.16)
$$
The field $ \varphi= - i t$ is a free scalar boson with background charge
$Q_{bk}=-i\raduek$. Indeed from the action (4.14), using the general
recipe provided by eq. (4.8), we obtain the following stress-energy tensor:
$$
T_t(\xp)=-\unmezzo (\derp t)^2 - \o 1{\sqrt{2k}} \derp \derp t
\eqno(4.17)
$$
which corresponds to a central charge
$$
c_t=1+\o 6k
\eqno(4.18)
$$
This follows from the general formula
$$ c~=~1~-~3 \, Q_{bk}^{2} \eqno(4.19)$$
substituting the value of the background charge.

As one sees the \simod on the configuration (4.9) is exactly conformal
invariant
at the quantum level and leads to a solvable conformal field theory: namely a
tensor product
of a Feigin-Fuchs model with a WZW-model (see ref.s [24,25,26] ).
This is the reason why we are particularly interested in this specific
instanton
that provides a good toy-model.

Actually, in order to discuss superstring theory, we rather need  the
supersymmetric
version of the just described \simod.  Strictly speaking, heterotic string
theory would
require a (1,0) supersymmetrization; however, in view of the $h$-map, we can go
one step
beyond and consider the case of (1,1) local supersymmetry.
Therefore we recall now some essential features of the geometrical
formulation of (1,1) supersymmetric \simod [23] and we include dilaton
contributions.
\svsk
\noi{\bf The  (1,1) locally supersymmetric \simod}
\svsk
One realizes a classical superconformal invariant theory in terms of
fields
living on a super-world-sheet with two fermionic coordinates
$\theta$ and $\bar \theta$ besides the two bosonic ones $z$ and
$\bar z$. The cotangent basis on the super world-sheet(the
``supervielbein'')
is given by the already introduced 1-forms $e^+, e^-$ and two
bidimensional
gravitinos $\zeta, \chi$.\para
The structure equations (4.3) are enlarged by the appearence of two
fermionic torsion 2-forms:
$$
\eqalign{
&\Tbu =d\zeta -\unmezzo \omedue \zeta \cr
&\Tci =d\chi +\unmezzo \omedue \chi
        }
\eqno(4.20)
$$
The ``curvatures'' $T^+, T^-, \Tbu, \Tci, \Rdue$ must satisfy the Bianchi
identities obtained by exterior differentiation of eqs.(4.3) and (4.20).
 This imposes a certain form for their parametrization, whose most relevant
part is:
$$
\eqalign{
&T^+ =\o i2 \zeta\zeta \cr
&T^- =-\o i2 \chi\chi
        }
\eqno(4.21)
$$
The superconformal invariance of this construction allows for the choice of a
``special superconformal gauge'' where
$$
\eqalign{
e^+=dz+\o i2 \theta d\theta \msk &; \msk e^-=d\bar z+ \o i2 \bar \theta
d\bar\theta \cr
\zeta =d\theta \msk &; \msk \chi=d\bar\theta
         }
\eqno(4.22)
$$
This is the choice we always use in \second order formalism (see
discussion after eq.(4.4)).\para
We describe superstring propagation on an arbitrary target manifold
$\Mtar$
by means of an embedding function $X^{\mu}(z,\bar z, \theta,\bar\theta)$
mapping the super world-sheet into $\Mtar$. We consider the
quantities
defining the geometry of $\Mtar$, such as its vielbeins and
spin-connection,
as superfield on the super world-sheet, and thus they can be expanded on the
cotangent
basis of this latter. In particular we set
$$
\V a =\Vp a e^+ + \Vm a e^- + \lambda^a \zeta + \mu^a \chi
\eqno(4.23)
$$
Also the torsion and curvature 2-forms of $\Mtar$ can be
expanded
in the various ``sectors'' on the super world-sheet. For example, the torsion,
defined by:
$$
d\V a + \omega^{ab} \V b = T^a = T^{abc} \V b \V c \eqno(4.24a)
$$
yields
$$
\eqalign{
&e^+ e^-: \msk -\nabla_{-} \Vp a + \nabla_{+} \Vm a - 2 T^{abc} \Vp b
\Vm c=0 \cr
&\hbox to3cm{\dotfill}
         }
\eqno(4.24b)
$$
(relations that we are always free to use because they are just
the ``pull-back'' of the original definitions).\para
The key point are the Bianchi identities of $\Mtar$ which become
differential equations
for $\V a$ as a super-wordlsheet function; that is, they determine the
eqs. of motion for $\Vp a, \Vm a, \lambda^a, \mu^a$ [23].
 The B.I. for the torsion
of
$\Mtar$  is $\nabla T^a = \nabla^2 \V a = R^{ab} \V b$ or, explicitly
$$
\eqalignno{
&\nabla^2 \Vp a = R^{ab} \Vp b &(4.25.a) \cr
&\nabla^2 \Vm a = R^{ab} \Vm b &(4.25.b) \cr
&\nabla^2 \lambda^a = R^{ab} \lambda^b &(4.25.c) \cr
&\nabla^2 \mu^a = R^{ab} \mu^b &(4.25.d)
         }
$$
Each of these equations can be analized in its various sectors. In
particular the $\lambda^a$ field equation, setting $\nabla_o\lambda^a=
0$, constraint compatible with the Bianchi identity:
$$
-\o i2 \nabla_{-} \lai a =- \curv abcd \lai b \mui c \mui d
\eqno(4.26a)
$$
is retrieved in the $\chi\chi$ sector of eq.(4.25.c)
and the $\mu^a$ field equation
$$
\o i2(\nabla_{+} \mui a + 2T^{abc}\mui b \Vp c)=-\curv abcd \mui b
\lai c\lai d
\eqno(4.26b)
$$
is retrieved in
the $\zeta\zeta$ sector of eq.(4.25d). Bianchi identities for the
curvature $R^{ab}$ do not give any new information.\par
Next one tries to write down an action defined on super world-sheet from which
both the
definitions (4.24) and the field equations follow as variational
equations.
To this purpose one starts writing down the most general geometrical
action defined on the super world-sheet which
respects invariance under Weyl rescalings and two-dimensional Lorentz
transformations, with undetermined coefficients; these latter are
fixed by
comparing the variational equations with parametrizations (4.24) and field
equations.\para
It turns out that  the projections of
the variational equations in $\varia \lambda^a$ and $\varia\mu^a$ is
sufficient
to fix all the coefficients. The super world-sheet action takes then the form
$$
\eqalign{
S=\1su4p \intb &(\V a- \lai a \zeta - \mui a \chi ) (\Ppp a e^+ - \Pmm a
e^-)
+ \Ppp a \Pmm a e^+ e^- + 2i \lai a \delm\lai a e^+ + \cr
&+ 2i\mui a \delp\mui a e^- + \lai a \V a \zeta - \mui a \V a \chi -
\lai a\mui a\zeta\chi +\o 43 i T_{abc}\lai a\lai b\lai c\zeta e^+ -\cr
&- \o 43 i T_{abc} \mui a\mui b\mui c\chi e^- + 4 R_{abcd} \lai a\lai b
\mui c\mui d e^+ e^- +\cr
&-2\Phi\Rdue +p_{+} T^+ + p_{-} T^- + p\bud \Tbu +
p\cid \Tci + \1su4p \intm H
         }
\eqno(4.27)
$$
The variation in $\varia X^{\mu}$, restricted to the sectors $\zeta\zeta, \sssk
\chi\chi$,
where it really corresponds to a supersymmetry variation, fixes
$$
T_{abc}~=~-3 H_{abc}~
\eqno(4.28)
$$
justifying our assumption that $T_{abc}$ is completely antisymmetric in its
indices.\para
The action (4.26) is a geometrical one on the super world-sheet, and is
therefore
invariant
against super-world-sheet diffeomorphisms. Its expression is however
uniquely
determined by its ``bosonic'' section $\zeta =\chi =0$, due to the
fact that the
components of the curvatures along the ``fermionic'' directions are
expressed
by eqs.(4.24) in terms of those along the ``bosonic'' (or ``inner'') ones.
This property is called ``rheonomy''. One can forget, if he wants to,
about
the super world-sheet and then the would-be diffeomorphisms in fermionic
directions appear
as supersymmetry transformations. For $\zeta =\chi =0$ the action
reduces to
$$
\eqalign{
S=\1su4p \intb &\V a (\Ppp a e^+ - \Pmm a
e^-)
+ \Ppp a \Pmm a e^+ e^- + 2i \lai a \delm\lai a e^+ +\cr
&+2i\mui a \delp\mui a e^- + 4\curv abcd \lai a\lai b
\mui c\mui d e^+ e^- +\cr
&-2\Phi\Rdue +p_{+} T^{+} + p_{-} T^{-} + \1su4p \intm H
         }
\eqno(4.29)
$$
The above action possesses a global (1,1) supersymmetry that is the remainder
of the local
one present when the gravitino fields are switched on. In the next section we
recall how,
for suitable target manifolds, this global (1,1)  SUSY extends to a global
(4,4) supersymmetry.

{}From the complete form (4.26) of the action, one can derive the super-\se
(i.e. the \se
and the supercurrent) extending eq.(4.6) to
$$
\varia S =\o{-1}{2\pi} \int \cT_{+} \varia e^+ + \cT_{-} \varia e^- + \cT\bud
\varia{\zeta} + \cT\cid\varia{\chi}
\eqno(4.30)
$$
Superconformal invariance requires
$$
\eqalign{
&\cT_{+-} =\cT_{-+}=\cT_{\bu -}=\cT_{-\bu}=\cT_{\ci +}=\cT_{+\ci}=0\cr
&\cT_{+\bu}=\unmezzo \cT_{\bu +} \msk ; \msk \cT_{-\ci}=
-\unmezzo\cT_{\ci -}
        }
\eqno(4.31)
$$
The surviving four independent components define the classical
holomorphic and antiholomorphic parts of \se and supercurrent:
$$
\eqalign{
&T(z)=\cT_{++}\cr
&G(z)=2\sqrt{2}\emp4\cT_{+\bu}
         };\lsk
\eqalign{
&\wt T(\zbar)=\cT_{--} \cr
&\wt G(\zbar)=2\sqrt{2} e^{-\o{3i\pi}{4}}\cT_{-\ci}
         }
\eqno(4.32)
$$
In the action (4.26) or (4.28) two different covariant derivatives appear,
$\delp$ and $\delm$, constructed with the two spin-connections $\omepm$,
defined as
$$
\eqalign{
&\omemi ab=\omerid ab - T_{abc} \V c\cr
&\omepi ab=\omerid ab + T_{abc} \V c=\omemi ab+2T_{abc} \V c
         }
\eqno(4.33)
$$
where $\omerid ab$ is the Riemannian connection, i.e. is such that
$d\V a + \omerid ab \V b=0$. The connection appearing in eq.(4.23) (the one
for which the torsion is $T^a$) is $\omega_{ab}=\omemi ab$. These connections
play an important role in the sequel.
\vfill\eject
\centerline{\bf 5.  Extended  global supersymmetry}
\centerline{\bf of the \simod and classical supercurrents}
\svsk
In the first part of this section we review the conditions for the existence of
additional
global supersymmetries in the (1,1)-locally supersymmetric \simod [27].

In the second part we discuss the specific conditions for (4,4) supersymmetry:
they imply a very particular structure of the target manifold that corresponds
to a generalization
with torsion of HyperK\" ahler geometry. This structure implies self-duality
and antiself-duality,
respectively, for  the two curvatures $R(\omem)$ and $R(\omep)$ and, as such,
it is the proper geometry for an instanton with torsion.

Finally in the third part we show how to construct the classical supercurrents
generating these additional supersymmetries.
\svsk
\noi{\bf Complex structures and extended SUSY}
\svsk
To discuss the additional supersymmetries, we formally introduce new fermionic
directions of the super world-sheet, adding to
the cotangent basis new ``gravitinos'' $\z x$ and $\c x$ so that the
parametrization (4.21) is extended to
$$
\eqalign{
&T^+=\o i2 (\zeta \zeta + \z x \z x)\cr
&T^-=-\o{i}{2} (\chi\chi +\c x \c x)
         }
\eqno(5.1)
$$
while the embedding of the extended super world-sheet in $\Mtar$ is described
by expanding the target-space vielbeins as follows:
$$
\V a=\Vp a e^+ +\Vm a e^- +\lai a\zeta +\cmJi xab\lai b\z x +\mui a
\chi +
\cpJi xab\mui b\c x
\eqno(5.2)
$$
(Note that the new terms do not introduce  any new dynamical quantities).
\para
 Consistency with the torsion definition and implementation of the
Bianchi Identities leads to constraints on the tensors $\cpmJi xab$, and
therefore to a
characterization of $\Mtar$.
\para
The torsion definition (4.24a): $\delm \V a = T_{abc} \V b\V c$ can now be
expanded in many sectors
\footnote*{\small  From now on we drop in all calculations
the superscript $\scriptstyle -$ for $\scriptstyle{\cmJ x, \delm} $ etc.}.
Using the sectors
$$
\eqalign{
\zeta\zeta &:\msk\o i2\Vp a +\nabla\bud\lai a + T_{abc}\lai b\lai c=0\cr
\zeta\z x &:\msk\nabla\bud(\cJi xab\lai b)+\nabla\bud^x\lai a+2T_{abc}
\lai b\cJi xcr\lai r=0\cr
\z x\z y &:\msk i \Vp a \delta^{xy}+\nabla\bud^x(\cJi yab\lai b) +
\nabla\bud^y
(\cJi xab\lai b)+2T_{abc}\cJi xbr\cJi ycs \lai r\lai s=0
         } \eqno(5.3)
$$
by looking at terms containing $\Vp a$, one finds:\para
for $x=y$
$$
\cJi xab \cJi xbr =-\delta_{ar} \msk {\rm i.e.}\msk (\cJ x)^2=-\un
\eqno(5.4a)
$$
for $x\not= y$
$$
\acomm{\cJ x}{\cJ y}_{ar}=0
\eqno(5.4b)
$$
It follows that the $\cJ x$ form a representation of the Clifford algebra.
{}From the
remaining terms in these equations, after some manipulations, one gets :
\para
for $x=y$, the condition  that the usual \nt relative to each  $\cJ x$ should
vanish:
$$
N_{abn}(\cJ x, \cJ x)=\delr_m \cJi xa{[b}\hskip 2pt
\cJi xm{n]}+\cJi xam \delr_{[b} \cJi xm{n]}=0
\eqno(5.5)
$$
and for $x\not= y$ analogous non-diagonal \ni conditions [27].
\para
{}From sectors $\chi\chi, \chi\c x,\c x\c y$ the same relations for
$\cpJ x$
are retrieved:
$$
\eqalign{
&(\cpJ x)^2=-\un\lsk ;\lsk \acomm{\cpJ x}{\cpJ y}=0\cr
&N_{abc}(\cpJ x, \cpJ y)=0
        }
\eqno(5.6)
$$
\para
Starting from the sector
$$
\z x\c  y :\msk\nabla\bud^x(\cpJi yab\mui b)+\nabla\cid^y(\cJi xab\lai b)
+2T_{abc}\cJi xbr\lai r\cpJi ycs\mui s=0
$$
and substituting the relations that follows from the other sectors
$\zeta\c x,\chi\z,\zeta\chi$
by considering the terms that contain  $\nabla\bud\mui a$ we come to the
conclusion that the two set of tensors should  commute:
$$
\comm{\cJ x}{\cpJ y}=0
\eqno(5.7)
$$
Now we can also consider the various sectors of the torsion Bianchi identities.
In particular from eq.(4.25.c) in the sector $\z x\z x$:
$$
\o i2\nabla_{+}\mui a+\nabla\bud^x\nabla\bud^x\mui a=-\curv abcd \cJi xcr
\cJi xds\lai r\lai s\mui b
$$
looking at the terms involving $\Vp r$ and using the field equation (4.26a) one
ends with
$$
\nabla_m \cJi xab=0
\eqno(5.8)
$$
while the other terms impose the condition:
$$
R={\cJ x}^T R \cJ x
$$
$R^{ab}=\curv abcd \V c\V d$ being the curvature two-form. This
coincides  with the integrability condition for eq.(5.8), namely
$$
\comm R{\cJ x}=0
\eqno(5.9)
$$
if
$$
{\cJ x}^T =-\cJ x
\eqno(5.10)
$$
which is just the hermiticity condition expressed in tangent indices.\para
Considering  the sector $\c x\c x$ of eq.(4.26.d) and  analizing terms
proportional to
$\Vm a$ one sees that the torsion terms are such that the analogue
of eq.(5.8)
is given by:
$$
\delp_m \cpJi xab=0
\eqno(5.11)
$$
At the same time in order for the other terms to reproduce the integrability
condition
$$
\comm {\sopra R+}{\cpJ x}=0
\eqno(5.12)
$$
(where ${\sopra R+}_{ab}$ is the curvature 2-form of the connection
$\omepi ab$) the hermiticity condition
$$
{\cpJ x}^T=-\cpJ x
\eqno(5.13)
$$
must be verified.
\svsk
Summarizing: {\it The condition to have $(N,N)$ supersymmetries is the
existence
of two sets of $N-1$ complex structures on the target space
(whose \ni tensors vanish), each  set realizing
a representation of the Clifford algebra, and the two sets commuting with
each other.
One  of the two sets, namely
$\cmJ x$, must be covariantly constant with respect to the connection $\omem$,
while the other one, $\cpJ x$  is covariantly constant with respect to $\omep$.
The target space metric should be hermitean with respect to all complex
structures.}
\svsk

\noi{\bf $(4,4)$ SUSY and generalized HyperK\" ahler Manifolds}
\svsk
Consider the case of exactly 3+3 additional supersymmetries. It is
easy to
see that if $\cJ 1$ and $\cJ 2$ are two complex structures satisfying
the above
requirements then $\cJ 3=\cJ 1 \sssk \cJ 2$ is another one.
Due to
the Clifford algebra requirement the set $\cJ x$ closes a quaternionic algebra:
$$
\cJ x \cJ y=-\delta^{xy}+\epsu xyz \cJ z
\eqno(5.14)
$$
The same holds true for $\cpJ x$.\para
 In the case of zero torsion,
a manifold $\Mtar$ with three  covariantly constant complex structures that
realize a quaternionic algebra, and respect to which the metric is hermitean
is said
to be  a  Hyper\K manifold. Indeed on $\Mtar$ there exixt three globally
defined 2-forms $\Ome x=\cJi xab
\V a\V b$ which are closed: $d\Ome x=0$. The role of these forms is
the generalization of that played for a \K space by the \K form $ \Omega ={\cal
J}_{ab} \V a\V b$.\para
Choosing a well-adapted basis of vielbeins one
can show that the holonomy group  ${\cal H}ol(H\!K_m)$ of a Hyper\K space
$H\!K_m$
with
dimension $4m$ is  contained in $Sp(2m)$ [28].  In particular a
four-dimensional
Hyper\K space has
a holonomy group contained in \su: the curvature 2-form is selfdual
or antiselfdual. Note that this is the requirement a manifold must
satisfy in order to be a gravitational instanton.\para
Let us what happens when we introduce torsion in the game.\para
Recall that the $\cJ x$ and $\cpJ x$  complex structures with both indices
lowered  are antysymmetric matrices.\para
In 4-dimensions we can construct a basis for  $4\times 4$  antisymmetric
matrices made by the following two
sets of three constant matrices, respectively named $\hJ x$ and $\tJ x$
($x=1,2,3$)\footnote*{\small $\scriptstyle \epsilon$ symbol vanishes on the
index 0}:
$$
\eqalign{
&\hJi xab=-(\delta_{a0} \delta_{bx}-\delta_{b0}\delta_{ax}+\epsd xab)\cr
&\tJi xab=(\delta_{a0} \delta_{bx}-\delta_{b0}\delta_{ax}-\epsd xab)
         }
\eqno(5.15)
$$
that is
$$
\eqalign{
&\hJ 1=\twomat {0}{i\si 2}{i\si 2}{0} \msk ; \msk
 \hJ 2=\twomat {0}{\un}{-\un}{0} \msk ; \msk
 \hJ 3=\twomat {-i\si 2}{0}{0}{i\si 2} \cr
&\tJ 1=\twomat {0}{-\si 1}{\si 1}{0} \msk ; \msk
 \tJ 2=\twomat {0}{\si 3}{-\si 3}{0} \msk ; \msk
 \tJ 3=\twomat {-i\si 2}{0}{0}{-i\si 2}
         }
\eqno(5.16)
$$
These matrices have the following properties:\para
$\bud$ each set $\hJ x$, $\tJ x$ gives a representation of quaternionic
algebra (5.14)\para
$\bud$ they are selfdual (resp anti-selfdual):
$$
\eqalign{
\hJi xab=\unmezzo \epsild abcd \hJi xcd \lsk &\leftrightarrow\lsk
\epsd ijk \hJi k0
=- \hJi xij\cr
\tJi xab=-\unmezzo \epsild abcd \tJi xcd \lsk &\leftrightarrow\lsk
\epsd ijk \tJi k0
=\tJi xij
         }
\eqno(5.17)
$$
$\bud$ all the $\hJ x$ commute with all the $\tJ x$.\para
\ssvsk
On a 4-manifold with (4,4) extended SUSY,
we have two sets of complex structures, $\cJ x$ and $\cpJ x$, that are
covariantly constant under the connection $\omem =\omer - T$ and
$\omep =\omer +T$, respectively, different for non-zero torsion,
so that $\cJ x$ and $\cpJ x$ cannot coincide. A priori the
matrices of both these sets can be expanded along the basis given by
$\hJ x, \tJ x$:
$$
\eqalign{
&\cJ x = \sxy xy \hJ y + \axy xy \tJ y \cr
&\cpJ x = \spxy xy \hJ y + \apxy xy \tJ y
        }
\eqno(5.18)
$$
For all the coefficients in the expansion (5.18) there are two
possibilities: they can be zero or, in order for the $\cJ x$ and
$\cpJ x$ to satisfy the quaternionic algebra, must be such that
$$
\eqalignno{
\sxy xp \sxy yp &=\delta^{xy} &(5.19)\cr
\epsu xyz \sxy zt &=\epsu pqt \sxy xp \sxy yq &(5.20)
           }
$$
The same  conditions hold for $\axy xy, \spxy xy, \apxy xy$.
Relations (5.19-20) mean that each of these $3\times 3$ matrices is
orthogonal, namely they belong to the adjoint representation of $SO(3)$.
\para
We can use a vector notation $\svec x$ ($\spvec x$) for the rows of
the matrix $\sxy xy$ ($\spxy xy$); if they are non zero,
these vectors constitute an orthonormal
basis in  three-dimensional space.\para
Let us  then consider the consequences of the fact that all the $\cJ x$
must commute with all the $\cpJ x$. Using the expansion (5.18) this
means that
$$
\svec x \wedge \spvec y =0 \msk ;\msk \avec x \wedge\apvec y =0 \msk
\forall\sssk x,y
\eqno(5.21)
$$
(here the symbol $\wedge$ denotes the usual exterior product of
three-dimensional vectors).
We can  expand the $\spvec y$ in the basis $\{\svec x\}$:
$$
\spvec y =c^y_p \svec p
$$
Suppose now that the $\svec x$ are different from zero. Then the
condition (5.21), upon use of eq.(5.20), states that
$$
c^y_p\svec x\wedge\svec p =\epsu xpq c^y_p \svec q =0
$$
implying $c^y_p=0$, that is $\spvec y=0$.\para
If the $\avec x$ were non-zero, then an analogous argument
would constrain the $\apvec y$ to vanish as well; then the $\cpJ x$ would
just be zero, which cannot be. The only allowed situation is the following
$$
\eqalign{
\cJ x&=\sxy xy \hJ y\cr
\cpJ x&=\apxy xy \tJ y
         }
\eqno(5.22)
$$
that is, the $\cJ x$ are selfdual while the $\cpJ x$ antiselfdual (or
viceversa).\para
Consider the curvature 2-form $R^{ab}$ relative to the connection
$\omemi ab$. Let $R$ be the matrix of components $R^{ab}$. It is an
antisymmetric matrix and as such it can be expanded as follows:
$$
R=A_x \hJ x+B_x \tJ x
$$
It must satisfy the integrability condition (5.9) for the covariant
constancy of $\cJ x$,
$$
\comm R{\cJ x}=0
$$
Inserting the expansions of $R$ and $\cJ x$ this means
$$
A_p \sxy xy \comm{\hJ p}{\hJ y}=2\epsu pyt A_p\sxy xy \hJ t=0
$$
The unique solution of this constraint is $A_p=0$; this
implies that $R$ is antiselfdual.\para
Repeating an analogous argument for the curvature ${\sopra R+}_{ab}$ of the
connection $\omep_{ab}$ we find that $\sopra R+$ is selfdual.
\svsk
\para
{\sl Summarizing:} {\it a 4-dimensional  target manifold $\Mtar$ of a
$(4,4)$-supersymmetric \simod is what we name a generalized
an Hyper\K manifold with torsion}
\svsk
{\sl DEFINITION:}
{\it A Generalized Hyper\K manifold with torsion admits two sets of  mutually
commuting complex structures that separately close the quaternionic algebra
(5.14) and that are covariantly constant, one set with
respect to the $\omem$ connection, the other set with respect to the $\omep$
connection.}
\svsk
In 4-dimensions the above definition implies that the curvatures constructed
from $\omem$ and $\omep$
are one antiselfdual and the other selfdual (or viceversa). Because of this
fact we can {\it identify} the concept of {\it 4-dimensional generalized
Hyper\K manifolds with torsion} with the concept of {\it axionic gravitational
instantons}
\svsk
\noi{\bf The classical supercurrents}
\svsk
Suppose that  a $(1,1)$ \simod on a manifold $\Mtar$
admits an extended $(4,4)$ supersymmetry. The 3+3 additional
supersymmetries are just global ones. The action on the bosonic world-sheet,
namely
eq.(4.29) is not modified at all: we
just find that it is invariant against additional transformations. The novelty
is
that we can now search for the complete form of the
action on the extended super world-sheet, i.e. the analogue of eq.(4.27)
including
terms proportional to $\z x$ and $\c x$. One should repeat the same
steps needed to fix the form (4.27) taking into account all the possible
new  terms. Since from our point of view the only relevance of such
an expression would be its use in the derivation of the classical
supercurrents,
we will confine ourselves to the terms involved in this derivation. Let us
note that the ``dilatonic'' terms will be enlarged to
$$
\Phi \Rdue + p_{+} T^{+} + p_{-} T^{-} +p\bud \Tbu +p\bud^x \Tbul x + p\cid
\Tcir
 + p\cid^x \Tcir x
\eqno(5.23)
$$
where (in perfect analogy with eq.(4.20)) $\Tbul x, \Tcir x$ are the
fermionic torsion two-forms relative to the new super world-sheet gravitinos:
$$
\Tbul x =d\z x -\unmezzo \omedue \z x \lsk ;\lsk \Tcir x =d\c x+\unmezzo
\omedue \c x
\eqno(5.24)
$$
Variations in the \first order fields $p$'s sets all the torsions to
zero. This allows the choice of an ``enlarged'' special superconformal
gauge (the extension of eq.(4.22)).\para
Variation in the two-dimensional spin-connection $\omedue$ yields
$$
\eqalign{
p_{+}=-2\sssk\partial_a \Phi \Vp a \lsk &;\lsk p_{-}=2\sssk
 \partial_a \Phi \Vm a \cr
p\bud=-4\sssk\partial_a\Phi\lai a\lsk &;\lsk p\cid=4\sssk
\partial_a\Phi \mui a\cr
p\bud^x=-4\sssk\partial_a\Phi(\cJ x\lambda)^a \lsk &; \lsk
p\cid^x=4\sssk\partial_a\Phi (\cpJ x \mu)^a
         }
\eqno(5.25)
$$
The fermionic torsion terms in (5.23) will contribute to the variation
of the action in the  new gravitinos, as it is seen from expression (5.24).
After variation we  make use of eqs.(5.25).\para
The supercurrents are obtained by obvious extensions of eqs.(4.30)
and following ones. Let
$$
\varia S=\o{-1}{2\pi}\int \cT_{+} \varia {e^+} + \cT_{-} \varia{e^-} +
\cT\bud
\varia{\zeta} + \cT\cid\varia{\chi}+\cT\bud^x\varia{\z x}+\cT\cid^x
\varia{\c x}
\eqno(5.26)
$$
Then superconformal invariance imposes on the  1-forms $\cT\bud^x$ and
$\cT\cid^x$
the analogue of conditions (4.31), namely:
$$
\cT_{+\bu}^x=\unmezzo\cT_{\bu +}^x \lsk ;\lsk\cT_{-\ci}^x=-\unmezzo
\cT_{\ci -}^x
$$
All the other components are zero.\para
Definition (4.32) is enlarged to include also the supercurrents $G^x$:
$$
G^x(z)=2\sqrt{2}e^{{-i\pi\over 4}}\cT_{+\bu}^x\lsk ;
\lsk \Gt^x(\zbar)=-2\sqrt{2} e^{-\o{3i\pi}{4}}
\cT_{-\ci}^x
\eqno(5.27)
$$
{}From the action (4.27) we can extract $G^0(z)=G(z)$ and $\Gt^0(\zbar)
=\wt G(\zbar)$. For example, to get $G^0\propto \cT_{\bu +}$ we vary  in
$\varia{\zeta}$ and  we look for the terms proportional to $e^+$; the
relevant terms are :
$$
\eqalign{
\varia S\sssk\rightarrow\sssk&\1su4p\intb\varia{\zeta}\left\{-\lai a
\Ppp a e^+ -\lai a \Vp a e^+
+\o 43 i T_{abc}\lai a\lai b\lai c \right\} +\varia (p\bud\Tbu)+...=\cr
&=\o{-1}{2\pi}\intb\varia{\zeta}\left\{\lai a\Vp a e^+ -\o 23 i T_{abc}
\lai a\lai b\lai c e^+ -\unmezzo \partial_+ p\bud e^+ +...\right\}
         }
$$
where we have integrated by parts the last term after use of the
definition (4.20). Using  eq.(5.25) we get
$$
\cT_{+\bu}=\unmezzo\cT_{\bu +}=\unmezzo\lai a\Vp a -\o i3 T_{abc}\lai a
\lai b\lai c +\partial_+ (\partial_r \Phi\lai r)
\eqno(5.28)
$$
so we finally  obtain the expression for $G^0=-2\sqrt{2}\emp4\cT_{+\bu}$.
In a similar way one obtains $\Gt^0(\zbar)$.\para
To derive the other supercurrents
we must analize the possible new terms that contribute to
the relevant variations, and fix their coefficients by comparing the
variational equations with the projections of the equation defining the target
torsion (4.24a).\para
For example to get $G^x(z)$ through the computation of $\cT_{\bu +}^x$
the relevant terms in the extended super world-sheet action are (compare with
eq.(4.27)):
$$
\eqalign{
S=&\intb(\V a-\lai a\zeta-(\cJ x\lambda)^a\z x-...)(\Ppp a e^+ -...)+
2i\lai a
\nabla\lai a e^+ +...+\lai a\V a\zeta +\cr
&+(\cJ x\lambda)^a\V a\z x+...
+\o 43 i T_{abc}\lai a\lai b\lai c\zeta e^+ + n_1 T_{abc}
(\cJ x\lambda)^a
\lai b\lai c\z x e^+ +...\cr
&+p\bud \Tbu +p\bud^x\Tbul x+...
         }
\eqno(5.29)
$$
 A priori, besides the term of the form $T({\cal J}\lambda)\lambda\lambda$,
 we could add to eq. (5.29) also two other kind
of terms, namely $T({\cal J}\lambda)({\cal J}\lambda)\lambda$ and
$T({\cal J}\lambda) ({\cal J}\lambda)({\cal J}\lambda)$. The reason why
it suffices to add only the first term is the vanishing of the \ni tensor.
Indeed the diagonal \nt constructed from $\cJ x$ or $\cpJ x$,
(see eq.(5.5)), upon use of the covariant constancy  condition $\delm_m \cJi
xab=0$, or
$\delp_m \cpJi xab=0$ can be rewritten as follows:
$$
N_{abc}({\cal J},{\cal J})=3T_{rm[a} {\cal J}_{rb} {\cal J}_{mc]}-T_{abc}
\eqno(5.30)
$$
(the antisymmetrization in  $abc$ is understood). By use of the \ni
condition $N_{abc}=0$ it is  easy to show that
$$
\eqalign{
&T({\cal J}\lambda)({\cal J}\lambda)\lambda \propto T\lambda\lambda
\lambda\cr
&T({\cal J}\lambda)({\cal J}\lambda)({\cal J}\lambda)
\propto T({\ca J}\lambda)\lambda\lambda
         }
$$
Hence there is only one coefficient  to fix in (5.29),  namely $n_1$. To obtain
its value, we consider the equation that follows from varying  the action
(5.29) in
$\varia \lai a$.  Focusing  on its   $\z x e^+$ sector and comparing with  the
$\z x\z x$
sector of the torsion definition (see eq.(5.3)),  we obtain
$$
n_1=4i
$$
 Varying now (5.29) in
$\varia\z x$ and searching for $\cT_{\bu +}^x$,  in analogy with the procedure
utilized for $G^0$, we get

$$
\cT\bud^x=\left\{{1\over 2}(\cJ x\lambda)^a\Ppp a e^+
+{1\over 2}(\cJ x\lambda)^a \Vp a e^+
-2iT_{abc}
(\cJ x\lambda)^a\lai b\lai c e^+  +\derp p\bud^x e^+ +...\right\}
$$
$$
\cT_{\bu +}^x=\left\{(\cJ x\lambda)^a\Vp a-2iT_{abc}(\cJ x\lambda)^a
\lai b \lai c +2\derp (\partial_a \Phi(\cJ x\lambda)^a)\right\}
$$
Thus $G^x(z)=2\sqrt{2}\emp4\cT_{+ \bu}^x$ is determined.\para
In a similar way one can calculate $\Gt^x(\zbar)$.
\para
\svsk
{\sl Summarizing:} {\it  when a $(1,1)$ supersymmetric \simod
described by the action (4.27) admits a global $(4,4)$ supersymmetry,
its classical supercurrents have the following expression in terms of the  3+3
complex
structures  of  $\Mtar$ }:
$$
\eqalign{
G^0(z)&=\sqrt{2}\emp4 \left\{\lai a \Vz a - \o 23 i T_{abc} \lai
a\lai b\lai c +2 \derp \left[\partial_a \Phi \lai a\right]\right\}\cr
G^x(z)&=\sqrt{2}\emp4 \left\{(\cJ x\lambda)^a \Vz a - 2i T_{abc}
(\cJ x\lambda)^a
\lai b\lai c +2 \derp \left[\partial_a \Phi (\cJ x\lambda)^a\right]
\right\}
         }
\eqno(5.31a)
$$
$$
\eqalign{
\Gt^0(\zbar)&=\sqrt{2} e^{-\o{3i\pi}{4}}
\left\{\mui a \Vzb a - \o 23 i T_{abc} \mui a
\mui b\mui c +2 \derm \left[\partial_a \Phi \mui a\right]\right\}\cr
\Gt^x(\zbar)&=\sqrt{2}e^{-\o{3i\pi}{4}}
\left\{(\cpJ x\mu)^a \Vzb a - 2i T_{abc} (\cpJ x\mu)^a
\mui b\mui c +2 \derm\left[\partial_a \Phi (\cpJ x\mu)^a\right]
\right\}
         }
\eqno(5.31b)
$$
\svsk


\centerline{\bf 6. The limit case of the  solvable $SU(2) \;\otimes \; R$
instanton.}
\svsk
Equipped with the general results of the previous sections, we now focus on the
$DRCHS$ instanton (3.28) and we consider the limit $A \; \longrightarrow \; 0$,
where asymptotic flatness is lost but superconformal solvability is gained.
This limit corresponds to the (1,1) locally supersymmetric \simod on the
background (4.9),
that can be advantegeously rewritten as in eq.s (4.10-4.12), leading to the
vielbein (4.13).
 As already pointed out, the \simod that emerges from this choice describes
the direct product
of a supersymmetric Feigin-Fuchs
model with a supersymmetric WZW model of the group $SU(2)$. This theory has
a (4,4) global supersymmetry since it satisfies all the conditions described in
the
previous sections. Let us see how this happens.
\svsk
\noi{\bf The classical \simod }
\svsk

Using the vierbein (4.13), we  can write the \mce of the group-manifold
\suR as follows:
$$
d\V a=\unmezzo \raduek \f abc\V b\V c \msk a=1,2,3,0
\eqno(6.1)
$$
where the totally antisymmetric structure constants $\f abc$ are given by
$$
\eqalign{
\f 0ab&=0\cr
\f ijk&=\epsd ijk
         }
\eqno(6.2)
$$
With these notations,  an  \suR element  in the adjoint representation is given
by the  $4 \times 4$ matrix
$$
\Gai ab=\twomat{\Gai ij}{0}{0}{1}
\eqno(6.3)
$$
 the $3\times 3$ submatrix $\Gai ij$ being an  \su element in its own   adjoint
representation.
 As such the matrix $\Ga$ has the properties that
$$
\eqalign{
\Ga^T \Ga&=\un\cr
(\Ga^T \sssk d\Ga)_{ab}&=\raduek \f abc\V c
         }
\eqno(6.4)
$$
In \second order formalism the action (4.29) for the (1,1)-supersymmetric
\simod
on a generic manifold is written as follows:
$$
\eqalign{
S=\meno1su4p\intb\misura &\left\{\Vz a\Vzb a +2i\lai a\delm_z \lai a -2i
\mui a \delp_{\zbar}\mui a -4\curv abcd \lai a\lai b\mui c\mui d
\right\} +\cr &+\1su4p\intm H
         }
\eqno(6.5)
$$
For our particular background, as for any other group-manifold, this
expression simplifies, due to the existence of two non-Riemannian
spin-connections (the ``zero'' and  the ``one'' connection
in  Cartan terminology [13])
that are proportional to the structure constants and that
parallelize the manifold. These two connections coincide exactly with
the  $\omem$ and $\omep$ discussed in the previous section.
Indeed, utilizing the expression of the torsion that follows from  eq.(6.1), we
find:
$$
\eqalign{
&\omemi ab=0\cr
&\omepi ab=\raduek \f abc\V c
         }
\eqno(6.6)
$$
so that
$$
\Rm ab =\Rp ab =0
\eqno(6.7)
$$
The ``minus'' covariant derivative is just an ordinary derivative, so
the fermions $\lai a$ are just free left-moving fermions. The $\mui a$,
instead, are neither free nor right-moving. However we can  rewrite the
action in terms of right-moving quantities, using  the 1-forms
$\Vt a =\Gamma^{ab} \V b$,
that provide an  alternative set of vielbein for our manifold.
They are given (compare with eq.(4.13)) by:
$$
\eqalign{
&\Vt 0 =dt\cr
&\Vt i =-\sqrt{k}\Omet i
         }
\eqno(6.8)
$$
where the forms $\Omet i$ are the components, along a Lie-algebra basis, of the
right-invariant form on the group  manifold: $\Omet =dg\ginv$. We expand these
right-moving vielbeins on the superworld-sheet as follows:
$$
\Vt a ={\Vt a}_+ e^+ +{\Vt a}_- e^- + \lati a e^+ +
\muti a e^-
$$
Relying on the relation
$$
\muti a =\Gamma^{ab} \mui b ,
\eqno(6.9)
$$
on the definition of $\delp$ and on the properties (6.4) of the adjoint matrix
we find that
$$
-2i\mui a\delp_z\mui a=-2i\muti a\derm\muti a
$$
Hence for group-manifolds the action (6.5) can be rewritten in such
a way that involves only free fermions:
$$
S=\meno1su4p\intb\misura\left\{\Vz a\Vzb a +2i \lai a\derp\lai a
-2i\muti a\derm\muti a\right\}+\o 1{24\pi} \raduek \intm \f abc
\V a\V b\V c
\eqno(6.10)
$$
On the four-dimensional group-manifold \suR it is  now easy to show that
the conditions for $(4,4)$ supersymmetry are matched.
Due to the vanishing of the $\omem$
connection, the set
of complex structures $\cJ x$ must be constant, and we can choose them to
coincide with the $\hJ x$ of eqs.(5.15-16):
$$
\cJ x=\hJ x
\eqno(6.11)
$$
The complex structures  $\cpJ x$, that commute with the previous set
and are covariantly constant
with respect to $\omep$  connection, are  given by
$$
\cpJ x=\Gamma^T\sssk\tJ x\sssk\Gamma
\eqno(6.12)
$$
This easily  follows from the properties of the adjoint matrix.
Substituting eqs.(6.11-6.12) into the general expression (5.31),
 we can write down
the explicit classical expression of  the
supercurrents in the case of the \suR background.
\para
Before doing this, we find it convenient to reformulate the theory in terms
of free fermions $\ps a(z) , \psit a(\zbar)$ that satisfy  the standard OPEs:
$$
\ps a(z)\ps b(w)=-\unmezzo\o{\delta^{ab}}{z-w}
\eqno(6.13)
$$
(and the same for the $\psit a(\zbar)$). This involves a simple
renormalization  of the original free fermions. Indeed from the classical
 Dirac brackets of the  fields $\lai a$ and the
 $\muti a$, that translate into their quantum OPEs, we have:
$$
\eqalign{
\lai a(z)\lai b(z) &=-\o i2\o{\delta^{ab}}{z-w}\cr
\muti a(\zbar)\muti b(\zbar) &=\o i2\o{\delta^{ab}}{z-w}
         }
$$
Hence it suffices to set:
$$
\lai a=e^{i\pi/4} \ps a\msk ;\msk\muti a=e^{i3\pi/4}\psit a
\eqno(6.14)
$$
For the left supercurrents, recalling the
form of the dilaton, (eq.(4.12)), which implies that
$$
\partial_a \Phi =\o{\delta_{a0}}{\sqrt{2k}}
$$
we immediately obtain the following expressions
$$
\eqalign{
G^0(z)&=\sqrt{2}\left\{\ps a\Vz a+\o 13\raduek\epsd ijk\ps i\ps j\ps k
+\raduek\derp\ps 0\right\}\cr
G^x(z)&=\sqrt{2}\left\{(\hJ x\psi)^a\Vz a+\raduek\epsd ijk
(\hJ x\psi)^i\ps j\ps k +\raduek\derp (\hJ x\psi)^0\right\}
         }
\eqno(6.15)
$$
For the right supercurrents, we must, first of all, give their expression
in terms of right-moving quantities. To this purpose it suffices
to make use of the properties (6.4) of the adjoint matrix and of the
additional one
$$
\f abc \Gai ar\Gai bs\Gai ct=\f rst
\eqno(6.16)
$$
corresponding to the invariance of the group structure constants.
These properties imply
$$
\eqalign{
\mui a\Vzb a=\muti a{\t V}^a_{\zbar}\msk &;\msk (\cpJ x\mu)^a
\Vzb a=(\tJ x\t\mu)^a{\t V}^a_{\zbar}\cr
\epsd ijk\mui i\mui j\mui k=\epsd ijk \muti i\muti j\muti k \msk &;
\msk \epsd ijk(\cpJ x\mu)^i\mui j\mui k=\epsd ijk (\tJ x\t\mu)^i
\muti j\muti k\cr
(\cpJ x\mu)^0&=(\tJ x\t\mu)^0
         }
$$
so that, in our case, from eq.(5.31) we obtain, in terms of the fermions $\psit
a$:
$$
\eqalign{
\Gt^0(\zbar)&=\sqrt{2}\left\{\psit a
{\t V}^a_{\zbar}-\o 13\raduek\epsd ijk\psit i\psit j\psit k
+\raduek\derm\psit 0\right\} \cr
\Gt^x(\zbar)&=\sqrt{2}\left\{(\tJ x\t\psi)^a
{\t V}^a_{\zbar}-\raduek\epsd ijk(\tJ x\t\psi)^i\psit j\psit k
+\raduek\derm (\tJ x\t\psi)^0\right\}
          }
\eqno(6.17)
$$


\svsk
\noi{\bf Quantization and abstract conformal field-theory }
\svsk
In the case of  supersymmetric WZW models [29],
 the  analysis of extended global SUSY can be also
performed in purely  algebraic terms; a complex structures is
in  one-to-one corrispondence
with a Cartan decomposition of the Lie algebra. The group $SU(2)
\times U(1)$ (this is our case) has actually three complex structures
and so $N$=4 SUSY follows.
We arrive at this algebraic description by quantizing  our theory.
\par
The quantization of the supersymmetric WZW on  any group manifold
and in particular on $SU(2)\times U(1)$ is
straightforward [13,23].
 Focusing on the left sector (we write the formulas for the right
sector only when some difference is present) and using the currents $J^a$
such that
$$\derp t=-i\sqrt 2J^0\eqno(6.18a)$$
$$\Omega^i={i\sqrt 2\over k}J^i\eqno(6.18b)$$
we find, as result of a standard procedure,
$$ J^i(z)J^j(w) = {k\over 2}{\delta^{ij}\over (z-w)^2} + {i
\epsilon^{ijk}J^k\over z-w}\eqno(6.19a)$$
$$J^0(z)J^0(w)={1\over 2(z-w)^2}\eqno(6.19b)$$
We will use also the notation $j^a=(J^0,\okp{J^i})$.\para
The correct quantum expression for the \se  includes the Sugawara form
for the level $k$ \su WZW model, and is given by
$$
 T(z)=J^0 J^0+\o 1{k+2}J^i J^i+\okp i \derp J^0+\ps a\derp \ps a
\eqno(6.20)
$$
Comparing this expression to eq.(4.17) we see that at quantum level a shift
$k\rightarrow k+2$ is necessary in the background charge term.
This shift of two unities
in the value of $k$ can be understood in the following way.\para
The term responsible for the background charge couples to the
supersymmetrized  version of the WZW-model at level $k$.
   From a purely algebraic point of view it is well known
that a super Kac-Moody  algebra of level $k$ corresponds to an ordinary bosonic
Kac-Moody algebra of level $k-C_V$ (where $C_V$ is the value of
the quadratic Casimir) plus a set of free fermions having regular OPEs with the
Kac-Moody currents.  The shift in $k$ is due to this fact: the relevant value
of $k$
 for the computation of the background charge is
the central charge of the super Kac-Moody currents:
$$j^{a}_{super}~=~j^{a} \; + \; const. \; f^{abc} \;\psi^{b} \;\psi^{c}$$
and not the central charge of the Kac-Moody currents $j^{a}$.\para
The central charge attribuited to the Feigin-Fuchs boson $t$ is shifted to
$c_{FF}=1+6/(k+2)$, the only value for which the total central charge sums up
to 6, the correct one for a four dimensional supersymmetric solution:
$$
c=c_{FF}+c_{WZW}+c_{ff}=1+\o 6{k+2}+\o{3k}{k+2}+2=6
\eqno(6.21)
$$
where $c_{WZW}$ is the ordinary central charge of the bosonic \su WZW at level
$k$ and $c_{ff}$=2 is the contribution of the four free fermions.\para
In other words we have a
\theory 64  in agreement with the general set up of section 2.
Note that the dilaton, not
necessary to obtain  $N$=4 supersymmetry at the classical level,
is essential at the quantum level to fix  the  central charge
to its correct value.\par
The quantum expressions of the supercurrents (the classical ones were given in
eq.(6.15-17))  are:
$$
\eqalign{
&G^0(z)=2\left\{-i j^a\ps a+\o 13 \okp{\epsd ijk}\ps i\ps j\ps k
+\okp{\derp \ps 0}\right\}\cr
&G^x(z)=2\left\{-i j^a\hJi xab\ps b+\okp{\epsd ijk}(\hJ x\psi)^i
\ps j\ps k+\okp{\derp (\hJ x\psi)^0}\right\}
         }
\eqno(6.22a)
$$
$$
\eqalign{
&\Gt^0(z)=2\left\{-i \t j^a\psit a-\o 13 \okp{\epsd ijk}\psit i\psit j\psit k
+\okp{\derp \psit 0}\right\}\cr
&\Gt^x(z)=2\left\{-i \t j^a\tJi xab\psit b-\okp{\epsd ijk}(\tJ x\t\psi)^i
\psit j\psit k+\okp{\derp (\tJ x\t\psi)^0}\right\}
         }
\eqno(6.22b)
$$
Without the dilatonic contributions (the last terms in the above eqs.), as
already stressed,
$N$=4 symmetry would be still present, but the supercurrents would not close
the standard algebra;
they would rather close the so called  $N$=4 extended algebra [30],
 based on the Kac-Moody algebra of $SU(2)\times SU(2)
\times U(1)$. The canonical way to reduce this extended algebra to
the standard one is to add a background charge with a particular value.
The solution we are considering automatically performs this reduction,
assigning the needed background charge to the field $t$.\para
The supercurrents (6.22) close thus the standard algebra, which requires $c$
to be a multiple of six:
$$T(z)T(w) = {c\over 2}{1\over (z-w)^4} + {2T(w)\over (z-w)^2}
 + {\partial T(w)\over (z-w)} + reg.\eqno(6.23a)$$
$$T(z)\cG a(w) = {3\over 2}{\cG a(w)\over (z-w)^2} + {\partial \cG a(w)
\over (z-w)} + reg.\eqno(6.23b)$$
$$T(z)\cGb a(w)
 = {3\over 2}{\cGb a(w)\over (z-w)^2} + {\partial\cGb a(w)
\over (z-w)} + reg.\eqno(6.23c)$$
$$T(z)A^i(w) = {A^i(w)\over (z-w)^2} + {\partial A^i(w)\over (z-w)} +
reg\eqno(6.23d).$$
$$A^i(z)\cG a(w) = {1\over 2}{\cG b(w)(\sigma^i)^{ba}\over (z-w)} +
reg.\eqno(6.23e)$$
$$A^i(z)\cGb a(w)
 = {-1\over 2}{\cGb b(w)(\sigma^i)^{ab}\over (z-w)} + reg\eqno(6.23f).$$
$$\cG a(z)\cGb b(w) = {2c\over 3}{\delta^{ab}\over (z-w)^3} +
{4(\sigma^*_i)^{ab}A^i(w)\over (z-w)^2} + {2\delta^{ab}T(w) + 2
\partial A^i(w)(\sigma^*_i)^{ab}\over (z-w)} + reg.\eqno(6.23g)$$
$$A^i(z)A^j(w) = {c\over 12} {\delta^{ij}\over (z-w)^2} +
{i\epsilon^{ijk}A^k(w)\over (z-w)} + reg.\eqno(6.23h)$$
The same holds for the right sector.\para
The $SU(2)_1$ currents of the two sectors are realized entirely in terms
of free fermions:
$$
\eqalign{
&A^i(z)=-\o i2\ps a\hJi iab\ps b=i(\ps 0\ps i+\um\epsu ijk\ps j\ps k)\cr
&\t A^i(z)=-\o i2\psit a\tJi iab\psit b=-i(\psit 0\psit i-\um\epsu ijk
\psit j\psit k)
         }
\eqno(6.24)
$$
i.e. they have the same expression
as for the flat space (see eq.(2.44c)), except that, due to
the non-vanishing torsion we are forced to use the two different sets
of complex structures in the two sectors.
The supercurrents $\cG a$, $\cGb a=(\cG a)^*$, organized in \su doublets
as dictated by the above OPEs, are given by
$$
\cGd=\o 1{\sqrt{2}}\twovec{G^0-iG^3}{-(G^2+iG^1)}\msk;\msk
\cGbd=\o 1{\sqrt{2}}\twovec{G^0+iG^3}{-(G^2-iG^1)}
\eqno(6.25)
$$
for the left sector, and by the same tilded expressions in the right
one.\para
Subsituting the explicit form (5.15-16) of the complex structures into
eqs.(6.22) we get
$$
\eqalign{
&G^0 = 2\left[-iJ^0\psi^0 -\okp i J^i\psi^i + {2\over \sqrt{k+2}}\psi^1
\psi^2\psi^3 + {\partial\psi^0\over \sqrt{k+2}}\right]\cr
&G^1 = 2\left[iJ^0\psi^1 -\okp i(J^1\psi^0 - J^2\psi^3 +J^3\psi^2)
 + {2\over \sqrt{k+2}}\psi^0
\psi^2\psi^3 - {\partial\psi^1\over \sqrt{k+2}}\right]\cr
&G^2 = 2\left[iJ^0\psi^2 -\okp i(J^1\psi^3 + J^2\psi^0 -J^3\psi^1)
 + {2\over\sqrt{k+2}}\psi^1
\psi^0\psi^3 - {\partial\psi^2\over \sqrt{k+2}}\right]\cr
&G^3 = 2\left[iJ^0\psi^3 -\okp i(-J^1\psi^2 + J^2\psi^1 +J^3\psi^0)
 + {2\over \sqrt{k+2}}\psi^1
\psi^2\psi^0 -{\partial\psi^3\over \sqrt{k+2}}\right]
          }
\eqno(6.26)
$$
while the $\t G$ have analogous but slightly different expressions.
This algebra was first obtained by Kounnas, Porrati and Rostand [31] and
used in this specific framework by Callan, Harvey and Strominger [6].\para
The doublets of supercurrents can be written as
$$
\eqalign{
\cGd=&\left[-i \sqrt{2} (j^0-i\sssk j^3)+2\raduekp i(\ps 0\ps 3-
\ps 1\ps 2)+\raduekp
\partial\right] \twovec{\ps 0+i\ps 3}{\ps 2+i\ps 1}+\cr
&-i\sqrt{2}(j^2+i\sssk j^1)\twovec{\ps 2-i\ps 1}{-(\ps 0-i\ps 3)}\cr
\cGbd=&(\cGd)^*
          }
\eqno(6.27a)
$$
and by
$$
\eqalign{
\cGtd=&\left[i \sqrt{2} (\t j^0+i\sssk \t  j^3)
+2\raduekp i(\psit 0\psit 3+\psit 1\psit 2)-\raduekp
\derm\right] \twovec{-(\psit 0-i\psit 3)}{\psit 2+i\psit 1}+\cr
&-i\sqrt{2}(\t j^2+i\sssk\t j^1)\twovec{\psit 2-i\psit 1}{\ps 0+i\psit 3)}\cr
\cGbtd=&(\cGtd)^*
          }
\eqno(6.27b)
$$
The relevant point is that we can easily obtain now the explicit form of the
moduli
operators for the conformal field theory we have just described.
We need primary fields of dimension
one which are the same time last  components of an $N$=4 representation,
namely we have to find solutions to the OPEs (2.43).
Remarkably, in our case the solution of these OPEs is very similar in form
to the solution (2.45) one obtains in the flat space case.
Indeed consider the SU(2) doublets:
$$
\Psi_1(z)=\emenradp\twovec{\ps 0+i\ps 3}{\ps 2+i\ps 1}\ssk ;\ssk
\Psi_2(z)=\emenradp\twovec{\ps 2-i\ps 1}{-(\ps 0-i\ps 3)}
\eqno(6.28a)
$$
$$
\t\Psi_1(\zbar)=\emenradpt\twovec{-(\psit 0-i\psit 3)}
{\psit 2+i\psit 1}\ssk ;\ssk
\t\Psi_2(\zbar)=\emenradpt\twovec{\psit 2-i\psit 1}{\ps 0+i\psit 3)}
\eqno(6.28b)
$$
These operators satisfy eq.s (2.43) with as last components the operators
$$\Phi_1(z)=\emenradp\left\{i\sqrt{2}(j^2+i j^1)+2\raduekp (\ps 0+i\ps 3)
(\ps 2+i\ps 1)\right\}$$
$$\Pi_1(z)=\emenradp\left\{i\sqrt{2}(j^0+i j^3)+2i\raduekp (\ps 0\ps 3-\ps 1\ps
2)
\right\}$$
$$\Phi_2(z)=\emenradp\left\{-i\sqrt{2}(j^0-i j^3)+2i\raduekp (\ps 0\ps 3-\ps
1\ps 2)
\right\}$$
$$\Pi_2(z)=\emenradp\left\{i\sqrt{2}(j^2-i j^1)+2\raduekp (\ps 0-i\ps 3)
(\ps 2-i\ps 1)\right\}\eqno(6.29a)$$
and
$$\t{\Phi}_1(z)=\emenradpt\left\{i\sqrt{2}(\t j^2+i \t j^1)+2\raduekp (\psit
0-i\psit 3)
(\psit 2+i\psit 1)\right\}$$
$$\t{\Pi}_1(z)=\emenradpt\left\{_i\sqrt{2}(\t j^0-i \t j^3)+2i\raduekp (\psit
0\psit 3
+\psit 1\psit 2)\right\}$$
$$\t{\Phi}_2(z)=\emenradpt\left\{i\sqrt{2}(\t j^0+i \t j^3)+2i\raduekp (\psit
0\psit 3
+\psit 1\psit 2)\right\}$$
$$\t{\Pi}_2(z)=\emenradpt\left\{i\sqrt{2}(\t j^2-i \t j^1)+2\raduekp (\psit
0+i\psit 3)
(\psit 2-i\psit 1)\right\}\eqno(6.29b)$$
Note that, as expected from the purely fermionic form of the currents
of
$SU(2)$, the doublets are quite completely expressed in terms of the
free fermions, the exponential term being only needed to cancel some
unwanted poles. We stress that, due to the existence of the background
charge, the operator of the F.F. theory
$$:\emenradp :\eqno(6.30)$$
has conformal dimension zero.
Indeed in a F.F. theory with \se $T(z)=-\um\derp t\derp t-\o i2
Q_{bk}\derp^2 t$ the vertex operators $:\exp (i\alpha t):$ have a
conformal weight $\Delta_{\alpha}=\um \alpha(\alpha +Q_{bk})$ and in our case
$Q_{bk}=-i\raduekp$.

This factor is the counterpart of the plane-wave factor $\eipx$ appearing
in the flat space case. Also there the exponential factor has conformal
weight zero since $k^2=0$. Indeed we can say that $k_0=\sqrt{\o{2}{k+2}}$
is the energy component of the four-momentum. It is fixed to a constant value
in terms of the space-like components ${\bf k}$.  The difference
resides in that ${\bf k}$ is a continuos variable for flat space, while its
analogue
 is quantized to fixed values for the \suR background, namely there is a finite
number
 of zero-mode operators rather then
a continuous infinity as in flat-space. This difference follows from the
different topology of the constant-time slices in the two cases:
noncompact ${\bf R}^3$ for flat-space, compact $S^3$ for the case under
consideration.
\par
The four fields $\Phi_a,\Pi_a,\ssk a=1,2$ are the moduli
of our conformal theory. Combining left and right fields, we find 16
infinitesimal deformations of our theory  that preserve the $N$=4
superconformal
algebra. These combinations are formally the same as the combinations
(2.45). Moreover it is possible to construct two abstract (0,1)-forms
$\Psi^{\star}_{\ca A^{\star}}\sp{0,\um}{0,\um}$, analogously to eq.(2.47).\para
As an abstract  \theory 64 -theory  the \suR background has
the same Hodge-diamond as flat space (compare with eq.(2.46)).
However since the torsion is different from zero, these abstract Hodge numbers
are
not the usual ones of compactified version of the underlying manifold
$S^1\times S^3$, whose Betti numbers
$$
b^0=1\ssk ,\ssk b^1=1\ssk ,\ssk b^2=0\ssk ,\ssk b^3=1\ssk ,\ssk b^4=1
$$
are obviously incompatible with such an Hodge decomposition.
\svsk


\centerline{\bf 7. Deformations of $\Mtar$ geometry in the \suR case }
\svsk
The existence of non-trivial $N$=4 moduli implies that the geometrical data
of the \simod, namely its metric $g_{\mu\nu}$  and torsion (related to the
axion
$B_{\mu\nu}$) can be deformed in such a way as to mantain $N$=4 supersymmetry.
In other words the existence of \ho 11 moduli implies that the generalized
HyperK\" ahler manifold we have
considered is just an element in a continuous family of generalized
HyperK\" ahler manifolds, parametrized by 4 \ho 11
parameters. For instance in the case of the $K_3$ manifold the existence of
20 $N$=4 moduli follows from the fact that, as an algebraic surface, $K_3$ is
described by a homogeneous
equation with 19 nontrivial complex coefficients fixing the
complex structure and that, for fixed complex structure, we  still have a one
parameter family
of deformations for the K\" ahler class.  These deformations of
the metric and of the torsion fill an  $80$-dimensional moduli space whose
global structure turns out to be
$ {\cal M}_{K_3}=SO(4,20)/SO(4)\times SO(20)/SO(4,20;{\bf Z})$. In a similar
way flat space has four $N$=4 moduli because
the constant metrics and constant torsions fills a space of dimension 16,
namely the space of all $4 \times 4$ matrices (the symmetric parts is the
metric,
the antisymmetric part is the axion).
\para
The geometrical interpretation of the moduli and  the knowledge of the moduli
space is very
important  because in the functional integral we are supposed to integrate over
all geometries.
In practice the use of the instanton conformal
field theory to calculate physical amplitudes is the following. Given a
scattering
process with $N$ external legs  ($i\;=\;1,.....,N$) there is an expression for
the emission vertex
of the $i$-th particle in each conformally flat background $Bk$ corresponding
to a specific \theory 64:
let us name this vertex $V_{Bk}(i)$. At every number of loops in string
perturbation theory the
true scattering amplitude is obtained by calculating the correlators for a
fixed bakground,
by integrating on the Riemann surface and on its moduli space and then by
summing over the backgrounds.
Schematically, if we disregard the Riemann surface integration we can write:
$$A(1,2,....,N)~=~\sum_{Bk} \; < V_{Bk}(1)\; V_{Bk}(2)\; ........\;
V_{Bk}(N)\;>$$
The sum on $Bk$ has a discrete and a continuous part. On one side we have to
sum over the various topologies, namely flat space and all the possible
instantons.
On the other side at fixed topology we have to integrate on the instanton
moduli
space.
\svsk
For the limit case of the \suR instanton we have discovered from the algebraic
approach that there are four $N$=4 moduli just as for flat space.
Their geometrical interpretation, however, is less clear. In this section we
explore the consequences
of the $N$=4 moduli on the geometry of the target space. Namely we calculate
the
explicit form of the infinitesimal deformations of the metric and of the
torsion
due to these moduli. We show that the deformed space is still generalized
HyperK\" ahler as expected:
the curvatures of $\omep$ and $\omem$ are no longer zero but still
self-dual (respectively antiselfdual) after the deformation and there exist
deformed complex structures fulfilling all the requirements.
A global characterization of
this space of metrics and torsions is still an open and interesting  problem.
Let us discuss the infinitesimal deformations obtained by inserting the moduli
operators in the \simod Lagrangian.
\svsk
We focus on the bosonic sector which suffices to give us informations
about the new metric, new torsion and  new complex structures. The bosonic
parts
of the moduli, reshifting the background charge to its classical value,
are expressed, for the left sector, in terms of the components of the
left-moving vielbeins:
$$
\eqalign{
\Phi_1(z)&=(\Vz 2+i\Vz 1)\emenrad\cr
\Pi_1(z)&=-(\Vz 0-i\Vz 3)\emenrad
         }
\lsk\msk
\eqalign{
\Phi_2(z)&=(\Vz 0+i\Vz 3)\emenrad\cr
\Pi_2(z)&=(\Vz 2-i\Vz 1)\emenrad
         }
\eqno(7.1a)
$$
and for the right sector, in terms of the right-moving ones:
$$
\eqalign{
\t\Phi_1(\zbar)&=(\Vtzb 2+i\Vtzb 1)\emenrad\cr
\t\Pi_1(\zbar)&=(\Vtzb 0+i\Vtzb 3)\emenrad
         }
\lsk\msk
\eqalign{
\t\Phi_2(\zbar)&=-(\Vtzb 0-i\Vtzb 3)\emenradt\cr
\t\Pi_2(\zbar)&=(\Vtzb 2-i\Vtzb 1)\emenradt
         }
\eqno(7.1b)
$$
Now we can construct conformal operators of weights $(1,1)$ to insert into
the Lagrangian combining these $(1,0)$ and $(0,1)$ ones in all
possible ways.
Hence  the most general expression we can add  to the Lagrangian is
simply
$$
e^{- \raduek t(\xp, \xm)} \Vz a \sssk M_{ab}\ssk\Vtzb b
\eqno(7.2)
$$
$M_{ab}$ being a constant matrix. The reality condition for this expression
imposes $M\in GL(4,\rea)$. Thus our deformations depend  on 16 real
parameters as anticipated from the abstract counting.
\para
In terms of the components of the undeformed vielbein, which
 we have chosen to be the left-moving ones, the term in (7.2)
has the form:
$$
\emenrad \Vz a (M \Ga)_{ab} \Vzb b
\eqno(7.3)
$$
$\Ga$ being the variable \suR element (point in the manifold) in the  adjoint
representation.
\para
It is useful to separate the  symmetric and antisymmetric part of the matrix
$M \Ga$ and to this purpose we introduce the notation
$$
\eqalign{
&h_{ab} = {1 \over 2}\emenrad (M \Ga + \Ga^T M^T)_{ab} \cr
&b_{ab} = -{1 \over 2}\emenrad (M \Ga - \Ga^T M^T)_{ab}
        }
\eqno(7.4)
$$
The overall normalization of the new term is of course irrelevant, since
$M$ is arbitrary, and we choose it in such a way  that the bosonic part of the
deformed \simod action is:
$$
S= \meno1su4p \int_{\partial \cM} d\xp d\xm \left\{\Vz a \Vzb a
+ 2 \Vz a h_{ab} \Vzb b - 2 \Vz a b_{ab} \Vzb b \right\}
+ \1su4p \int_\cM H
\eqno(7.5)
$$
The torsion deformation (parametrized by the antisymmetric matrix $b$)
can be recast in a shift of the 3-form $H$:
$$
S= \meno1su4p \int_{\partial \cM} d\xp d\xm \left\{\Vz a \Vzb a + 2 \Vz a
h_{ab} \Vzb b \right\} + \1su4p \int_\cM (H + \varia H)
\eqno(7.6)
$$
where $\delta H = dB$ with $B= b_{ab} \V a \V b$.\para
Following [32] it is also convenient to use the combinations:
$$
\eqalign{
& \Gpi ab = h_{ab} + b_{ab}=\emenrad (\Ga^T M^T)_{ab}=(\Gm )^T_{ab} \cr
& \Gmi ab = h_{ab} - b_{ab}=\emenrad (M \Ga)_{ab}
        }
\eqno(7.7)
$$
Relying on the properties of the adjoint matrix (see eq.s(6.4)) simple
expressions
are obtained for the derivatives of the above matrices:
$$
\eqalign{
&\partial_a \Gmi bc =\raduek(-\delta_{a0}\Gmi bc +\Gmi br \f rca)\cr
&\partial_a \Gpi bc =-\raduek(\delta_{a0}\Gpi bc +\f abr \Gpi rc)
        }
\eqno(7.8)
$$
\ssvsk
\noi So far we have  identified the deformation of the vielbein
(i.e. of the metric):
$$
{V'}^a = \V a + \varia {\V a} =\V a+  h_{ab} \V b
\eqno(7.9.a)
$$
and the components of the new torsion in the old basis which are given
(this follows from
the same supersimmetry variation argument as in the undeformed case)
as
$$
(T+ \delta T)_{abc} = -3(H + \delta H)_{abc}
\eqno(7.9.b)
$$
Now we must solve the relevant torsion equations for the two
non-Riemannian connections we are interested in,
these latter, in the undeformed situation, being given  by eq.(6.6)
The two torsion equations are, working at \first order in the moduli:
$$
d{V'}^a + ( \omepm + \varia {\omepm} )^{ab}\ssk {V'}^b = \mp (T')^a
\eqno(7.10)
$$
The solutions of eqs.(7.10) are given by
$$
\varia {\omepm_{ab|c}} = -2 \delpm _{[a} \Gpmi {b]}{c} \mp 4 \Gpmi
{[a}{r} T_{rb]c} - \delpm _c \Gpmi {[a}{b]} \pm 2 \Gpmi rc T_{rab}
\eqno(7.11)
$$
Making  the covariant derivatives explicit,
using eqs.(7.8) and the undeformed connections, we  finally get
$$
\varia {\omepm_{ab|c}} = \raduek \left\{ 2 \delta_{[a0} \Gpmi {b]}{c}
+ \delta_{c0} \Gpmi {[a}{b]} \pm \Gpmi {[a}{r} \f r{b]}{c} \pm
\Gpmi rc \f rab \right\}
\eqno(7.12)
$$
\ssvsk
\noi Next we look for the deformations of the associated curvatures.
{}From the general formula $\varia R=\nabla\varia{\omega}$
we have
$$
\eqalign{
&\varia {\Rm ab} = d \varia {\omemi ab} = (\partial_p
\varia {\omemti ab{|q}}
+ \unmezzo \raduek \varia {\omemti ab{|r}} \f rpq ) \V p \V q \cr
&\varia {\Rp ab} = (\partial_p \varia {\omepti ab{|q}}
+ \unmezzo \raduek \varia {\omepti ab{|r}} \f rpq
+2 \raduek \f {[a}{r}{p} \varia {\omepti r{b]}{|q}}) \V p \V q
         }
\eqno(7.13)
$$
Using eqs.(7.12) and (7.8), after some algebra one ends up with the
following results:
$$
\eqalign{
&\varia {\Rpm 0i} = - \o 2k \left\{ \Gpmi ij \V 0\V j \mp
\Gpmi il \epsd ljk \V j\V k\right\} \cr
&\varia {\Rpm jk} = \pm \epsd ijk \varia {\Rpm 0i}
         }
\eqno(7.14)
$$
We have that {\it the curvature of $\omep + \varia{\omep}$} (which is
$\varia {R^+}$) {\it is selfdual}, while {\it that of $\omem
+\varia{\omem}$} (which is $\varia{R^-}$){\it is antiselfdual}:
$$
\varia{\Rpm ab}=\pm\unmezzo\epsild abcd\varia{\Rpm cd}
\eqno(7.15)
$$
Recall that this is a necessary condition for the deformed $\Mtar$ to
have $N$=4 supersymmetry, as we discussed in sec.5.
\svsk
\noi{\bf Deformations of the Complex Structures}
\svsk
Having singled out the deformations of the vielbein, of the torsion and,
consequently, of the two non-Riemannian connections, our aim is now
to find  the deformations of the complex structures corresponding to
the insertion of the $N$=4 moduli in  the lagrangian. Indeed they must
exist since $N$=4 symmetry is mantained.
\para
Any infinitesimal deformation of one of the sets of complex structures
must be such that the quaternionic algebra is preserved:
$$
(\cJ x+\varia{\cJ x})(\cJ y+\varia{\cJ y})=-\delta^{xy}+\epsu xyz
(\cJ z+\varia{\cJ z})
$$
that is
$$
\varia{\cJ x}\cJ y +\cJ x\varia{\cJ y}=\epsu zxy\varia{\cJ z}
\eqno(7.16)
$$
The general ansatz solving this requirement is
$$
\varia{\cJ x}=\comm{\cJ x}{F}+\sum_z M_z\epsu zxy\cJ y
\eqno(7.17)
$$
$F$ being a generic (infinitesimal) matrix and $M_z$ generic
infinitesimal parameters.\para
We have to impose the ``deformed'' covariant-constancy conditions,
different for the two sets of complex structures relevant in  the left and in
the
right sector:
$$
\delpm\sssk\varia{\cpmJi xab}+2\varia{\omepmi{[a}{r}}\cpmJi xr{b]}=0
\eqno(7.18)
$$
Inserting the ansatz (7.17) with $M_z$=0 into eq.(7.18) we get
$$
\cpmJi x{[a}{r}(\delpm F^{\pm}-\varia{\omepm})_{rb]}=0
\eqno(7.19)
$$
Note that the deformations of the connections, (see eq.s(7.12)) can also be
written as
$$
\eqalign{
\varia{\omepi ab}&=-\delp\Gpi{[a}{b]}-\raduek{\hJ x}_{ab}
\Gpi xr\V r\cr
\varia{\omemi ab}&=-\delp\Gmi{[a}{b]}+\raduek{\tJ x}_{ab}
\Gmi xr\V r
         }
\eqno(7.20)
$$
where $\hJ x$ and $\tJ x$ are the two sets of constant complex structures
introduced in sec. 5. (see eq.(5.16)). Therefore if we start from
the ansatz (17) with
$$
F^{\pm}_{ab}=\Gpmi{[a}{b]}
\eqno(7.21)
$$
and $M_z$=0, the requirement (7.19) reduces to
$$
\eqalign{
&\cpJi x{[a}{r}\hJi yr{b]}\sssk\Gpi yr\sssk\V r=0\cr
&\cmJi x{[}{ar}\tJi yr{b]}\sssk\Gmi yr\sssk\V r=0
         }
\eqno(7.22)
$$
Recall that for our undeformed manifold, $\cmJ x\equiv\hJ x$. The above
equations hold then true due to the commutations relations (see sec. 5)
$$
\eqalign{
\left [ \; {\cpJ x} \; , \; {\cmJ y} \; \right ]&=0\cr
\left [ \; {\hJ x}\; , \; {\tJ y}\; \right ]&=0
         }
\lsk\forall x,y
\eqno(7.23)
$$
Summarizing, we have obtained that the deformations of the left- and
right-moving complex structures due to the insertion of the moduli in
the original $N$=4 theory are given by
$$
\varia{\cpmJ x}=\left [ \; {\cpmJ x}\; , \; {\pm b}J\; \right ]
\eqno(7.24)
$$
where
$$
b_{ab}=\mp\Gpmi{[a}{b]}=\emenrad(M\sssk\Ga-\Ga^T\sssk M^T)_{ab}
$$
\svsk
\noi{\bf Breaking of the old isometries}
\svsk
As in general the deformed curvatures differ (as forms) from zero, the effect
of the deformations cannot be trivially reabsorbed by a coordinate
change. The deformed space is actually a new kind of manifold: it is no
longer a group-manifold. Due to the exponential factor,
there is no longer a direct product
between a ``time'' coordinate and three ``spatial'' ones. The
``radius'' of the constant-time slices increases as $t\rightarrow
-\infty$; at the same time these slices get more and more deformed
respect to a three-sphere along some appropriate harmonics of the group
$SU(2)$ (recall the presence of the adjoint matrix in the deformed
expressions). The deformations of the ``radius'' and of the ``shape'' of
the constant-time slices interplay so as to mantain the properties
characterizing the space as  Generalized Hyper\K. The undeformed
situation (the ``tube'') is recovered as $t\rightarrow +\infty$.\para
In agreement with this,
we show now that apparently none of the isometries is conserved by the above
infinitesimal deformations. However, as stressed in the introduction, further
study is needed to discuss the possibility, for some particular choice
of the moduli, of modifying the old Killing vectors
in such a way to become Killing vectors of the new metric, corresponding to the
possibility of readsorbing the effect of thedeformation by coordinate changes.
\par
The group of isometries of a group-manifold $G$ is, for $G$ a non-abelian Lie
group,
$G\times G$, corresponding to the existence of two basis of Killing vectors,
the left-invariant
ones $\kvec A$, generating right translations, and the right-invariant ones
$\kvect A$, generating
left translations. The two sets are related by
$$
\kvect A=\Gai AB \kvec B
\eqno(7.25)
$$
$\Ga$ being as usual the adjoint matrix of the L.A. representing the group
element $g$.\para
The vector fields $\kvec A$ are dual to the group-manifold vierbeins:
\footnote*{\small We use the geometric formalism extensively developed, for
instance,
in [13]; we indicate in particular with $\scriptstyle{i_{{\bf k}}V}$ the
``contraction'' between vectors
and forms, and with $\scriptstyle{\ell_{{\bf k}}}$the Lie derivative along the
vector $\scriptstyle{{\bf k}}$ }
$$\contr{\kvec A}{\Ome B}=\delta_A^B\eqno(7.26)$$
For $G$ abelian, the two translations coincide, and the isometry group is
simply
$\rea$ or $U(1)$.\para
The isometry group of the manifold \suR is therefore
$$
SU(2)\times SU(2)\times \rea
$$
and it is generated by the Killing vectors $\kvec i, \kvect i,\ssk i=1,2,3$ and
$\kvec 0$, which
can be normalized so that their non-zero contraction with the vielbeins of the
manifold are
$$
\eqalign{
&\contr{\kvec i}{\V j}=\delta_{ij}\cr
&\contr{\kvect i}{\V j}=\Gai ij\cr
&\contr{\kvec 0}{\V 0}=1
         }
\eqno(7.27)
$$
To check explicitely that these vectors correspond to isometries of the
manifold it is sufficient
to compute the Lie derivative of the line element $
ds^2=\V a\otimes\V a$ along each of them,
finding in all cases that it vanishes.\para
To perform the computation one uses the fact that, from the formula for the
Lie-derivative
$$
\ell_{\bf k}V =d(\contr{{\bf k}}{V})+\contr{{\bf k}}{dV}
$$
one gets
$$
\eqalign{
&\ell_{\kvec 0}\V a=\ell_{\kvect i}\V a=\ell_{\kvec i}\V 0=0\cr
&\ell_{\kvec i}\V j=-\raduek\epsd ijk\V k
         }
\eqno(7.28)
$$
Now we raise the question whether any of these isometries remains an isometry
of the deformed
manifold. To see if this is the case, we have simply to compute the Lie
derivative along the above
Killing vectors of the deformed line element:
$$
ds'^2=ds^2+\varia{ds^2}=\V a\otimes\V a+\emenrad\V a\otimes(M\Ga+\Ga^T
M^T)_{ab}\V b
\eqno(7.29)
$$
By explicit computation we find:\para
$$\ell_{\kvec 0}\varia{ds^2}=-\raduek\emenrad (M\Ga +\Ga^T M^T)_{ab}\V
a\otimes\V b
\eqno(7.30a)$$
$$\ell_{\kvec i}\varia{ds^2}=-2\raduek\emenrad\epsd ilk (M\Ga)_{la}\V a
\otimes\V k
\eqno(7.30b)$$
$$\ell_{\kvect i}\varia{ds^2}=2\raduek\emenrad\epsd pni\Gai pj M_{an}\V
a\otimes\V j
\eqno(7.30c)$$
Since in general none of these expression vanishes, none of the isometries is
mantained after deformation.
This does not exclude that some modified Killing vectors exist, as discussed
before.
\svsk


\def\Psic#1{\Psi_{\ca #1}}
\centerline {\bf A. List of the vertices for a generic (6,6) solution}
\svsk
We list now the vertices that correspond to emission of particle zero-modes
for the various fields appearing in the effective 4-dimensional lagrangian
in terms of the conformal fields of the generic "instantonic'' CFTs
decribed in sec. 2.
The notations used in this list are essentially all explained in
sec.2, for what concernes the space-time operators, that in the
following expressions are distinguished by the square brackets.
The $\Psi_{\cal A}, \Psi^*_{\cal A^*}$ are the operators correspondent to
the $(1,0)$ and $(0,1)$ forms, $\Phi_{\cal A}, \Pi_{\cal A}$ (and the starred
analogues) the correspondent ``upper components''.
$\unhat$ is the identity times maybe the dimension zero operator that plays
the same role as
$e^{i k\cdot X(z,\bz)}$ in the flat space case, and whose presence
depends on the uncompactified geometry represented by the abstract \theory 64.

The operators in the internal theory are labeled by their left and right
conformal weights and $U(1)$ charges. In particular, the (chiral,
chiral) and (chiral, antichiral) fields $\Psi_k^{\pm}$ are lowest
components of short $N$=2 reps. and play the role of abstract (1,1) and
(2,1) forms of the compactifying \cy manifold.
The internal fields $\Omega_i$ are all the possible primary fields with
the specified weights and charges, including more than the
$\Psi_k^{\pm}$. We refer for more extensive
exposition of the notation to [13, sec.VI.10].\para
The branchings of the $SU(6)$ reps into $SO(6)\times SU(2)\times U(1)$
which explains how the $SU(6)$ index of the charged
vertices is reconstructed are indicated in the form
$$
rep_{SU(6)}=(rep_{SO(6)}\sssk,\sssk rep_{SU(2)}\sssk,\sssk \t q)
$$
(we omit the $U(1)$ charge $\t q$ when it is equal to zero).
We list the vertices referring to their interpretation as zero-modes of
the various $E_6$-neutral and $E_6$-charged fields of the effective
$d$=4 theory (see sec.2), just to facilitate the comparison with
the zero modes counting of that section; we count them for each
field of the effective lagrangian of a certain kind; the number of these fields
depends of course in the usual way from the topological numbers of the internal
CY manifold.\par
In the specific case of $K3$ the abstract (0,1)-forms
are not in the spectrum of the theory (\ho 01=0).
\ssvsk\noi
{\it Gravitational multiplet:}
\vskip 0.1cm
{\bf Graviton}
$$ e^{i\phi_{sg}(z)}\Phi_A\stwomat {1/2}{1}{1/2}{0}^a{\bf 1}\twomat
{0}{0}{0}{0}$$
$$ e^{i\phi_{sg}(z)}\Pi_A\stwomat {1/2}{1}{1/2}{0}^a{\bf 1}\twomat
{0}{0}{0}{0}$$
\vskip 0.1cm
\centerline{4\ho 11 zero modes}\par
\vskip 0.1cm
{\bf Gravitino}
$$ e^{{i\over 2}\phi_{sg}(z)}
\Phi_A\stwomat {1/4}{1}{0}{0}{\bf 1}\twomat
{3/8}{0}{-3/2}{0}$$
$$ e^{{i\over 2}\phi_{sg}(z)}
\Pi_A\stwomat {1/4}{1}{0}{0}{\bf 1}\twomat
{3/8}{0}{-3/2}{0}$$
$$ e^{{i\over 2}\phi_{sg}(z)}
\Phi^*_{\cal A^*}\stwomat {1/4}{1}{1/2}{0}
{\bf 1}\twomat
{3/8}{0}{3/2}{0}$$
$$ e^{{i\over 2}\phi_{sg}(z)}
{\bf 1}\sp {1/4}{1/2}^a
\Pi^*_{\cal A^*}\stwomat {1/4}{1}{1/2}{0}
{\bf 1}\twomat
{3/8}{0}{3/2}{0}$$
\vskip 0.1cm
\centerline{2\ho 11
 zero modes of (+) chirality and 4\ho 01 zero modes of (-) chirality}
\ssvsk
\noi{\it Neutral WZ multiplets}
\vskip 0.1cm
{\bf SU(6)-singlet scalars}
$$e^{i\phi_{sg}(z)} \unhat\stwomat {0}{0}{0}{0} \Omega_i
\twomat {1/2}{1}{1}{0}$$
\vskip 0.1cm
\centerline{One zero mode.}
{\bf SU(6)-singlet fermions}
$$e^{{i\over 2}\phi_{sg}(z)} \unhat\stwomat {1/4}{0}{1/2}{0}^a \Omega_i
\twomat {3/8}{1}{-1/2}{0}$$
$$e^{{i\over 2}\phi_{sg}(z)} \Psic A\stwomat{1/4}{0}{0}{0} \Omega_i
\twomat {3/8}{1}{1/2}{0}$$
\vskip 0.1cm
\centerline{2 zero modes of (-) chirality and \ho 10 of (+) chirality.}
\ssvsk\noi
{\it $E_6$ Gauge bosons:}
\vskip 0.1cm
{\bf Vertices in the 35 of SU(6), ``SU(6)-gauge bosons''}
$$
35 = (15,1) + (1,1) + (1,3) + (4,2,\t q=3/2) + (\overline 4,2,
\t q=-3/2)
$$
\vskip 0.1cm
$$ e^{{i\over 2}\phi_{sg}(z)} \Psic A\stwomat{1/2}{0}{1/2}{0}^a
{\bf 1}\twomat {0}{0}{0}{0} J^A(\overline z)$$
$$ e^{{i\over 2}\phi_{sg}(z)} \Psic A\stwomat{1/2}{0}{1/2}{0}^a
{\bf 1}\twomat {0}{0}{0}{0} \partial\phi(\overline z)$$
$$ e^{{i\over 2}\phi_{sg}(z)}\Psic A\stwomat{1/2}{0}{1/2}{0}^a
{\bf 1}\twomat {0}{0}{0}{0} {\t A}^i(\overline z)$$
$$ e^{{i\over 2}\phi_{sg}(z)}\Psic A\stwomat{1/2}{1/4}{1/2}{1/2}^{a
\t a}
{\bf 1}\twomat {0}{3/8}{0}{3/2} \Sigma_{\alpha} (\overline z)$$
$$ e^{{i\over 2}\phi_{sg}(z)}\Psic A\stwomat{1/2}{1/4}{1/2}{1/2}^{a
\t a}
{\bf 1}\twomat {0}{3/8}{0}{-3/2} \Sigma_{\dot\alpha} (\overline z)$$
\vskip 0.1cm
\centerline {2\ho 10 zero modes.}
\vskip 0.1cm
$J$ are the currents in the adjoint (15) of $SO(6)$, $\t A$ the currents
of the $SU(2)$ of the right $N$=4 algebra, $\partial \phi$ is
the internal $U(1)$ current expressed in term of a free boson.
\vskip 0.1cm
{\bf SU(6)-singlet scalars}
$$e^{i\phi_{sg}(z)} \Omega_i \stwomat{1/2}{1}{1}{0}
{\bf 1}\twomat {0}{0}{0}{0} $$
\vskip 0.1cm\noi
The $\Omega_i$ are all the ``space-time'' fields with the indicated weights
and isospins, so they are more than the $\Psi_A$ (see sec.2) and
their number should correspond to $\#End(T_{K3})$
\vskip 0.1cm
{\bf``Scalar'' vertices in the 20 of SU(6)}
$$
20 = (6,2) + (4,1,\t q=3/2) + (\overline 4,1,\t q=-3/2)
$$
\vskip 0.1cm
$$ e^{i\phi_{sg}(z)}
\Psi_A\stwomat {1/2}{1/2}{1/2}{1/2}^{a\tilde a}{\bf 1}\twomat
{0}{0}{0}{0} \theta_P(\overline z)$$
$$ e^{i\phi_{sg}(z)}
\Psi_A\stwomat {1/2}{1/4}{1/2}{0}^{a}{\bf 1}\twomat
{0}{3/8}{0}{3/2} \Sigma_{\alpha}(\overline z)$$
$$ e^{i\phi_{sg}(z)}
\Psi_A\stwomat {1/2}{1/4}{1/2}{0}^{a}{\bf 1}\twomat
{0}{3/8}{0}{-3/2} \Sigma_{\dot\alpha}(\overline z)$$
\vskip 0.1cm
\centerline{$2h^{(1,1)}$ zero modes.}
\vskip 0.1cm\noi
The heterotic fermions $\theta_P(\zbar)$ transform in the fundamental of
SO(6); $\Sigma_{\alpha}$ and $\Sigma_{\dot\alpha}$ are the SO(6) spin
fields of the two chiralities.
\ssvsk\noi
{\it $E_6$ Gauginos:}
\vskip 0.1cm
{\bf Vertices in the 35, ``SU(6) gauginos''}
$$ e^{{i\over 2}\phi_{sg}(z)} \unhat\stwomat {1/4}{0}{1/2}{0}^a{\bf 1}
\twomat {3/8}{0}{-3/2}{0} J^A(\overline z)$$
$$ e^{{i\over 2}\phi_{sg}(z)} \unhat\stwomat {1/4}{0}{1/2}{0}^a{\bf 1}
\twomat {3/8}{0}{-3/2}{0} \partial\phi(\overline z)$$
$$ e^{{i\over 2}\phi_{sg}(z)} \unhat\stwomat {1/4}{0}{1/2}{0}^a{\bf 1}
\twomat {3/8}{0}{-3/2}{0} {\t A}^i(\overline z)$$
$$ e^{{i\over 2}\phi_{sg}(z)} \unhat\stwomat
{1/4}{1/4}{1/2}{1/2}^{a\t a}
{\bf 1}
\twomat {3/8}{3/8}{-3/2}{3/2} \Sigma_{\alpha} (\overline z)$$
$$ e^{{i\over 2}\phi_{sg}(z)} \unhat
\stwomat {1/4}{1/4}{1/2}{1/2}^{a\t a}
{\bf 1}
\twomat {3/8}{3/8}{- 3/2}{-3/2} \Sigma_{\dot\alpha} (\overline z)$$
\vskip 0.1cm\noi
\centerline{Two zero modes of (-) chirality.}
$$
e^{{i\over 2}\phi_{sg}(z)} \Psic A\stwomat{1/4}{0}{0}{0}
{\bf 1}\twomat {3/8}{0}{3/2}{0} J^A(\overline z)
$$
\centerline{\hbox to 5cm{\dotfill}}
\centerline{\hbox to 5cm{\dotfill}}
\vskip 0.1cm\noi
\centerline{\ho 10 of (+) chirality.}
\vskip 0.1cm
{\bf SU(6) singlets fermions)}
$$e^{i\phi_{sg}(z)} \Omega_i \stwomat{1/4}{1}{0}{0}
{\bf 1}\twomat {3/8}{0}{3/2}{0} $$
\vskip 0.1cm
\centerline{
The numbers of these vertices is again related to $\# End(T_{K3})$.
            }
\vskip 0.1cm
{\bf Fermions in the 20 of SU(6)}
$$ e^{{i\over 2}\phi_{sg}(z)}
\Psi_A\stwomat {1/4}{1/2}{0}{1/2}^{\tilde a}{\bf 1}\twomat
{3/8}{0}{3/2}{0} \theta_P(\overline z)$$
$$ e^{{i\over 2}\phi_{sg}(z)}
\Psi_A\stwomat {1/4}{1/4}{0}{0} {\bf 1}\twomat
{3/8}{3/8}{3/2}{3/2} \Sigma_{\alpha}(\overline z)$$
$$ e^{{i\over 2}\phi_{sg}(z)}
\Psi_A\stwomat {1/4}{1/4}{0}{0} {\bf 1}\twomat
{3/8}{3/8}{3/2}{-3/2} \Sigma_{\dot\alpha}(\overline z)$$
\vskip 0.1cm
\centerline{\ho 11 zero modes, all of (+) chirality.}
\ssvsk\noi
{\it 27-charged Scalars:}
\vskip 0.1cm
{\bf Scalars in the 15 of SU(6)}
$$15 = (6,1,\t q=1) + (4,2,\t q=-\unmezzo) + (1,1,\t q=-2)$$
\vskip 0.1cm
$$ e^{i\phi_{sg}(z)} \unhat\stwomat {0}{0}{0}{0} \Psi_k^+
\twomat {1/2}{1/2}{1}{1} \theta_P (\overline z)$$
$$ e^{i\phi_{sg}(z)} \unhat\stwomat {0}{1/4}{0}{1/2}^{\tilde a}
 \Psi_k^+
\twomat {1/2}{3/8}{1}{-1/2} \Sigma_{\alpha}(\overline z)$$
$$ e^{i\phi_{sg}(z)} \unhat\stwomat {0}{0}{0}{0} \Psi_k^+
\twomat {1/2}{1}{1}{-2}$$
\vskip 0.1cm
\centerline{one zero mode.}
\vskip 0.1cm
{\bf scalars in the 6 of SU(6)}
$$6 = (1,2,\t q=1) + (4,1,\t q=-\unmezzo)$$\par
\vskip 0.1cm
$$ e^{i\phi_{sg}(z)} \Psi^*_{\cal A^*}
\stwomat{0}{1/2}{0}{1/2}^{\t a}\Psi_k^+
\twomat {1/2}{1/2}{1}{1} $$
$$ e^{i\phi_{sg}(z)}\Psi^*_{\cal A^*}
\stwomat{0}{1/4}{0}{0}
\Psi_k^+\twomat {1/2}{3/8}{1}{-1/2} \Sigma_{\alpha}(\overline z)$$

\vskip 0.1cm
\centerline {\ho 01 zero modes.}
\ssvsk\noi
{\it 27-charged fermions}
\vskip 0.1cm
{\bf fermions in the 15}
$$ e^{{i\over 2}\phi_{sg}(z)} \unhat\stwomat {1/4}{0}{1/2}{0}^a \Psi_k^+
\twomat {3/8}{1/2}{-1/2}{1} \theta_P (\overline z)$$
$$ e^{{i\over 2}\phi_{sg}(z)} \unhat\stwomat {1/4}{1/4}{1/2}{1/2}^{a\tilde a}
 \Psi_k^+
\twomat {3/8}{3/8}{-1/2}{-1/2} \Sigma_{\alpha}(\overline z)$$
$$ e^{{i\over 2}\phi_{sg}(z)} \unhat\stwomat {1/4}{0}{1/2}{0}^a \Psi_k^+
\twomat {3/8}{1}{-1/2}{-2}$$
\vskip 0.1cm
\centerline{Two zero modes of (-) chirality.}
\vskip 0.1cm
$$ e^{{i\over 2}\phi_{sg}(z)} \Psi_{\cal A}\stwomat {1/4}{0}{0}{0}
\Psi_k^+
\twomat {3/8}{1/2}{1/2}{1} \wt\theta_P (\overline z)$$
$$ e^{{i\over 2}\phi_{sg}(z)} \Psi_{\cal A}\stwomat {1/4}{1/4}{0}{1/2}^{\tilde
a}  \Psi_k^+
\twomat {3/8}{3/8}{1/2}{-1/2} \wt\Sigma_{\alpha}(\overline z)$$
$$ e^{{i\over 2}\phi_{sg}(z)} \Psi_{\cal A}\stwomat {1/4}{0}{0}{0} \Psi_k^+
\twomat {3/8}{1}{1/2}{-2}$$
\vskip 0.1cm
\centerline{\ho 10 zero modes of (+) chirality.}
\vskip 0.1cm
{\bf fermions in the 6}
$$ e^{{i\over 2}\phi_{sg}(z)} \Psi_A\stwomat {1/4}{1/2}{0}{1/2}^{\tilde
a}
\Psi_k^+
\twomat {3/8}{1/2}{1/2}{1} $$
$$ e^{{i\over 2}\phi_{sg}(z)} \Psi_A
\stwomat {1/4}{1/4}{0}{0}
 \Psi_k^+
\twomat {3/8}{3/8}{1/2}{-1/2} \Sigma_{\alpha}(\overline z)$$
\vskip 0.1cm
\centerline{\ho 11 zero modes of (+) chirality.}
\vskip 0.1cm
$$ e^{{i\over 2}\phi_{sg}(z)} \Psi^*_{\cal A^*}\stwomat {1/4}{1/2}{1/2}{1/2}^{
a\tilde a} \Psi_k^+
\twomat {3/8}{1/2}{-1/2}{1} $$
$$ e^{{i\over 2}\phi_{sg}(z)} \Psi^*_{\cal A^*}
\stwomat {1/4}{1/4}{1/2}{0}^a
 \Psi_k^+
\twomat {3/8}{3/8}{-1/2}{-1/2} \wt\Sigma_{\alpha}(\overline z)$$
\vskip 0.1cm
\centerline{2\ho 01 zero modes of (-) chirality.}
\ssvsk\noi
{\it $\overline {27}$-charged Scalars:}
\vskip 0.1cm
{\bf scalars in the $\overline{15}$ of SU(6)}
$$\overline{15} = (6,1,\t q=-1) + (\overline 4,2,\t q=\unmezzo)
+ (1,1,\t q=2)$$
\vskip 0.1cm
$$ e^{i\phi_{sg}(z)} \unhat\stwomat {0}{0}{0}{0} \Psi_k^-
\twomat {1/2}{1/2}{1}{-1} \theta_P (\overline z)$$
$$ e^{i\phi_{sg}(z)} \unhat\stwomat {0}{1/4}{0}{1/2}^{\tilde a}
 \Psi_k^-
\twomat {1/2}{3/8}{1}{1/2} \Sigma_{\dot\alpha}(\overline z)$$
$$ e^{i\phi_{sg}(z)} \unhat\stwomat {0}{0}{0}{0} \Psi_k^-
\twomat {1/2}{1}{1}{2}$$
\vskip 0.1cm
\centerline{one zero mode.}
\vskip 0.1cm
{\bf scalars in the $\overline 6$ of SU(6)}
$$6 = (1,2,\t q=-1) + (\overline 4,1,\t q=\unmezzo)$$\par
\vskip 0.1cm
$$ e^{{i\over 2}\phi_{sg}(z)} \Psi^*_{\cal A^*}
\stwomat{0}{1/2}{0}{1/2}^{\t a}
\Psi_k^-\twomat {3/8}{1/2}{1/2}{-1} $$
$$ e^{{i\over 2}\phi_{sg}(z)}\Psi^*_{\cal A^*}
\stwomat{0}{1/4}{0}{0}
\Psi_k^-\twomat {3/8}{3/8}{1/2}{1/2} \Sigma_{\alpha}(\overline z)$$
\vskip 0.1cm
\centerline {\ho 01 zero modes}
\ssvsk\noi
{\it $\overline{27}$-charged fermions:}
\vskip 0.1cm
{\bf fermions in the $\overline{15}$}
$$ e^{{i\over 2}\phi_{sg}(z)} \unhat\stwomat {1/4}{0}{1/2}{0}^a
\Psi_k^-
\twomat {3/8}{1/2}{-1/2}{-1} \theta_P (\overline z)$$
$$ e^{{i\over 2}\phi_{sg}(z)} \unhat
\stwomat {1/4}{1/4}{1/2}{1/2}^{a\tilde a}\Psi_k^-
\twomat {3/8}{3/8}{-1/2}{1/2} \Sigma_{\alpha}(\overline z)$$
$$ e^{{i\over 2}\phi_{sg}(z)} \unhat\stwomat {1/4}{0}{1/2}{0}^a
\Psi_k^- \twomat {3/8}{1}{-1/2}{2}$$
\vskip 0.1cm
\centerline{Two zero modes of (-) chirality.}
\vskip 0.1cm
$$ e^{{i\over 2}\phi_{sg}(z)} \Psi_{\cal A}\stwomat {1/4}{0}{0}{0}
\Psi_k^-
\twomat {3/8}{1/2}{1/2}{-1} \wt\theta_P (\overline z)$$
$$ e^{{i\over 2}\phi_{sg}(z)} \Psi_{\cal A}\stwomat {1/4}{1/4}{0}{1/2}^{\tilde
a}  \Psi_k^-
\twomat {3/8}{3/8}{1/2}{1/2} \wt\Sigma_{\alpha}(\overline z)$$
$$ e^{{i\over 2}\phi_{sg}(z)} \Psi_{\cal A}\stwomat {1/4}{0}{0}{0}
\Psi_k^-
\twomat {3/8}{1}{1/2}{2}$$
\vskip 0.1cm
\centerline{\ho 10 zero modes of (+) chirality.}
\vskip 0.1cm
{\bf fermions in the $\overline 6$}
$$ e^{{i\over 2}\phi_{sg}(z)} \Psi_A\stwomat {1/4}{1/2}{0}{1/2}^{\tilde
a} \Psi_k^-
\twomat {3/8}{1/2}{1/2}{-1} $$
$$ e^{{i\over 2}\phi_{sg}(z)} \Psi_A
\stwomat {1/4}{1/4}{0}{0}\Psi_k^-
\twomat {3/8}{3/8}{1/2}{1/2} \Sigma_{\alpha}(\overline z)$$
\vskip 0.1cm
\centerline{\ho 11 zero modes of (+) chirality.}
\vskip 0.1cm
$$ e^{{i\over 2}\phi_{sg}(z)} \Psi^*_{\cal A^*}\stwomat {1/4}{1/2}{1/2}{1/2}^{
a\tilde a} \Psi_k^-
\twomat {3/8}{1/2}{-1/2}{-1} $$
$$ e^{{i\over 2}\phi_{sg}(z)} \Psi^*_{\cal A^*}
\stwomat {1/4}{1/4}{1/2}{0}^a
 \Psi_k^-
\twomat {3/8}{3/8}{-1/2}{1/2} \wt\Sigma_{\alpha}(\overline z)$$
\vskip 0.1cm
\centerline{2\ho 01 zero modes of (-) chirality.}
\vskip 1cm
\centerline {REFERENCES}
\def\pvn{P.van Nieuwenhuizen~}
\svsk
\item{[1]\sssk}{T.Eguchi, P.B.Gilkey, and A.J.Hanson, Phys. Rep. {\bf 66}
(1980) 213.}
\item{[2]\sssk}{T.Eguchi, A.J.Hanson Phys. Lett. {\bf 74B}(1978) 249; Ann.
Phys. {\bf 120} (1979) 82}
\item{[3]\sssk}{K.Konishi, N.Magnoli and H.Panagopoulis, Nucl. Phys. {\bf B309}
(1988) 201, Nucl. Phys {\bf B323} (1989) 441.}
\item{[4]\sssk}{E.Witten, Phys. Rev {\bf D44} (1991) 314; R.Dijkgraf,
E.Verlinde
and H.Verlinde {\it ``String propagation in a Black Hole geometry''} PUPT-1252;
S.Elitzur, A.Forge, E.Rabinovici {\it ``Some Global aspects of String
Compactifications''} RI-143/90; G.Mandal, A.Sengupta and S.Wadia {\it
``Classical
solutions of 2D string theory''} IASSNS-HEP-91/10; M.Rocek, K.Schoutens,
A.Sevrin {\it ``Off-Shell WZW Models in Extended Superspace''}
IASSNS-HEP-91/14; A.Giveon Mod. Phys. Lett. {\bf A6} No 31 (1991) 2843.}
\item{[5]\sssk}{M.Rocek, C.Ahn, K.Schoutens, A.Sevrin {\it ``Superspace WZW
Models and
Black Holes''} IASSNS-HEP-91/69; P.Ginsparg, F.Quevedo, {\it ``Strings on
Curved
Space-Time: Black Holes, Torsion, and Duality''} LA-UR-92-640,
hepth@xxx/9202092;
C.Kounnas, D.Lust {\it ``Cosmological String Backgrounds from Gauged WZW
Models''}
CERN-TH.6494/92}
\item{[6]\sssk}{C.G.Callan, J.A.Harvey, A.Strominger Nucl. Phys.
{\bf B359} (1991) 611\hfill;
C.G.Callan, Lectures at Sixth J.A.Swieca summer school, PUPT-1278.}
\item{[7]\sssk}{R.d'Auria, T.Regge Nucl. Phys. {\bf B195} 308  }
\item{[8]\sssk}{S.J.Rey, Phys. Rev. {\bf D43} (1991) 526; {\it ``On String
Theory and
Axionic Strings and Instantons''} Slac-PUB-5659 (Sept. 1991).}
\item{[9]\sssk}{P.Candelas, G.Horowitz, A.Strominger, E.Witten Nucl. Phys. {\bf
B258}
(1985) 46; For a review see M.Green, J.Schwarz, E.Witten {\it ``Superstring
Theory'' (Cambridge U.P.) Cambridge , 1987.}

\item{[10]\sssk}{F.Englert, H.Nicolai,
A.N.Schellekens, Nucl. Phys. {\bf B 274} (1986) 315; W.Lerche, D.Luest,
A.N.Schellekens, Nucl. Phys. {\bf B 287} (1987) 447, Phys Lett {\bf B
187} (1987) 45; D.Gepner, Trieste Lectures 1989; Nucl. Phys.
{\bf B296} (1988) 757; Phys. Lett. {\bf B199} (1987) 380.}

\item{[11]\sssk}{D.J.Gross, J.A.Harvey, E.Martinec and R.Rohm Nucl. Phys.
{\bf B256} (1985) 253.}

\item{[12]\sssk}{E.Verlinde, H.Verlinde Nucl. Phys. {\bf B288} (1987) 357.}

\item{[13]\sssk}{L.Castellani, R.d'Auria and P.Fr\'e, {\it ``Supergravity and
Superstrings: A geometric Perspective"} (World Scientific ed.), 1990}
\item{[14]\sssk} {T.Eguchi, H.Ooguri, A.Taormina and S.K.Yang, Nucl. Phys.
{\bf B315} (1989) 193.}
\item{[15]\sssk}{E.Cremmer, S.Ferrara, L. Girardello, A.van Proeyen Nucl.
Phys. {\bf B212} 413; Phys. Lett. {\bf 116B} (1982) 231 }
\item{[16]\sssk}{R. Barbieri, S.Ferrara, D.V.Nanopoulos, K.S. Stelle
Phys. Lett. {\bf 113B} (1982) 219 }
\item{[17]\sssk}{S.Cecotti, S.Ferrara, L.Girardello Int. Jour. Mod. Phys. {\bf
A10}
(1989) 2475;\hfill A.Strominger, Phys. Rev. Lett. {\bf 55} (1985) 2547;
A.Strominger, E.Witten Comm. Math. Phys. {\bf 101} (1985) 341; P.Candelas
Nucl. Phys. {\bf B298} (1988) 458.}
\item{[18]\sssk}{L.Dixon, V.S.Kaplunovski, J.Louis Nucl. Phys. {\bf B329}
(1990) 27.}
\item{[19]\sssk}{S. Ferrara, \pvn Phys. Lett. {\bf 74B}
(1978) 333; {\bf 76B} (1978) 404; {\bf 78b} (1978) 573; K.S.Stelle,
P.C.West Phys. Lett. {\bf 74B} (1978) 330; {\bf 77B} (1978) 376;
S. Ferrara, L.Girardello, T.Kugo, A.van Proyen Nucl. Phys. {\bf B233} (1983)
191. }
\item{[20]\sssk}{V.P. Akulov, D.V.Volkov and V.A.Suroka Teor. Math. Phys. {\bf
31} (1977) 285; M.F.Sohnius, P.C.West Phys. Lett. {\bf 105B} (1981) 353; see
also
R.D'Auria, P.Fr\'e, \pvn, P.K.Townsend Ann. Phys {\bf 155} (1984) 423 }
\item{[21]\sssk}{R.D'Auria, P.Fr\'e, G.De Matteis, I.Pesando, Int. Jour.
Mod. Phys. {\bf A4} No14 (1989) 3577 }
\item{[22]\sssk}{B.A.Ovrut, S.Kalyana Rama, Phys. Lett. {\bf B245}(1990)
429:; see also Nucl. Phys {\bf B343} (1990) 86. }
\item{[23]\sssk}{L.Castellani,R.d'Auria and D.Franco, Int.Jour.Mod.Phys.
{\bf A6} 4009}
\item{[24]\sssk}{E.Witten, Comm. Math. Phys. {\bf 92} (1984) 455}
\item{[25]\sssk}{V.Dotsenko and V. Fateev, Nucl.Phys. {\bf B240}
(1984) 312}
\item{[26]\sssk}{I.Antoniadis, C. Bachas, J.Ellis and D.Nanopoulos,
Phys. Lett. B211 (1988) 393; Nucl.Phys. {\bf B328} (1989) 117}
\item{
[27]\sssk}{L.Alvarez-Gaum\'e and D.Z.Freedman, Comm. Math. Phys.
{\bf 80} (1981) 443;  S.J.Gates, C.M.Hull, M.Rocek Nucl. Phys.
{\bf B248} (1984) 157;
P.S.Howe and G.Papadopoulos, Nucl. Phys. {\bf B289} (1987) 264,
Class. Quant. Grav. {\bf 5} (1988) 1647}
\item{[28]\sssk}{K.Galicki Comm. Math. Phys. {\bf 108} (1987) 117}
\item{[29]\sssk}{Ph.Spindel, A.Sevrin, W.Troost, A.Van Proeyen
Phys.Lett. {\bf B206} (1988) 71, Nucl.Phys. {\bf B308} (1988) 662}
\item{[30]\sssk}{A.Sevrin,W.Troost and A.Van Proeyen, Phys.Lett.
{\bf B208} (1988) 447}
\item{[31]\sssk}{C.Kounnas, M.Porrati and B.Rostand, Phys. Lett. {\bf B258}
(1991) 61}
\item{[32]\sssk}{R.d'Auria, P.Fr\'e , F.Gliozzi and A.Pasquinucci,
Nucl.Phys. {\bf B334} (1990) 24.}
\bye